\documentclass{aa}  

\usepackage[a-2u]{pdfx}
\usepackage[utf8]{inputenc}

\usepackage{graphicx}
\usepackage{txfonts}
\usepackage{siunitx}
\usepackage{multirow}
\usepackage{aas_macros}
\usepackage{natbib}

\newcommand{\Rnom}{\ifmmode{\rm R}_\odot\else R$_{\odot}$\fi}
\newcommand{\Mnom}{\ifmmode{\rm M}_\odot\else M$_{\odot}$\fi}
\newcommand{\Lnom}{\ifmmode{\rm L}_\odot\else L$_{\odot}$\fi}

\usepackage{hyperref}
\usepackage{csquotes}

\hypersetup{
    colorlinks=true,        
    linkcolor=blue,        
    citecolor=blue,        
    urlcolor=blue
}

\usepackage{xparse}

\NewDocumentCommand{\scnum}{ >{\SplitArgument{1}{e}}m }
  {\scnumaux#1}
\NewDocumentCommand{\scnumaux}{ m m }
  {$#1\,\mathrm{e}{#2}$}

\usepackage{txfonts}

\newcommand{\ks}{km\,s$^{-1}$}

\begin{document} 

\title{VLTI observations of Orion Belt stars}
\subtitle{I. $\varepsilon$~Orionis
\thanks{Based on ESO proposal 112.25JX, PI A. Oplištilová}
}

\author{
A.~Opli\v{s}tilov\'a\inst{1}\and
M.~Bro\v{z}\inst{1}\and
C.~A.~Hummel\inst{2}\and
P.~Harmanec\inst{1}\and
B.~N.~Barlow\inst{3}
}

\offprints{A.O.,\\ \email{alzbeta.oplistilova@gmail.com}}

\institute{
Charles University, Faculty of Mathematics and Physics, Astronomical Institute, 
V~Hole\v{s}ovi\v{c}k\'ach~2, CZ-180 00 Praha~8-Tr\'oja, Czech Republic
\and
European Southern Observatory, Karl-Schwarzschild-Str. 2, D-85748 Garching, Germany
\and
Department of Physics and Astronomy, University of North Carolina at Chapel Hill, Chapel Hill, NC 27599, USA}

\date{Received 27 June 2025}

\abstract
{   
Massive stars play a decisive role in the evolution of the Universe.
They are the primary sources of ionising radiation, generating strong stellar winds that affect the interstellar medium. They
 ultimately end their lives as supernovae, ejecting synthesised, r-process elements.
}
{
To constrain their current state and structure,
we need access to sufficiently complex models,
constrained by astrometric, interferometric, and spectroscopic observations.
However, such tools are not available for distant stars.
Therefore, we focused on the nearest massive stars in  Orion's Belt for the purposes of this work.
}
{
We obtained VLTI interferometric observations
of the stars of Orion's Belt and
calibrated visibility data from the GRAVITY and PIONIER instruments.
Additionally, we obtained spectroscopic data
from the CFHT and CTIO observatories.
For the modelling, we used a modified version of PHOEBE2,
extended with new interferometric and spectroscopic modules.
To describe non-spherical, rotating, or Roche-like stars,
we needed to compute integrals over triangular meshes,
using extensive grids of synthetic spectra
(OSTAR, BSTAR, ATLAS).
For the fitting, we used the simplex algorithm and $\chi^2$ mapping 
of the parameter space.
}
{
In this paper, the first in a series,
we present single-star models of the B0\,Ia supergiant $\varepsilon$~Ori.
Interferometric visibilities indicate that the star is not spherical,
but it is rotating close to its critical velocity.
The preferred distance,
$d = (384\pm 8)\,{\rm pc}$,
corresponds to the median of distances for 
the Orion OB1b association.
Specifically,
we obtained the following parameters:
mass of $m = (28.4\pm 2.0)\,\Mnom$,
equivalent radius of $R = (27.6\pm 1.5)\,\Rnom$, where the polar and equatorial values are
22.3\,\Rnom\ and 33.6\,\Rnom, respectively,
effective temperature of $T_\mathrm{eff} \simeq 25\,000\,{\rm K}$,
inclination of the rotation axis of $i \simeq 45\,{\rm deg}$,
longitude of the ascending node (of the equator) of $\Omega \simeq 300\,{\rm deg}$, and
a period of $P_\mathrm{rot} = 4.3^{+1.0}_{-0.0}\,{\rm d}$.
This \enquote*{compromise} model provides a reasonable fit to wind-free Balmer line profiles
(H$\gamma$, H$\delta$, H$\varepsilon$, etc.). However, there is still some tension between the interferometric and spectroscopic datasets when comparing a~faster rotating star versus a slower one.
}
{
Our fast-rotating model implies that circumstellar matter should naturally be present,
taking the form of wind or disk,
and ought to contribute to the continuum radiation.
The fast rotation of $\varepsilon$~Ori is compatible with a merger,
formed from a multiple system of comparable mass,
such as $\delta$, $\zeta$, or $\sigma$~Ori.
}

\keywords{
Stars: massive -- 
Techniques: interferometric -- 
Stars: individual: $\varepsilon$~Ori --
Stars: individual: Orion's Belt
}

\maketitle

\section{OB stars in  Orion's Belt}

O- and B-type (OB) stars have relatively short lifetimes, lasting only 
a few million or tens of millions of years,
yet they are crucial cosmic engines
with a long-lasting influence on the evolution of the Universe. 
They exert strong feedback on their environment,
not only as
sources of energy, 
ionising photons,
chemical elements through winds or 
core-collapse supernova (SN) explosions \citep{Clayton1983psen.book.....C}, but 
also as precursors of neutron stars, black holes, and, consequently, gravitational wave sources \citep{Marchant2021A&A...650A.107M}.

The most likely precursors include
overcontact binaries \citep{Almeida2015ApJ...812..102A},
X-ray binaries \citep{Atri2019MNRAS.489.3116A},
quiescent binaries \citep{Shenar2022NatAs...6.1085S}, or
pair-instability SNe \citep{Renzo2022RNAAS...6...25R}.
Various formation channels must be constrained by photonic observations.
The distribution of black hole masses \citep{Abbott_2021}
indicates a hierarchical merging.
This is complicated by \enquote*{kicks}
of the order of 100\,km/s,
during explosions \citep{Wongwathanarat2013A&A...552A.126W}
or mergers \citep{Shenar2022NatAs...6.1085S}.
The kicks can be compensated by
the escape velocity from nuclear clusters or
the escape velocity from multiple systems,
where the radial velocities (RVs) of components are comparable to the kick velocities.

One particularly characteristic aspect of massive stars is their multiplicity:
90\% of massive, hot OB stars possess at least one stellar companion
\citep{Sana2014ApJS..215...15S,
Sota2014ApJS..211...10S,
Apellaniz2019A&A...626A..20M,
Pauwels2023A&A...678A.172P}, 
which influences their evolution throughout their lives.
Yet, some stars remain single (e.g. $\varepsilon$~Ori),
even though they share the same birth environment with multiple systems.
Considering the possible interactions among stars,
they might be the result of a more complicated evolutionary process, including
mass transfer or
mergers resulting from close encounters between stars.

The evolution of massive stars still remains poorly constrained by observations. 
The primary kind of observation constraining diameters is long-baseline interferometry. 
While the most massive stars in the Large Magellanic Cloud,
such as BAT99-98 with $M{\approx}\,200\,\Mnom$ located in the Tarantula nebula 
\citep[30 Dor,][]{Hainich2014A&A...565A..27H,Kalari2022ApJ...935..162K},
are too distant (50\,kpc)
for interferometric measurements,
the Orion OB1 association
at about 400\,pc from us seems to be a perfect region
for the investigation of questions related
to the birth of OB stars and their evolution.

The brightest stars in  Orion's Belt, members of the OB1b Orion association, 
$\delta$~Ori, $\zeta$~Ori, $\sigma$~Ori, and $\varepsilon$~Ori
with individual component masses of up to $35\,\Mnom$
\citep{Hummel2013,
Schaefer2016AJ....152..213S,
Puebla2016MNRAS.456.2907P,
Oplistilova2023AA...672A..31O}
represent massive stars that are accessible to the most advanced 
optical interferometers.
The parallaxes of faint stars surrounding the aforementioned bright stars
were measured in Gaia DR3 \citep{Brown2021A&A...649A...1G};
the corresponding distances are all around 0.38\,kpc
\citep{Oplistilova2023AA...672A..31O}.
It is thus possible to measure their angular separations
and angular diameters.

In this first paper, we focus on
$\varepsilon$~Ori (Alnilam, 46~Ori, HD\,37128),
which is the largest and brightest star 
in  Orion's Belt.
Based on photometric measurements taken 
at the Hvar Observatory 
\citep{Hvar}, 
its standard magnitudes are:
$V=1.691$\,mag,
$B=1.509$\,mag, and
$U=0.495$\,mag.
In the near-infrared (NIR, the passband of PIONIER), the magnitude is
$H=2.07$\,mag \citep{2002yCat.2237....0D}.
It is classified as 
a B0\,Ia blue supergiant and
represents the only massive single star in Orion's Belt.
Its mass of about 30\,\Mnom\
\citep{Puebla2016MNRAS.456.2907P}
is similar to the total masses of the other multiple systems 
in Orion.
Also, it has an intense wind
\citep{Puebla2016MNRAS.456.2907P}
and a mass-loss rate of up to
$\Dot{M} = 5.25 \cdot 10^{-6}$\,\Mnom\,yr$^{-1}$ 
\citep{2005A&A...440..261R}.
It might be the case that
$\varepsilon$~Ori is a post-mass-transfer object,  
and represents a future evolved state 
of multiple stellar systems in Orion.
For this reason, we include a brief discussion of the relevant systems,
$\delta$, $\zeta$, and $\sigma$~Ori,
to enable a comparison of their masses, radii, spectral types, and so on.

$\delta$~Ori (Mintaka, HD\,36486, HR\,1852/1851)
is our closest massive multiple system 
consisting of five components in total:
[(Aa1\,+\,Aa2)\,+\,Ab]\,+\,(Ca\,+Cb).
The triple star A (O9.5\,II\,+\,B2\,V\,+B0\,IV)
has the periods
$P_1 = 5.732436$\,d,
$P_2 = 55\,450\,{\rm d}$
and the mass
$17.8 + 8.5 + 8.7 \simeq 35$\,\Mnom\
\citep{Oplistilova2023AA...672A..31O}.
The primary (Aa1) is an unusually evolved O-type star.
The star has been studied in a series of papers by
\citet{corcoran2015}, \citet{pablo2015}, \citet{nichols2015}, and \citet{shenar2015}. 
We have already constructed a model of $\delta$~Ori~A
\citep{Oplistilova2023AA...672A..31O},
based on diverse observational data
(photometry, astrometry, radial velocities, eclipse timings, 
eclipse duration, spectral line profiles, and spectral-energy distribution),
with the exception of interferometry, as 
previous VLTI/AMBER data were not usable.
One conclusion from that study was
that the compact binary Aa1+Aa2
is a pre-mass-transfer object, 
while the tertiary seems to be unusually inflated
(according to its $\log g$ and the HR diagram).

$\zeta$~Ori (Alnitak, 50 Ori, HR\,1948/1949)
consists of four components
[((Aa\,+\,Ab)\,+\,B]\,+\,C, 
also includes a triple star 
O9.7\,Ib\,+\,B0.5\,IV\,+\,B0\,III and, in particular, 
a double-lined spectroscopic binary with the mass
$33\,+\,14 \simeq 47$\,\Mnom\
\citep{Hummel2000ApJ...540L..91H}.
The 2\,mag fainter companion Ab 
with the period $P_1 \approx 2687$\,d, 
mean separation of 45\,mas, 
and eccentricity $e_1=0.338$
was discovered by \citet{Hummel2000ApJ...540L..91H}.
The primary Aa is the only known magnetic 
O-type supergiant.
Tertiary B has a fast rotation of 350\,km/s, 
separation of 2.4\,arcsec, 
period $P_2 = 1509$\,yr, and 
eccentricity $e_2 = 0.07$.

Finally, $\sigma$~Ori (48\,Ori, HD\,37468, HR\,1931) has six components
$\left[\left(\mathrm{Aa} + \mathrm{Ab}\right) + \mathrm{B}\right] + \mathrm{C} + \mathrm{D} + \mathrm{E}$ 
of spectral types 
$\left[\left(\mathrm{O9.5\,V} + \mathrm{B0.5\,V}\right) + \mathrm{A2V}\right] + \mathrm{B2V} + \mathrm{B2V} + \, ?$.
The triple star has masses of 
$17 + 13 + 12 \simeq 42\,\Mnom$.
Binary A ($P_1 \approx 143$\,d) and component B 
form a visual pair with $P_2 \approx 160$\,yr.
The angular separations of all components are
4.3\,mas, 260\,mas, 11\,arcsec, 13\,arcsec, and 42\,arcsec, respectively. 
Unlike $\delta$~Ori,
the inner orbit of $\sigma$~Ori is eccentric,
while the outer is circular
\citep{Schaefer2016}. 
According to \citet{Schaefer2016}, 
the expected angular diameters are 0.27 and 0.21\,mas 
for Aa and Ab, respectively. 
The system has already been observed by interferometers 
such as CHARA/MIRC, NPOI, and VLTI/AMBER;
however, the diameters were unresolved. 

In 2023, we succeeded with the ESO proposal (Programme ID: 112.25JX) to observe
$\varepsilon$, $\delta$, $\zeta$, and $\sigma$~Ori
with the Very Large Telescope Interferometer (VLTI).
The main goal was to resolve the angular diameters of individual components
to constrain complex models of the stellar systems.

\begin{figure*}[htbp]
\centering
\includegraphics[width=\textwidth]{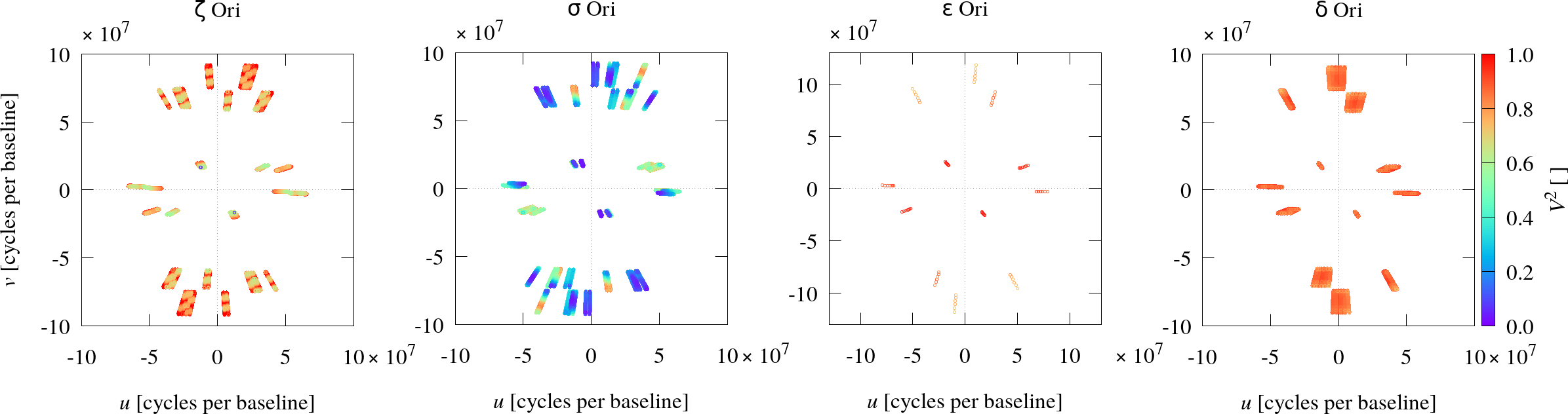}
\caption{
Coverage for interferometric measurements showing the squared visibility $V^2$ vs baselines $(u,v) \equiv \vec{B}/\lambda$ in cycles per baseline. Individual panels show four stars in Orion's Belt
($\zeta$, $\sigma$, $\varepsilon$, and $\delta$~Ori).
For each star, all nights are plotted.
Colours correspond to visibility values.
}
\label{uv_all} 
\end{figure*}

\begin{figure*}
\centering
\includegraphics[width=0.49\textwidth]{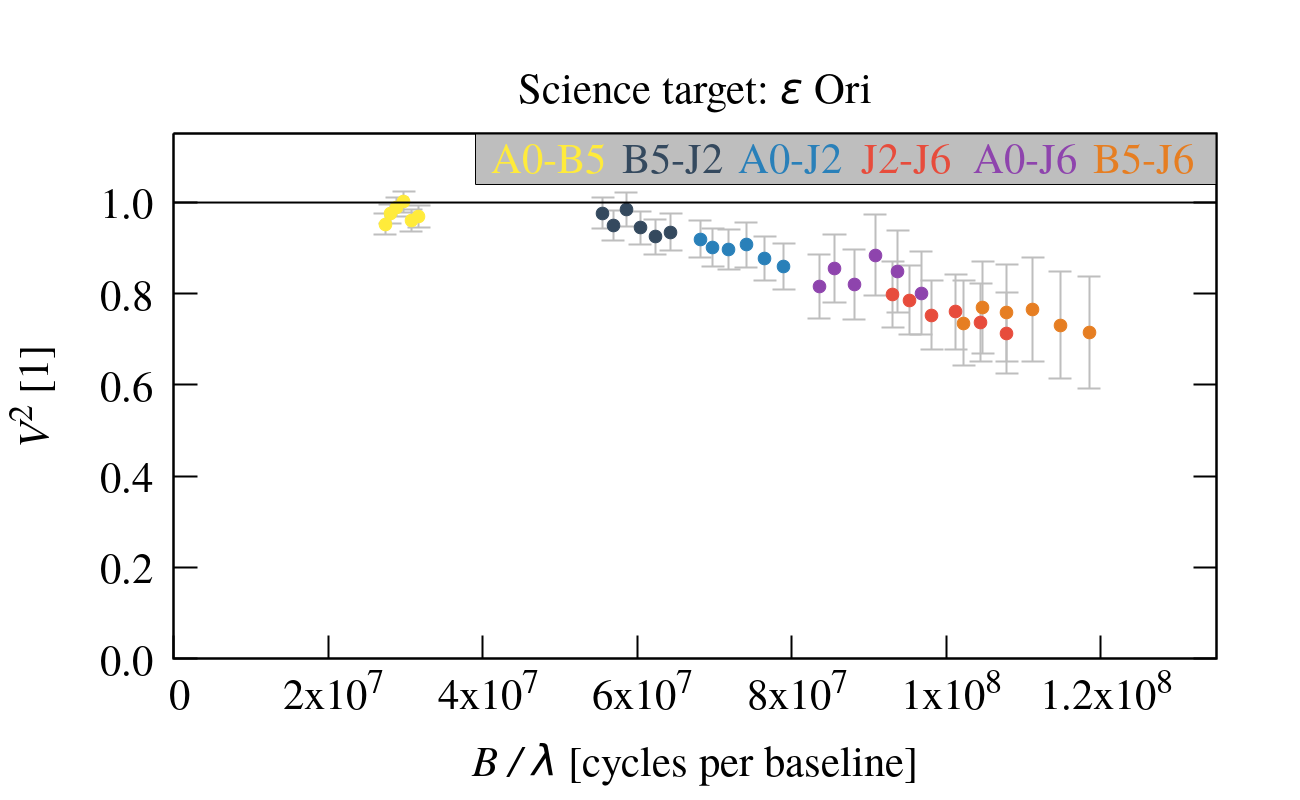}
\includegraphics[width=0.49\textwidth]{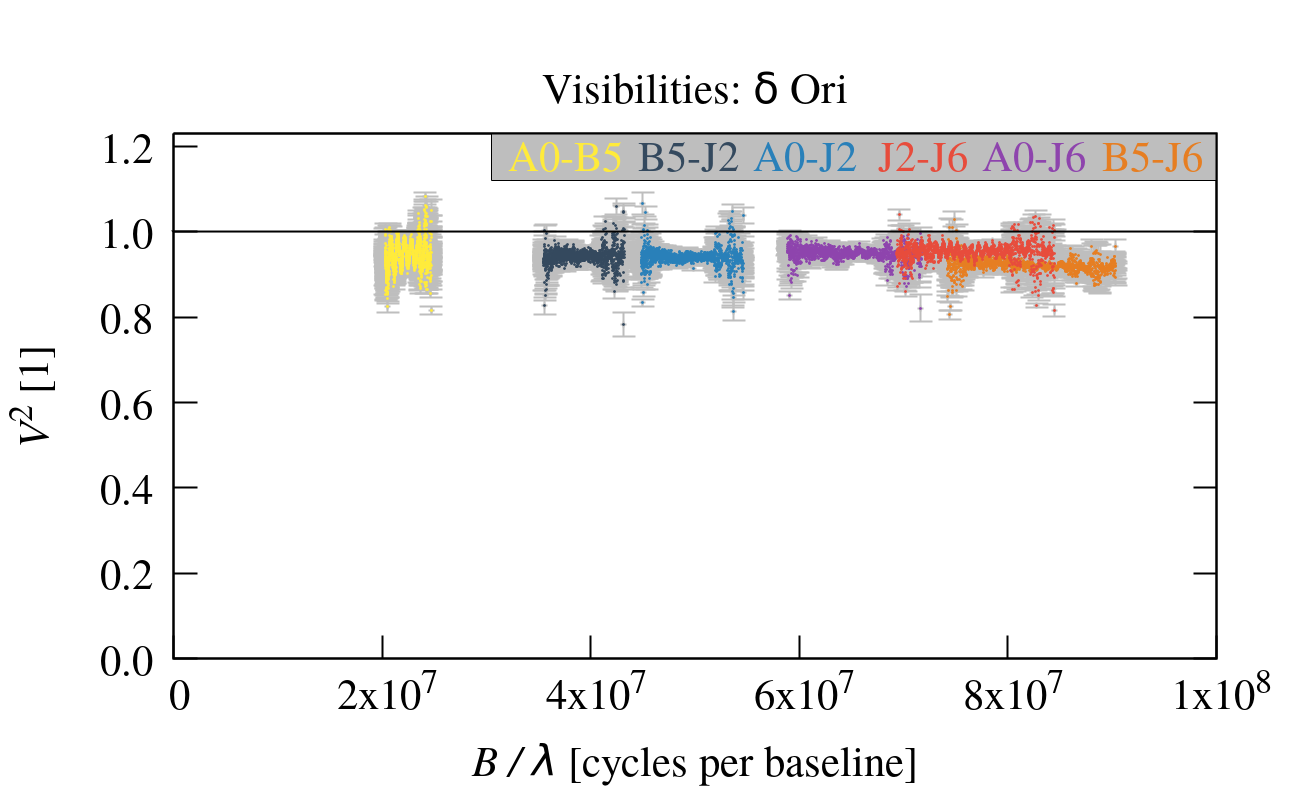}
\includegraphics[width=0.49\textwidth]{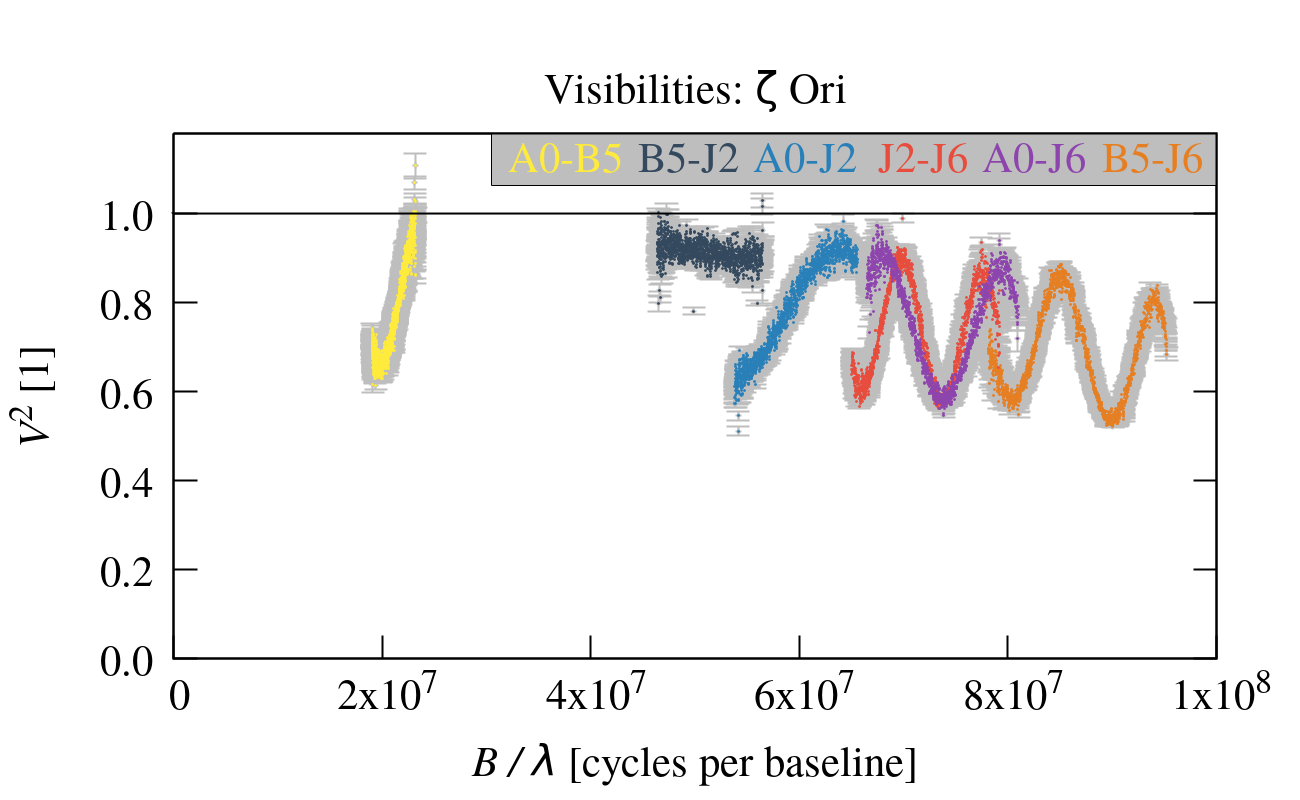}
\includegraphics[width=0.49\textwidth]{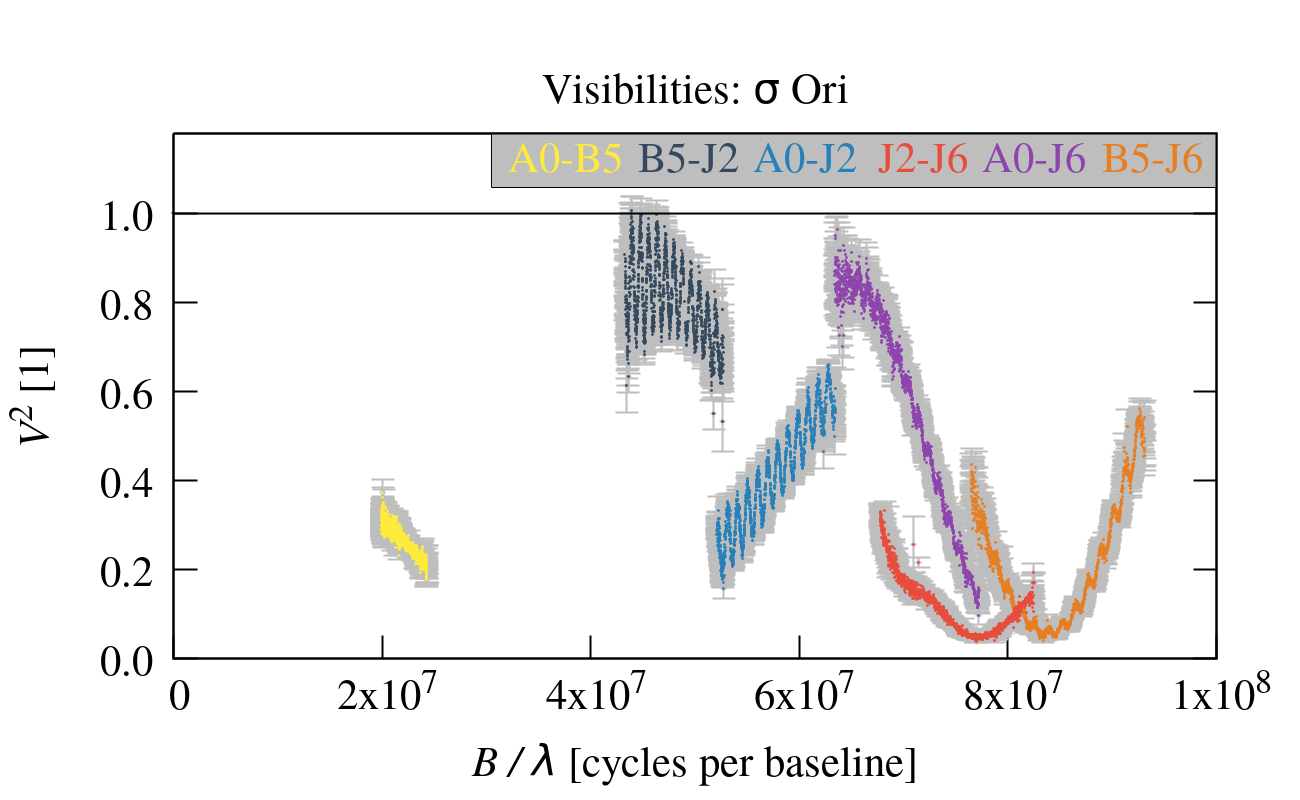}
\caption{
Examples of 
reduced squared visibilities, $V^2$, 
of the science targets,
$\varepsilon$, $\delta$, $\zeta$, and $\sigma$~Ori.
Measurements are from nights: 
20 November 2023, 
22 November 2023,
8 January 2024,
and 23 November 2023, respectively.
Science targets are with obvious signals from companions.
For the single star $\varepsilon$~Ori,
the signal suggests an elongated or non-spherical shape, unlike the calibrator $\zeta$~Lep, which
exhibits a perfectly spherical shape.
Colours correspond to individual baselines.
The squared visibilities of calibrators are in Fig.~\ref{reduced_visibility_cal}.
}
\label{reduced_visibility}
\end{figure*}

\begin{figure*}
\centering
\label{distance}
\includegraphics[width=\textwidth]{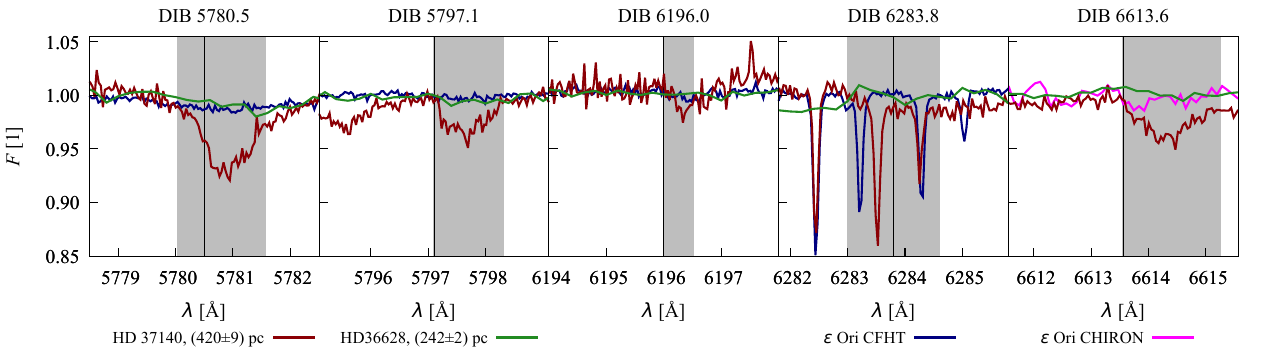}
\caption{
Comparison of DIBs' intensities for $\varepsilon$~Ori and stars which are close on the sky
(HD 37140 and HD 36628).
The most distant star has the deepest DIBs,
while the star with the lowest distance has very weak DIBs.
The spectrum of $\varepsilon$~Ori also shows weak DIBs,
which suggests its distance is lower than 420\,pc.
The spectra were taken from public archives of CFHT and ESO.
}
\end{figure*}

\section{Observational data}

Hereinafter, we describe the observational datasets, including
interferometry from VLTI instruments,
spectroscopy from CFHT and CTIO. We also describe the other datasets
we used for the modelling $\varepsilon$~Ori.

\subsection{VLTI/GRAVITY interferometry}

In our two runs of Programme ID: 112.25JX, 
(PI A. Oplištilová, 12\,h\,+\,1\,h) -- in phase 112, 
running between 1/10/23 and 31/3/24
at Cerro Paranal in northern Chile,
we obtained observations of 
$\zeta$, $\sigma$,
and $\delta$~Ori with the GRAVITY instrument
\citep{GRAVITY2011Msngr.143...16E, GRAVITY2017A&A...602A..94G}
and observations of $\varepsilon$~Ori
with the PIONIER instrument
\citep{Bouquin2011A&A...535A..67L}. Both runs were performed in service mode.
Such data should enable us to fit stars' angular diameters 
and angular positions
with up to an accuracy of 10 microarcseconds.

For observations with GRAVITY,
we requested 12 concatenations,
each including two observing blocks,
science target and calibrator (CAL-SCI).
Each concatenation lasted 1\,h; thus, we obtained 
observations on four different nights for each target, 
12 hours of observations in total. All our three targets were very bright;
therefore, we used the auxiliary telescopes (ATs) with the extended configuration, which made
these observations possible for the first time. 
We also permitted observations under the poorest conditions that still allow data acquisition; namely, 
a seeing of $<$1.4\,arcsec,
variable sky with thin cirrus, and an air mass of 1.6.
We chose the spectrometer with a high spectral resolution.
The observations (time series) were carried out in single on-axis mode, 
with automatic fringe tracking, 
adaptive optics with Coude guiding, 
and standard calibration.
For $\delta$ and $\sigma$~Ori, 
we set the Wollaston spectrometer out
and for $\zeta$~Ori, in. 
Otherwise, the brightest object $\zeta$~Ori 
would be saturating the instrument camera.

We performed the calibration using the EsoReflex 
GRAVITY pipeline
\citep{Freudling2013A&A...559A..96F}.
The essence of calibration is to compare 
the observed visibilities of (partially resolved) calibrators 
(see Fig.~\ref{reduced_visibility}),
with the theoretical visibilities of the corresponding uniform disk, 
which is described by
\begin{equation}\label{visibility}
\mu(u, v) = \frac{2\,J_1(\pi\,\theta \sqrt{u^2+v^2})}{\pi\,\theta \sqrt{u^2+v^2}},
\end{equation} 
where $J_1$ is the Bessel function of the first order,
the square root $\sqrt{u^2+v^2}$ corresponds to the length of a baseline,
and $\theta$ is the angular diameter of the calibrator.
The visibility function for a uniform disk is the simplest model,
but it is sufficient for the calibration process.
The larger the diameter, the faster the visibility drops
as a function of the spatial frequency.
The comparison of theoretical and observed visibilities 
is expressed by a transfer function
as the ratio of the calibrator’s squared visibilities
and theoretical squared visibilities,
\begin{equation}
    \mathrm{TF} = \frac{V^2_\mathrm{cal}}{V^2_\mathrm{UD}}
.\end{equation}
The transfer function is used to reduce raw uncalibrated data;
the reduced data, $V^2_*$, are the ratio of raw data
and the transfer function,
\begin{equation}
    V^2_* = \frac{V^2_\mathrm{raw}}{\mathrm{TF}}.
\end{equation}

For all three objects, we obtained squared visibilities 
measured on six baselines,
covering the wavelength range from 1.95\,$\mu$m to 2.45\,$\mu$m.
The coverage of measurements in $uv$-planes for each target
is shown in Fig.~\ref{uv_all}, 
and examples of reduced squared visibilities, $V^2$,
and the corresponding calibrators' visibilities
are in Fig.~\ref{reduced_visibility}.
We also obtained closure phases
for four triangles composed of these six baselines. 
We checked the closure phases of calibrators, which were close 
to zero as expected, indicating the central symmetry, characteristic 
of single, unspotted stars.

In the case of $\zeta$~Ori and its calibrator, 
we obtained two polarisation directions,
$P_1$ and $P_2$,
thanks to the Wollaston prism.  
However, since the object does not have a strong magnetic field,
we cannot use the polarimetric data to determine any properties.
The Wollaston prism was used just to prevent saturation
of this bright target. For more information on the 
calibration of $\zeta$\,Ori, we refer to Appendix \ref{Calibrator_zeta}.

\subsection{VLTI/PIONIER interferometry}

With the PIONIER instrument,
we proposed a single concatenation composed of
three observing blocks (CAL-SCI-CAL)
in an extended configuration with ATs.
We
obtained six interferometric measurements
of $\varepsilon$~Ori on six baselines,
within one hour of observing time.
This resulted in a total of 36 squared visibility 
and 24 closure phase measurements.
To obtain this snapshot, 
we applied for the grism as the disperser for PIONIER.
The PIONIER instrument works in the H-band
(1.52--1.76)\,$\mu$m, 
which delivers a better angular resolution than GRAVITY 
for measuring a single star diameter.
The target is very bright; thus, 
observations could be conducted under relaxed weather constraints:
seeing $<$1.15\,arcsec,
variable sky with thin cirrus, 
and an air mass of 2.0.
Again, the observations were obtained with automatic 
fringe tracking, 
adaptive optics with Coude guiding, 
and standard calibration.
Data reduction was performed using the \texttt{Pndrs} software \citep{Bouquin2011A&A...535A..67L}.
The $uv$-plane coverage 
is shown in Fig.~\ref{uv_all},
and the reduced squared visibilities, $V^2$,
together with those of the calibrator $\zeta$ Lep 
are in Fig.~\ref{reduced_visibility}.

\begin{table}
\centering
\small
\renewcommand{\arraystretch}{1.1}
\caption{Epochs of interferometric observations
and uniform-disk diameters (UDD) of calibrators.
}
\label{interfer_ob}
\begin{tabular*}{\hsize}{l@{\extracolsep{\fill}}ccrl@{\hspace{0.2cm}}}
\hline\hline\noalign{\smallskip}
Star& $T$ [HJD] &  Calibrators   & $\mathrm{UDD}_\mathrm{cal}$ [mas]\\
\noalign{\smallskip}\hline\noalign{\smallskip}
\multirow{4}{*}{$\delta$~Ori}    & 2460268.66081427 &                                      &              \\
                                 & 2460270.66197214 &      \multirow{2}{*}{HIP\,26149}     & \multirow{2}{*}{$0.6658\pm0.0492$}          \\
                                 & 2460271.64182168 &                                                \\
                                 & 2460300.59365039 &                                                \\
\noalign{\smallskip}\hline\noalign{\smallskip}
\multirow{4}{*}{$\zeta$~Ori}     & 2460269.65275756 &                                                \\
                                 & 2460271.75297908 &      \multirow{2}{*}{HIP\,26108}     &   \multirow{2}{*}{$1.9350\pm 0.1051$}        \\ 
                                 & 2460300.67146289 &                                                \\
                                 & 2460317.64698373 &                                                \\
\noalign{\smallskip}\hline\noalign{\smallskip}                                 
\multirow{4}{*}{$\sigma$~Ori}    & 2460268.82653117 &                                                            &                          \\ 
                                 & 2460271.72554332 &      \multirow{2}{*}{HIP\,26174}     &  \multirow{2}{*}{$0.6419\pm0.0120$}          \\ 
                                 & 2460317.61074530 &                                &                           \\ 
                                 & 2460318.55852424 &                                                \\
\noalign{\smallskip}\hline\noalign{\smallskip}                                              
$\varepsilon$~Ori                & 2460268.50080073 &       $\zeta$ Lep            & $0.7950\pm0.0472$                   \\
                           
\noalign{\smallskip}\hline\noalign{\smallskip}
\end{tabular*}
\tablefoot{
The values were taken from the VLTI or PIONIER pipeline
database (\texttt{GRAVI\_FAINT\_CALIBRATORS.fits}),
or recalibrated by us in the case of $\zeta$~Ori.
}
\end{table}

\subsection{CTIO/CHIRON and CFHT Spectroscopy}

We obtained four echelle spectra of the spectral range
(4504--8900)\,\AA \, at
the Cerro Tololo Inter-American Observatory (CTIO) 
with the 1.5-m reflector
using the highly stable cross-dispersed echelle spectrometer CHIRON. 
We used the fibre mode with the resolution of $R \approx 25\,000$.
A preliminary reduction to 1D spectra
was carried out at CTIO \citep{Tokovinin2013}.
We performed rectification using
the \texttt{reSPEFO2} software
written by A.~Harmanec%
\footnote{\url{https://astro.troja.mff.cuni.cz/projects/respefo}}.

Additionally, we used four spectra measured by 
the \mbox{3.6-m} Canada-France-Hawaii Telescope (CFHT), 
located near the summit of Mauna Kea on Hawaii.
The spectra from the \mbox{ESPaDOnS} instrument 
cover the region of (3815--6600)\,\AA \ and have the 
resolution of 68\,000.
These archival spectra have relatively low uncertainties
in the normalised intensity, less than 0.01,
corresponding to photon noise for bright stars.
Due to remaining rectification systematics,
we added a value of 0.01 to uncertainties.

\subsection{Spectral energy distribution (SED)}\label{SED_sect}

To compute the SED,
we downloaded absolute fluxes from the photometric catalogues in 
the \href{http://vizier.u-strasbg.fr/vizier/sed/doc/}{VizieR tool}
\citep{2014ASPC..485..219A}.
We selected all measurements within the wavelength range of
the BSTAR absolute synthetic spectra.
Specifically, we obtained measurements between
0.353 and 2.1\,$\mu$m
in the standard Johnson photometric system
\citep{Ducati_2002yCat.2237....0D}, along with measurements from
Gaia DR3 \citep{Gaia_2020yCat.1350....0G} and
2MASS \citep{Cutri_2003yCat.2246....0C}.
We omitted clear outliers and multiple entries.
For measurements without reported uncertainties,
we assumed the value of 0.02\,mag.
Hipparcos measurements were not used due to 
underestimated systematic uncertainties.
The final dataset contained 13 data points
see Table~\ref{SED_data}.

To compute the reddening, $E_{B-V}$, for $\varepsilon$~Ori, 
we used the differential photometry 
measured in October 2006 and February 2024 (2454015.6--2460362.4\,HJD) 
at the Hvar Observatory \citep{Hvar}; namely,
the observed colour index $(B-V)_0 = -0.192\,{\rm mag}$.
For the corresponding spectral type B0Ia,
the intrinsic colour is $(B-V) = -0.240$,
according to \citep{Golay1974ASSL...41.....G} and, consequently,
$E_{B-V}=0.05$\,mag and $A_V = 0.155$\,mag, which is
in agreement with \citet{Fan2017ApJ...850..194F}
and negligible in the infrared (IR).

For the construction of the SED, we also used calibrated 
UBV photometry from the Hvar
differential archive. It contains 21 observations 
secured in 2006 by Petr Harmanec,
Domagoj Ru\v{z}djak and Davor Sudar, and 4 observations 
secured during one night in 2024
by Hrvoje Bo\v{z}i\'c. We used the mean values of all 25 observations, 
$U=0.479\pm0.025$, $B=1.487\pm0.034$, and $V=1.679\pm0.018$.
The SED data were calculated using the
wavelengths from \citet{Bessell2000eaa..bookE1939B} and
calibration fluxes from \citet{Wilson2010ApJ...723.1469W}. 
The resulting values of SED are in Table~\ref{SED_data}.

\subsection{Parallax}\label{parallax}

As $\varepsilon$~Ori is too bright, it saturates Gaia’s detector,
designed primarily for stars fainter than $G = 6$\,mag,
and its parallax measurements are not reliable.
Hipparcos parallaxes were less precise;
\citet{Perryman1997A&A...323L..49P} estimated the distance 
of 412\,pc,
while \citet{vanLeeuwen2007A&A...474..653V} 606\,pc.
Instead, in \citet{Oplistilova2023AA...672A..31O},
we used Gaia DR3 parallaxes of the faint stars
in the same stellar association (Orion OB1b),
as these provide more precise and unbiased distance estimates than those available for the brightest members, which suffer from larger Gaia systematics. This approach yields a more reliable association distance.
We assumed that the most massive stars are located close
to the median distance \citep[][Sect. 9.3]{Kuhn2010ApJ...725.2485K}.
Then the distance of $\varepsilon$~Ori should be close to
$d \simeq 384$\,pc,
which is just intermediate between the other Orion Belt stars,
$\zeta$~Ori (386\,pc) and
$\delta$~Ori (381\,pc).

As a verification, we checked diffuse interstellar bands (DIBs).
Their strength depends on the interstellar medium
(and not on the spectral type)
and is correlated with the colour excess, $E_{B-V}$,
which quantifies the amount of dust and reddening along the line of sight
\citep{Friedman2011ApJ...727...33F}.
We assumed that the interstellar medium in the direction of $\varepsilon$~Ori
has similar properties to other stars close on the sky.
We used the same methodology as
\citet{Guinan2012A&A...546A.123G};
we selected two stars (Table~\ref{near_stars})
that are not in a dusty region according to the WISE map \citep{Aladin2022ASPC..532....7B}
and have both Gaia parallaxes and high-resolution spectra in public archives.
Then, we compared the strengths of their DIBs
(Fig.~\ref{distance}).
The star HD 37140 shows significant DIBs at the distance of ($420\pm9$)\,pc,
while HD 36628 shows no DIBs at ($242\pm2$)\,pc.
$\varepsilon$~Ori presents very low DIBs comparable to
HD 37140; thus, we verified that $\varepsilon$~Ori is likely located
at the distance ${<}420$\,pc (see Table~\ref{distance})
and our previous estimate seems to be reasonable.

Similarly, the colour excess $E_{B-V}$ is 
closely correlated with the distance.
We thus computed $E_{B-V}$, according to 
\citet{Johnson1953ApJ...117..313J},
using $U$, $B$, and $V$ magnitudes from 
\citet{Blanco1970pcmc.book.....B} and \citet{Mermilliod1994BICDS..45....3M},
see Table~\ref{near_stars}.
The colour excess of $\varepsilon$~Ori 
is computed in Sect.~\ref{SED_sect}.
Moreover, according to \citet{Green_2019ApJ...887...93G}%
\footnote{\url{http://argonaut.skymaps.info/}},
there is a sudden \enquote*{jump} in reddening in this direction,
occurring at a distance of 400\,pc.

We conclude that the previous distance estimate 
of about 600\,pc is certainly incorrect.
The relation of $E_{B-V}$, the strengths of DIBs,
and distance of $\varepsilon$~Ori corresponds to our estimate,
$(384\pm 8)\,{\rm pc}$.

\begin{table}[htbp]
\caption{
Distances and reddening of stars that are 
close to $\varepsilon$~Ori on the sky.
}
\label{near_stars}
\centering
\small
\renewcommand{\arraystretch}{1.1}
\begin{tabular}{llccccc}
\noalign{\smallskip}\hline\hline\noalign{\smallskip}
Star              &    $d$ [pc]        & $E_{B-V}$ [mag] & Spectral type  \\
\noalign{\smallskip}\hline\noalign{\smallskip}
HD 37140          &    $420\pm9$       & 0.256           & B8\,II$^*$        \\
$\varepsilon$~Ori &    $384$ (assumed) & 0.050           & B0\,Ia            \\
HD\,36628         &    $242\pm2$       & 0.030           & B9\,IV/V$^*$      \\  
\noalign{\smallskip}\hline\noalign{\smallskip}
\end{tabular}
\tablefoot{
$^*$ \citet{Houk1999MSS...C05....0H}. The corresponding DIBs observations are shown in Fig.~\ref{distance}.
}
\end{table}

\section{$\varepsilon$~Ori as a single star}\label{sec3}

We constructed a single-star model of 
the B0\,Ia supergiant $\varepsilon$~Ori
using PHOEBE2%
\footnote{\url{http://phoebe-project.org}}
\citep{Prsa_2016ApJS..227...29P,Horvat_2018ApJS..237...26H,Jones_2020ApJS..247...63J,Conroy_2020ApJS..250...34C}.
PHOEBE2 employs seamless triangular meshes,
closely following the generalised Roche potential.
Each triangular element of the mesh is assigned local quantities
(e.g. temperature, surface gravity, intensity, directional cosine).
The total flux is then computed by integrating over visible elements.
PHOEBE2 incorporates various stellar atmosphere models,
which enables a realistic, self-consistent treatment of
limb darkening,
gravity darkening, and
rotational distortion.
It is a powerful tool for solving an inverse problem,
constrained by light curves and radial velocity curves,
and, recently, it was extended to include
interferometry,
spectra,
and SED.

A new interferometric module in PHOEBE2%
\footnote{\url{https://github.com/miroslavbroz/phoebe2}}
\citep{Broz_2025a}
enables the computations of interferometric observables, squared visibility ($V^2$), 
closure phase ($\arg T_3$), and amplitude ($|T_3|$), by means of the Fourier transform approach.
Additionally, a new spectroscopic module
\citep{Broz_2025b}
utilises interpolations in grids of synthetic spectra,
assigning one spectrum to each element.
Again, the total spectrum is obtained
by integrating over all elements,
weighted by appropriate passband intensities.
It enables the fitting of detailed spectral line profiles.

\begin{figure}
\centering
\includegraphics[width=8.5cm]{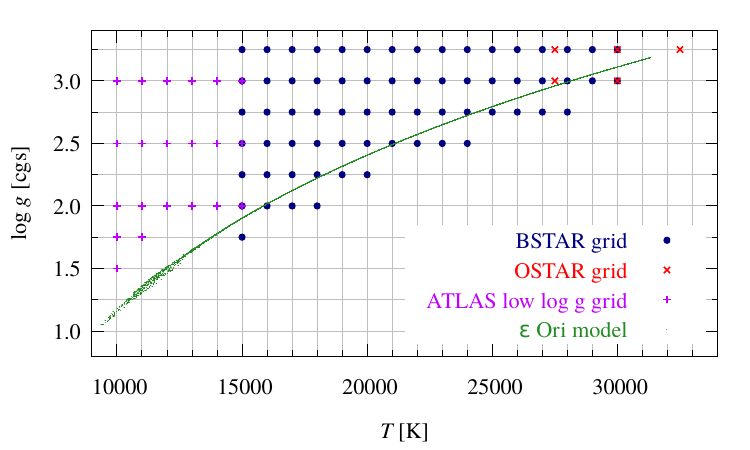}
\caption{
Grids of synthetic spectra BSTAR and OSTAR
\citep{Lanz_2003ApJS..146..417L,Lanz_2007ApJS..169...83L}
used in our spectroscopic models.
Each spectrum is parametrised by $\log g$ and $T$.
 To describe also critically rotating stars (cf. green points),
it was necessary to compute additional ATLAS spectra
\citep{Castelli2003IAUS..210P.A20C} for low values of $\log g$ and $T$.
}
\label{eps_grid}
\end{figure}

\begin{figure*}[htbp]
\centering
\includegraphics[width=0.46\textwidth]{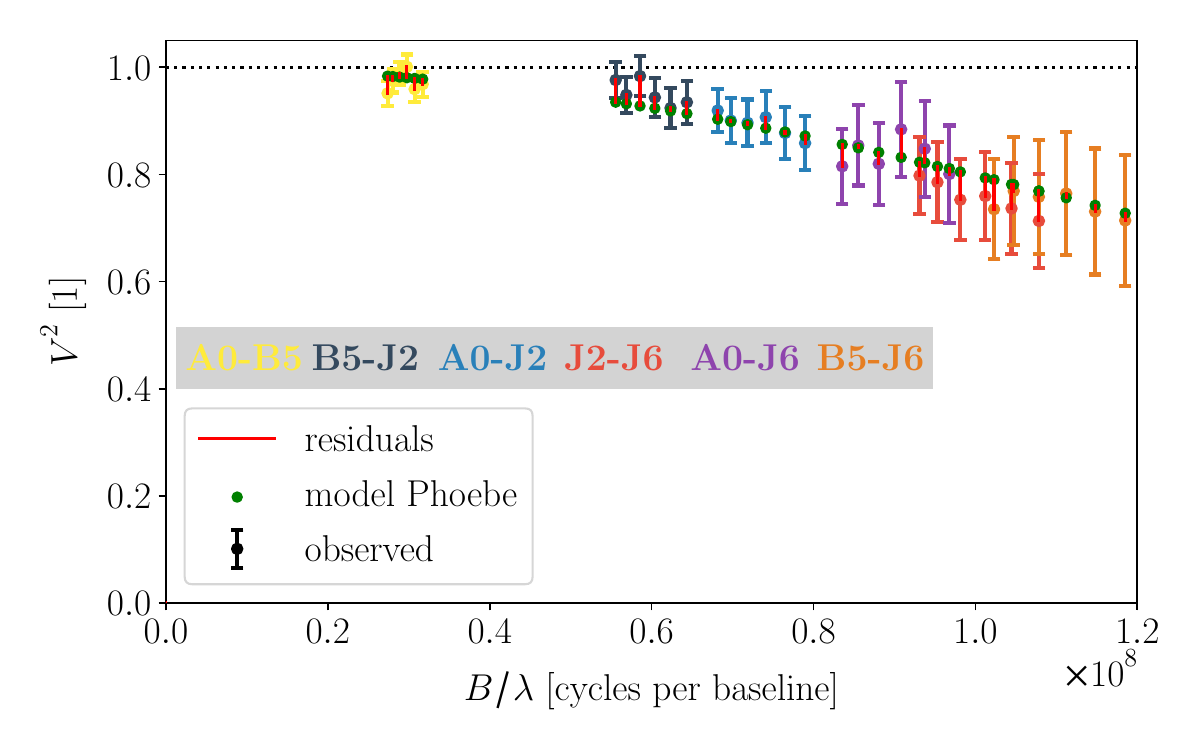}
\includegraphics[width=0.41\textwidth]{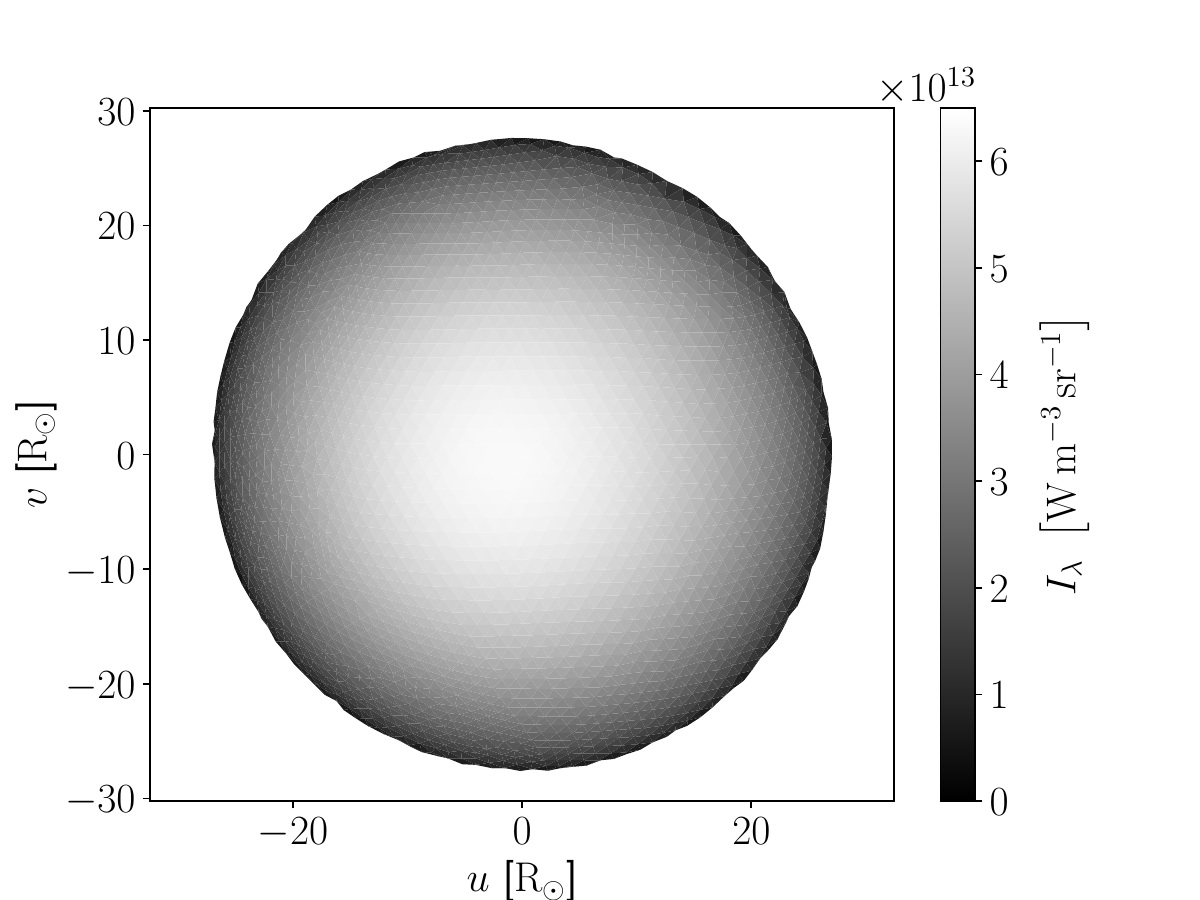}
\caption{
Almost spherical model of $\varepsilon$~Ori 
based on PIONIER observations in H-band with fixed $v\sin i$ \citep{Puebla2016}.
Left: Squared visibility vs projected baseline $B/\lambda$. Right: Corresponding triangular mesh with the passband intensities (greyscale).
The model was converged starting with the parameters from Table~\ref{vsini_fixed_table} (bold line).
The best-fit was with
$\chi^2_{\mathrm{VIS}} = 11.23$.
The free parameters were
$i = 13.6$\,deg,
$m = 20.3$\,\Mnom,
$\Omega = 293.7$\,deg,
$P_\mathrm{rot} = 4.01$\,d. The fixed parameters were
$v\sin i = 70\,{\rm km}\,{\rm s}^{-1}$,
$T = 25\,000$\,K,
$\gamma = 25.9$\,km\,s$^{-1}$ and
the derived parameters were
$v = 298\,{\rm km}\,{\rm s}^{-1}$,
$R_\mathrm{equiv} = 22.43$\,\Rnom \ (derived),
$R_{\mathrm{pole}} = 22.29$\, \Rnom \ (derived),
$R_{\mathrm{equ}} = 22.46$\, \Rnom \ (derived),
$\theta_\mathrm{equiv} = 0.543$\,mas.
The star is close to critical rotation and has
an almost pole-on orientation
in order to decrease projected rotation (cf. $v\sin i$).
}
\label{eps_best_fit_fixed_vsini} 
\end{figure*}

\begin{table*}[htbp]
\caption{
Grid of almost spherical models for $\varepsilon$~Ori
assuming a fixed value of
$v\sin i = 70$\,km\,s$^{-1}$.
}
\label{vsini_fixed_table}
\centering
\small
\begin{tabular}{cccccS[table-format=3.2]S[table-format=3.2]ccccc}
\noalign{\smallskip}\hline\hline\noalign{\smallskip}
$i$ [deg] &
$m$ [\Mnom] &
$\Omega$ [deg] &
$R_\mathrm{equiv}$ [\Rnom] &
$\theta_\mathrm{equiv}$ [mas] &
{$P_\mathrm{rot}$ [d]}  &
{$v$ [km s$^{-1}$]} &
$v \sin i$ [km s$^{-1}$] &
$\chi^2_{\rm VIS}$ [1] &
$T$ [K] &
$\chi^2_\mathrm{SPE}$ [$10^3$] \\
\\
\noalign{\smallskip}\hline\noalign{\smallskip}
14 & 22.90 & 289.7 & 24.68 & 0.598 & 4.31 & 289 & 70 & 11.451 & 25384 & 13.4 \\
15 & 23.99 & 289.9 & 25.26 & 0.612 & 4.72  & 270 & 70 & 11.532 & 25363 & 14.7 \\
20 & 26.06 & 289.9 & 26.32 & 0.638 & 6.51  & 205 & 70 & 11.658 & 26366 & 17.0 \\
30 & 27.12 & 290.5 & 26.85 & 0.650 & 9.70  & 140 & 70 & 11.715 & 26718 & 18.5 \\
40 & 27.43 & 299.4 & 27.00 & 0.654 & 12.55 & 109 & 70 & 11.732 & 26929 & 18.9 \\
50 & 27.58 & 290.2 & 27.08 & 0.656 & 14.99 & 91  & 70 & 11.740 & 27962 & 20.1 \\
60 & 27.66 & 299.1 & 27.12 & 0.657 & 16.97 & 81  & 70 & 11.743 & 27956 & 20.3 \\
70 & 27.68 & 290.5 & 27.13 & 0.657 & 18.43 & 74  & 70 & 11.745 & 27198 & 19.3 \\
80 & 27.68 & 290.1 & 27.13 & 0.657 & 19.31 & 71  & 70 & 11.747 & 28041 & 20.4 \\
90 & 27.71 & 290.0 & 27.14 & 0.657 & 19.62 & 70  & 70 & 11.746 & 28050 & 20.4 \\
\hline
\end{tabular}
\tablefoot{
$i$ denotes the inclination of the spin axis with respect to the sky plane;
$m$, mass; 
$\Omega$, the orientation of star's equator, as seen by the observer;
$R_\mathrm{equiv}$, equivalent radius;
$\theta_\mathrm{equiv}$, angular diameter;
$P_\mathrm{rot}$, rotational period;
$v$, circumference velocity.
Except for the temperature $T$, 
all the parameters were determined from interferometry.
The values of $\chi^2_\mathrm{VIS}$
do not exhibit a significant minimum,
because the model is too constrained by $v\sin i$.
Then, we kept the determined parameters fixed and fitted the temperature
according to the spectroscopic data to find $\chi^2_{\mathrm{SPE}}$.
However, synthetic line profiles were in all cases deeper than the observed ones.
The model with the lowest
$\chi^2_\mathrm{VIS}$ and $\chi^2_\mathrm{SPE}$ is for $i=14$\,deg.
For models based on interferometry, the parameters
$i$, $\log g = 3.0$, $v\sin i = 70$\,km\,s$^{-1}$, 
$T = 25\,000\,{\rm K}$
were fixed, while
$m$, $\Omega$ were free.
}
\end{table*}

\begin{figure*}[htbp]
\centering
\includegraphics[width=0.46\textwidth]{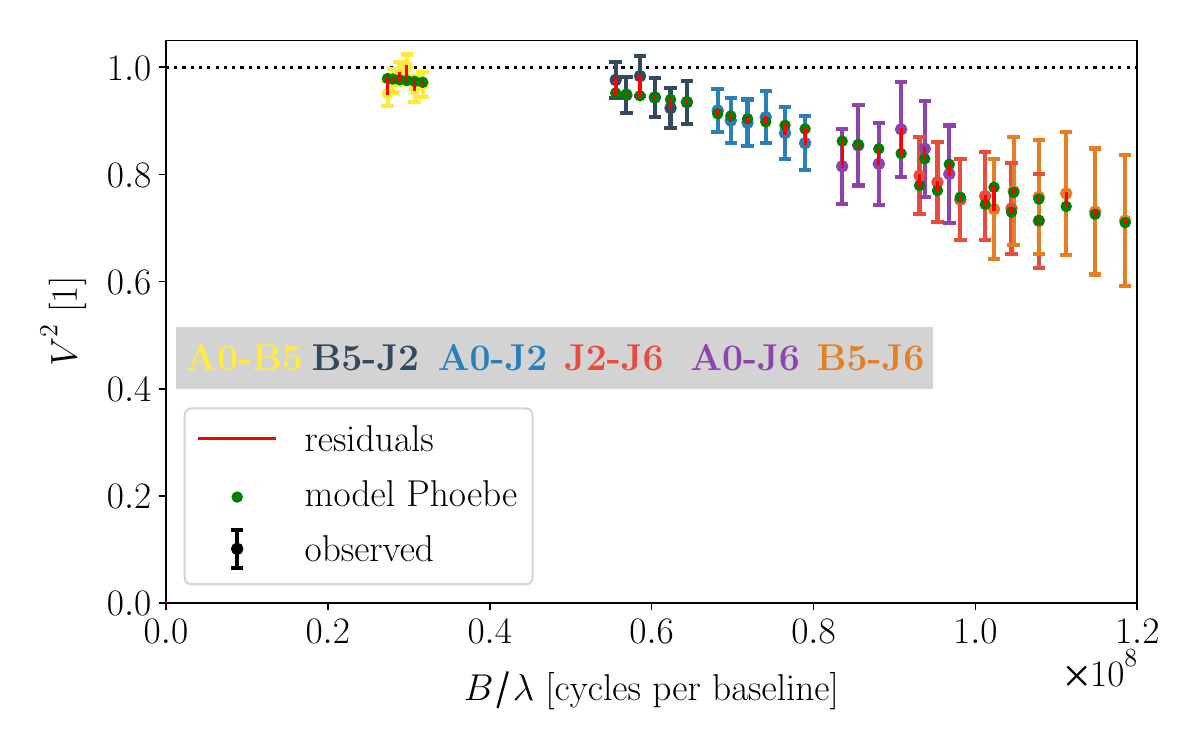}
\includegraphics[width=0.41\textwidth]{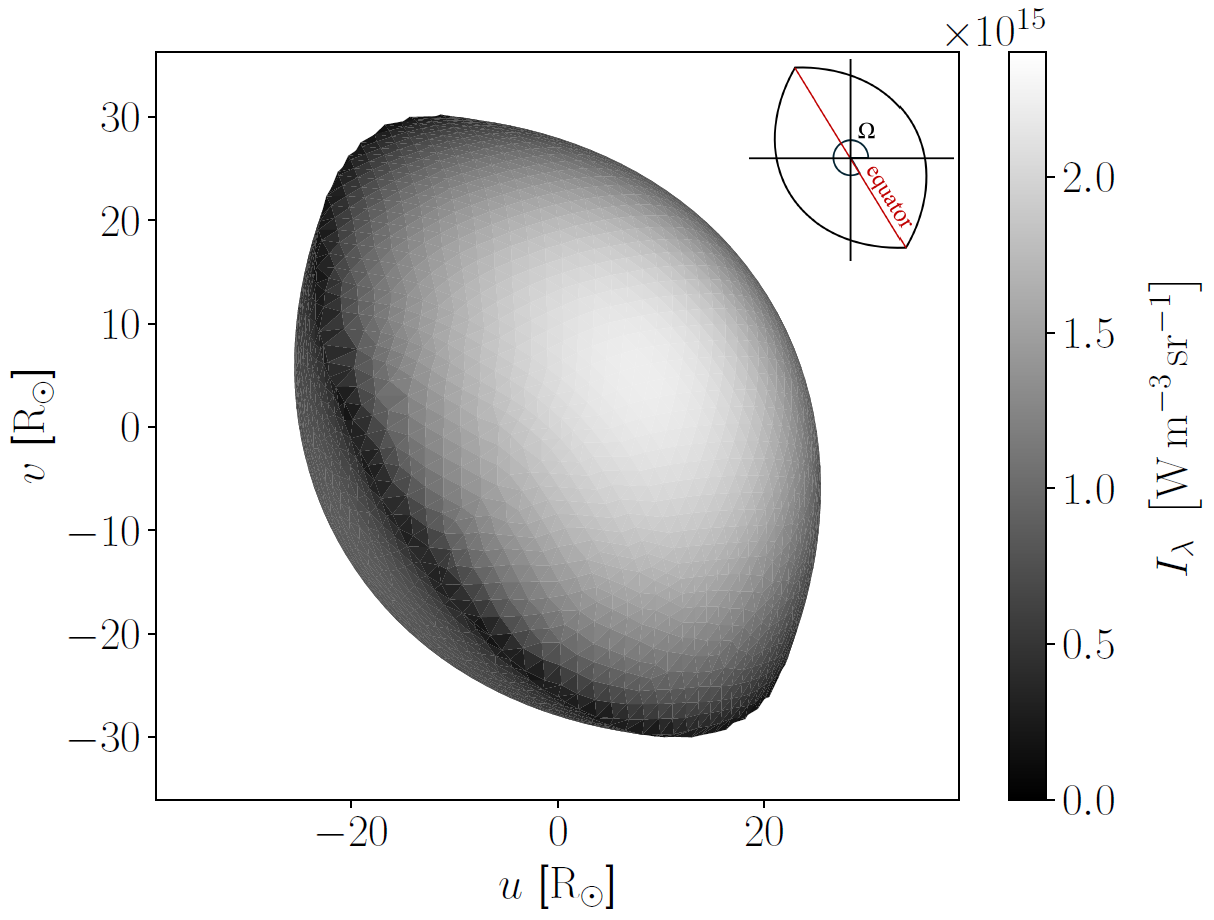}
\caption{
Same as Fig.~\ref{eps_best_fit_fixed_vsini}, 
but with free $v \sin i$.
The model was converged starting with the parameters 
from Table~\ref{vsini_free_table}.
The best fit model was with 
$\chi^2_{\mathrm{VIS}} = 6.928$.
The free parameters were
$i = 58.3^\circ$,
$m = 28.0$\,\Mnom,
$\Omega = 301.5$\,deg, and
$P_\mathrm{rot} = 4.26$\,d. The fixed parameters were
$T = 27\,000$\,K and
$\gamma = 25.9$\,km\,s$^{-1}$. 
The derived parameters are
$v\sin i = 277\,{\rm km}\,{\rm s}^{-1}$,
$v = 325\,{\rm km}\,{\rm s}^{-1}$,
$R_\mathrm{equiv} = 27.37\,\Rnom$, 
$R_\mathrm{pole}=22.29\,\Rnom$,
$R_\mathrm{equ}=33.61\,\Rnom$,
$\theta_\mathrm{equiv} = 0.667$\,mas,
$\theta_\mathrm{pole} = 0.540$\,mas, and
$\theta_\mathrm{equ} = 0.814$\,mas.
Again, the star is close to the critical rotation,
but with an oblique orientation.
This model better explains the visibilities
along different baselines $\vec{B}/\lambda$.
The meaning of $\Omega$ is illustrated.
}
\label{eps_best_fit_free_vsini} 
\end{figure*}

\begin{table*}[htbp]
\caption{
Grid of non-spherical models of $\varepsilon$~Ori
without a $v\sin i$ constraint.
}
\small
\label{vsini_free_table}
\centering
    \begin{tabular}{ccccccccS[table-format=3.2]cS[table-format=3.2]}
       \noalign{\smallskip}\hline\hline\noalign{\smallskip}
$i$ [deg] &
$m$ [\Mnom] &
$\Omega$ [deg] &
$R_\mathrm{equiv}$ [\Rnom]  &
$\theta_\mathrm{equiv}$ [mas] & $P_\mathrm{rot}$ [d]  &
$v$ [km s$^{-1}$] &
$v \sin i$ [km s$^{-1}$] &
{$\chi^2_\mathrm{VIS}$ [1]}  &
$T$ [K] &
{$\chi^2_\mathrm{SPE}$ [$10^3$]} \\
\noalign{\smallskip}\hline\noalign{\smallskip}
0  & 23.75 & 306.1 & 25.20 & 0.610 & 4.85 & None & 0   & 12.188 & 25300 & 71.7 \\
10 & 23.87 & 307.2 & 25.26 & 0.612 & 4.80 & 266  & 46  & 11.949 & 25283 & 31.7 \\
20 & 21.47 & 290.2 & 23.96 & 0.580 & 3.99 & 304  & 104 & 10.273 & 24537 & 1.9 \\
30 & 23.46 & 300.6 & 25.04 & 0.607 & 4.07 & 311  & 155 & 8.734  & 23970 & 15.4 \\
40 & 24.96 & 308.1 & 25.83 & 0.626 & 4.14 & 316  & 203 & 7.363  & 23378 & 43.0 \\
50 & 26.73 & 300.8 & 26.73 & 0.648 & 4.21 & 321  & 246 & 6.959  & 22747 & 68.1 \\
60 & 28.18 & 301.3 & 27.45 & 0.665 & 4.27 & 325  & 282 & 6.929  & 22282  & 85.8 \\
70 & 28.58 & 300.7 & 27.64 & 0.669 & 4.28 & 326  & 307 & 6.967  & 21997 & 96.1 \\
80 & 28.27 & 298.6 & 27.49 & 0.666 & 4.36 & 319  & 314 & 7.222  & 22018 & 100.0 \\
90 & 28.16 & 298.1 & 27.44 & 0.665 & 4.44 & 312  & 312 & 7.339  & 22192 & 100.0 \\
        \hline
    \end{tabular}
\tablefoot{
The parameters
$i$ (range of value), $\log g = 3.011$, and $T = 24\,985$\,K
were fixed, while
$m$, $\Omega$, $P_\mathrm{rot}$
free.
Otherwise, $\chi^2_\mathrm{VIS}$ and $\chi^2_\mathrm{SPE}$ were computed as in Table~\ref{vsini_fixed_table}.
The best-fit model based on interferometry ($i = 60$\,deg) is shown in Fig.~\ref{eps_best_fit_free_vsini},
while the best one based on spectroscopy ($i = 20$\,deg) is in Fig.~\ref{spe}.
}
\end{table*}

\begin{figure*}
\centering
\includegraphics[width=\textwidth]{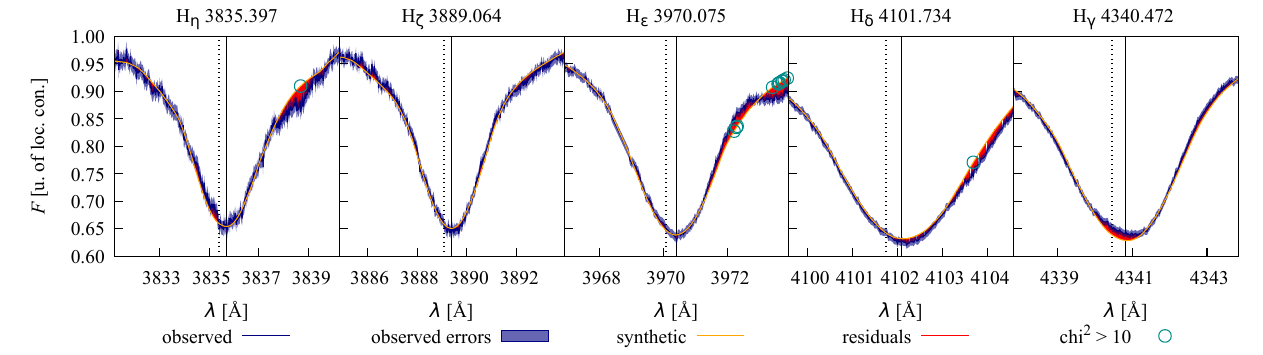}
\caption{
Almost spherical model of $\varepsilon$~Ori based on
CFHT spectroscopy. 
Hydrogen Balmer lines were fitted using 1443 data points,
excluding H$_\alpha$ and H$_\beta$,
which are \enquote*{filled} by wind-induced emission.
An interstellar calcium 3968\,\AA \ for H$_\varepsilon$ was masked.
The best fit resulted in
$\chi^2_\mathrm{SPE} = 1770$.
Free parameters were
$P_{\mathrm{rot}} = 4.551$\,d,
$T = 25\,037$\,K,
$\log g = 3.016$,
$i = 21.40$\,deg, and
$\gamma=25.00$\,km\,s$^{-1}$.
Other parameters were fixed and set according to interferometry.
The derived parameters were
$v \sin i = 111$\,km\,s$^{-1}$ and
$v = 305$\,km\,s$^{-1}$.
The dotted lines show laboratory wavelengths, while 
the solid vertical lines indicate the Doppler-shifted line centres, corresponding to the resulting $\gamma$.
The teal circles denote the points that contributed to 
$\chi^2$ with values greater than 10.
This model explains reasonably well the Balmer lines,
with a few systematics remaining in H$_\varepsilon$, H$_\eta$, 
possibly due to different metal abundances.
The temperature was constrained, and $\log g$ was slightly adjusted.
However, due to the difference in inclination
($i = 21$ vs $62$\,deg),
the interferometric model would become worse,
with $\chi^2_{\rm VIS} = 16.04$.
 The fit of \ion{He}{I} lines is given in Fig.~\ref{HeI_fit}.
}
\label{spe} 
\end{figure*}

\paragraph{Visibility.}\label{VIS_sect}
First, we modelled $\varepsilon$~Ori as a spherical star
with fixed radiative parameters according to
\cite{Puebla2016MNRAS.456.2907P}; namely,
$T = (27\,000 \pm 500)$\,K and
$\log g = (3.00 \pm 0.05)$ that they estimated
by the fitting of line profiles of
H$_\beta$, H$_\gamma$, H$_\delta$, and H$_\varepsilon$.
In PHOEBE2,
we set the Kurucz atmospheres,
but linear limb darkening coefficient
$c_1 = 0.01621$
interpolated from \citet{vanHamme_1993AJ....106.2096V} tables
for comparison purposes (see below).
We assumed the distance
$d \simeq 384$\,pc, as discussed in Sect.~\ref{parallax}.
We verified the computation of $V^2$
using the independent implementation of
\cite{Broz_2017ApJS..230...19B}.
The fit to $V^2$ from PIONIER and
the convergence of one free parameter ($m$)
using the simplex method \citep{Nelder_1965} 
resulted in exactly the same $\chi^2_{\rm VIS} = 12.1$
in both codes
(Fig.~\ref{eps_xitau_phoebe_VIS_12-1633_paper}).
The $\chi^2$ value is smaller than
the number of degrees of freedom
$\nu = N - M = 35$,
which most likely indicates overestimated uncertainties
of PIONIER observations.
In fact, previous PIONIER observations were more precise
\citep{Pietrzynski_2019Natur.567..200P}.
We rescaled the uncertainties down by the factor of 12.1/35 ($\chi^2/(N-M)$).
For the nominal uncertainties, we used the ones computed by the \texttt{Pndrs} pipeline, which calculates uncertainties of $V^2$ from the frame-to-frame scatter of the raw coherence factor, and then propagates transfer function uncertainties (including calibrator diameter errors) into the final calibrated 
errors reported in OIFITS outcome of \texttt{Pndrs} pipeline.
With these rescaled uncertainties,
the best-fit mass is
$m = (23.5\pm 0.5)$\,\Mnom,
corresponding to the angular diameter,
$\theta = (0.615 \pm 0.007)$\,mas,
and physical radius,
$R = (25.4\pm 0.3)$\,\Rnom.

Second, we modelled a rotating, almost spherical star.
For simplicity, we used black-body atmospheres
to avoid problems with too low $\log g$ values,
occurring when the star is
close to the critical rotation
(cf. Fig.~\ref{eps_grid}).
Nevertheless, we derived the limb darkening coefficients
from the Kurucz atmospheres
and set them manually,
with a power-law prescription
\citep{Prsa_2016ApJS..227...29P} of
\begin{equation}
\frac{I}{I_0} = 1 - c_1 (1-\mu^{\frac{1}{2}}) - c_2 (1-\mu) -c_3 (1-\mu^{\frac{2}{3}}) -c_4 (1-\mu^2),\,
\end{equation}
where
$c_1 = 1.698$,
$c_2 = -4.151$,
$c_3 = 5.887$,
$c_4 = -2.525$ 
are suitable for the temperature, $T = 27\,000$\,K and $\log g = 3.0$ 
\citep{Puebla2016MNRAS.456.2907P}.
We set the bolometric gravity brightening exponent
$\beta = 1.0$,
which corresponds to hot stars \mbox{($T > 8\,000$\,K)}. 
The exponent occurs in the implementation of local temperature in PHOEBE2 \citep{vonZeipel1924MNRAS..84..665V, Prsa2011PHOEBE, Prsa_2016ApJS..227...29P}, expressed as
\begin{equation}
T = T_{\rm pole}\,(g/g_{\rm pole})^{0.25\,\beta}\,,
\end{equation}
where
$g$ is the local gravitational acceleration.
To reveal all possible minima of $\chi^2$,
we computed a more detailed grid in the inclination of the rotation 
axis (hereafter only \enquote*{inclination}) $i$.
The orientation of the star is also determined by $\Omega$, 
longitude of the ascending node (of the equator).
We made the parameters $m$ and $\Omega$ free,
the parameters $\log g$, $d$, $v\sin i$ (and $i$) fixed,
while the dependent parameters $P_\mathrm{rot}$, $R_\mathrm{equiv}$, and $\theta_\mathrm{equiv}$
were constrained by the following relations,
\begin{equation}\label{Prot}
P_{\mathrm{rot}} = \frac{2\pi R_\mathrm{equiv}}{v\sin i} \sin i\,,
\end{equation}
\begin{equation}\label{Requiv}
R_\mathrm{equiv} = \sqrt{\frac{G m}{g}}\,,
\end{equation}
\begin{equation}\label{theta}
\theta_\mathrm{equiv} = \frac{2R_\mathrm{equiv}}{d}\,,
\end{equation}
where the value of $v\sin i$ was set up to
$70$\,km\,s$^{-1}$
\citep{Puebla2016MNRAS.456.2907P}.

The converged models for fixed values of $i$ are listed in
Table~\ref{vsini_fixed_table}.
Unfortunately, the models were too constrained by 
the fixed value of $v\sin\,i$;
the $\chi^2_\mathrm{VIS}$ dependence on $i$ 
remained flat over the full range of $i$.
Formally, the lowest $\chi^2_\mathrm{VIS}$ value
is for the lowest inclination, $i=14$\,deg.
Lower values of $i$ cause problems in the modelling of edges.
For the best-fit model,
converged from parameters with the lowest $\chi^2_\mathrm{VIS}$,
we refer to Fig.~\ref{eps_best_fit_fixed_vsini}.
It is an almost spherical star close to critical rotation,
with almost pole-on orientation.

To obtain a model that best corresponds to 
the measured squared visibilities, 
we relaxed the $v\sin i$ constraint
and computed another grid in $i$,
using still only the interferometric data.
In this case, there were three free parameters,
$m$, $\Omega$, and $P_\mathrm{rot}$.
Our results are summarised in Table~\ref{vsini_free_table}.
Fortunately, $\chi^2_{\rm VIS}$ substantially decreased,
from $12.2$ down to $6.9$,
and the best-fit model is again a critically rotating star,
seen at an oblique inclination of
$i = (58\pm 20)\,{\rm deg}$
(see Fig.~\ref{eps_best_fit_free_vsini}).
On the other hand, the nominal projected velocity,
$v\sin i = 277^{+30}_{-80}\,{\rm km}\,{\rm s}^{-1}$,
seems to be too high.

\begin{figure}[htbp]
\centering
\includegraphics[width=0.5\textwidth]{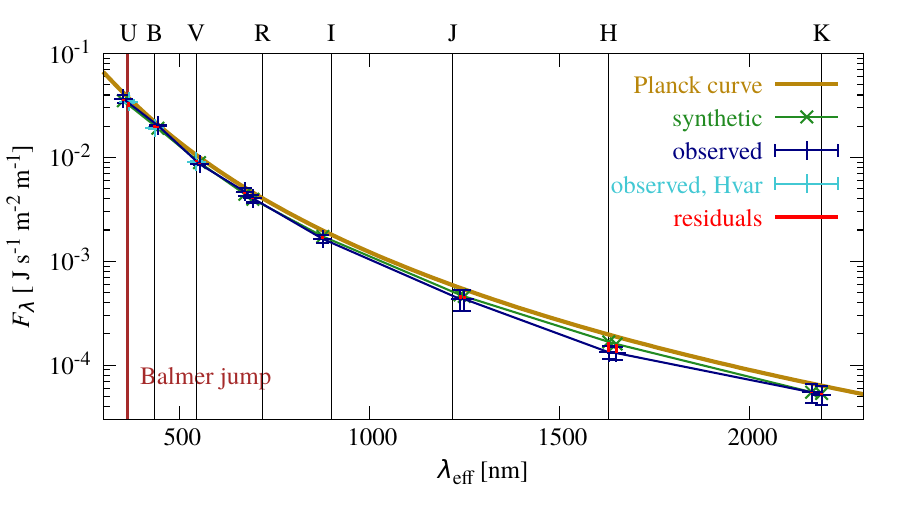}
\caption{
SED model of $\varepsilon$~Ori based on data from the Hvar Observatory and Vizier:\ the 
monochromatic flux, $F_\lambda$, vs wavelength, $\lambda$.
The best-fit value was $\chi^2_{\rm SED}= 52$ with the following parameters:
$T_\mathrm{eff}=24\,654$\,K (fitted),
$d = 384\,{\rm pc}$ (fixed),
$M = 28.43$\,\Mnom \ (fixed), and
$R = 27.45$\,\Rnom \ (derived);
other parameters were fixed and set according to interferometry.
This model served as an independent verification
of the temperature from the spectroscopic model.
}
\label{SED} 
\end{figure}

\paragraph{Spectroscopy.}

To constrain the radiative parameters, 
which were only assumed in the previous interferometric models,
we used the spectra from CFHT.
First, we checked the influence of $i$, i.e. $v\,\sin i$,
on synthetic line profiles,
by computing forward models for the Balmer lines,
H$_\gamma$, H$_\delta$, H$_\varepsilon$, H$_\zeta$, and H$_\eta$,
which are not significantly influenced by the wind
\citep{Puebla2016MNRAS.456.2907P}.
We used narrow ranges of wavelengths
($\pm 3, \pm 2.5, \pm 4, \pm 5, \pm 5$)\,\AA, 
respectively,
in order to eliminate neighbouring lines.
Also, we masked one remaining interstellar calcium line, 3968\,\AA, near H$_\varepsilon$.
In PHOEBE2, we set \texttt{spe\_method} to \enquote*{integrate} and the interpolation to nearest-neighbour,
to remain within the respective grid (Fig.~\ref{eps_grid})
and prevent excessive extrapolation.
Furthermore, we computed additional ATLAS synthetic spectra 
\citep{Castelli2003IAUS..210P.A20C}%
\footnote{\url{https://github.com/RozanskiT/vidmapy}}
for low values of $\log g$ and $T$,
to describe critically rotating stars.

The resulting $\chi^2_{\rm SPE}$ values for fixed $v\,\sin\,i$ 
are listed in Table~\ref{vsini_fixed_table}
and for relaxed $v\,\sin\,i$ in Table~\ref{vsini_free_table}.
The $\chi^2_{\rm SPE}$ contribution to the total $\chi^2$
is more than three orders of magnitude larger than $\chi^2_{\rm VIS}$
because of the number of measurements.
While $\chi^2_{\rm SPE}$ and $\chi^2_{\rm VIS}$ are the lowest
for $i=14$\,deg in the models with fixed $v\,\sin\,i$,
there is a tension between spectra and visibilities
for models with relaxed $v\,\sin\,i$,
which describes the interferometric data much better.

For comparison, we reconverged a model based only on spectroscopy,
starting from the parameters with the lowest $\chi^2_{\rm SPE}$
in Table~\ref{vsini_free_table} (bold);
again with relaxed $v\,\sin\,i$.
The best-fit model is shown in Fig.~\ref{spe},
corresponding to $\chi^2_\mathrm{SPE} = 1797$.
The free parameters and their resulting values were
$P_{\mathrm{rot}} = 4.535$\,d,
$T = 24915$\,K, and
$\log g =3.01$,
$i = 21.0$\,deg.
Other parameters were fixed and set according to interferometry.
The derived projected velocity,
$v\sin i = 111$\,km\,s$^{-1}$,
is substantially lower.
A forward model based on parameters from interferometry, 
with higher $v\sin i$,
gives smeared lines, while the observed lines are sharper.

\paragraph{Closure phase.}

Unfortunately, even though closure phases are very useful
to detect photocentre offsets,
the observed amplitude from PIONIER seems to be too high
(up to 4\,deg)
compared to any of our models.
Even critically rotating stars,
which have the largest differences of polar-to-equatorial temperatures (see Fig.~\ref{HeI_lines_plot}),
exhibit synthetic amplitudes less than 0.1\,deg
(Fig.~\ref{CLO}).
Nevertheless,
taking into account only the signs and trends of
$\arg T_3$ versus $\lambda$,
synthetic closure phases should change in the same sense as observed ones,
which allowed us to resolve the $\Omega$ ambiguity
(120 vs 300\,deg) in Sect.~\ref{VIS_sect}.

\paragraph{SED.}

As an independent check, we computed the PHOEBE2 model
for the SED (using again the new module).
As synthetic spectra,
we used the absolute fluxes from the
BSTAR,
OSTAR, and 
ATLAS
grids.
The observed SED values were discussed in Sect.~\ref{SED_sect}.
We converged only the effective temperature $T_\mathrm{eff}$.
Other parameters were fixed according to the model
based on interferometry
(from Fig.~\ref{eps_best_fit_free_vsini}). 
The resulting value was $T_\mathrm{eff} = 24\,880$\,K,
which is very close to the value inferred from the 
fitting of Balmer lines.
Our model was in good agreement with the observed SED
(see Fig.~\ref{SED}).

\paragraph{Compromise model.}

\begin{figure*}
\centering
\begin{tabular}{@{}c@{}c@{}c@{}}
\includegraphics[width=0.3\textwidth]{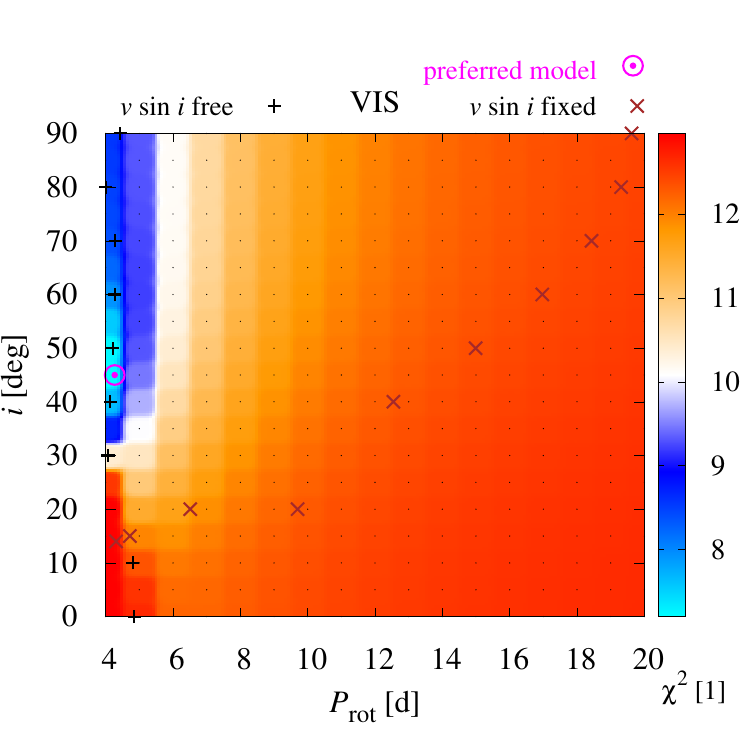} &
\includegraphics[width=0.3\textwidth]{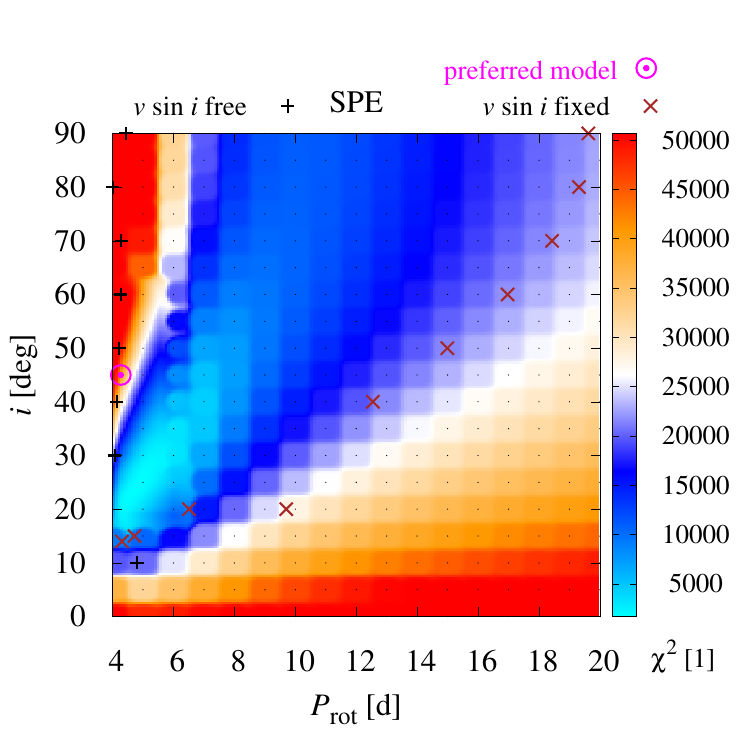} &
\includegraphics[width=0.3\textwidth]{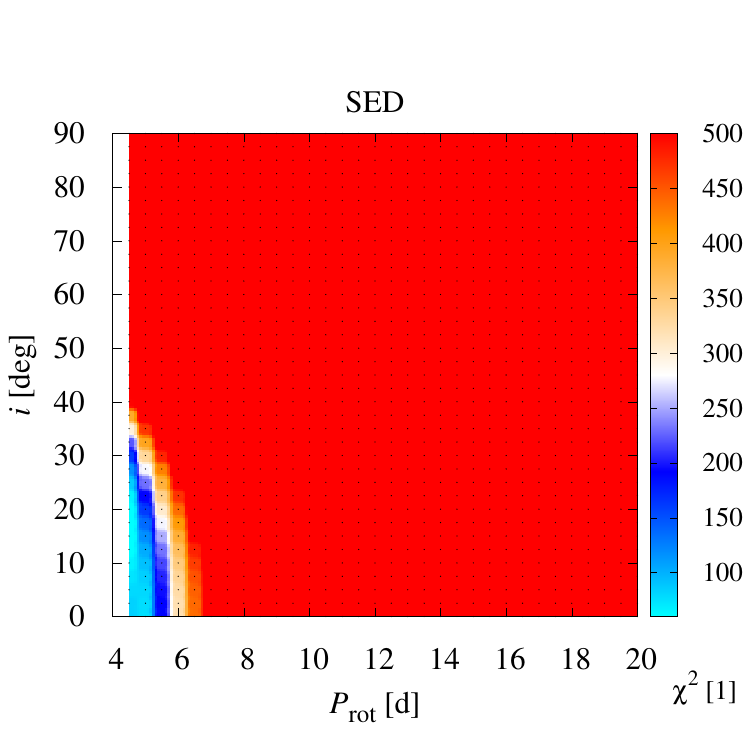} \\[-0.6cm]
\includegraphics[width=0.3\textwidth]{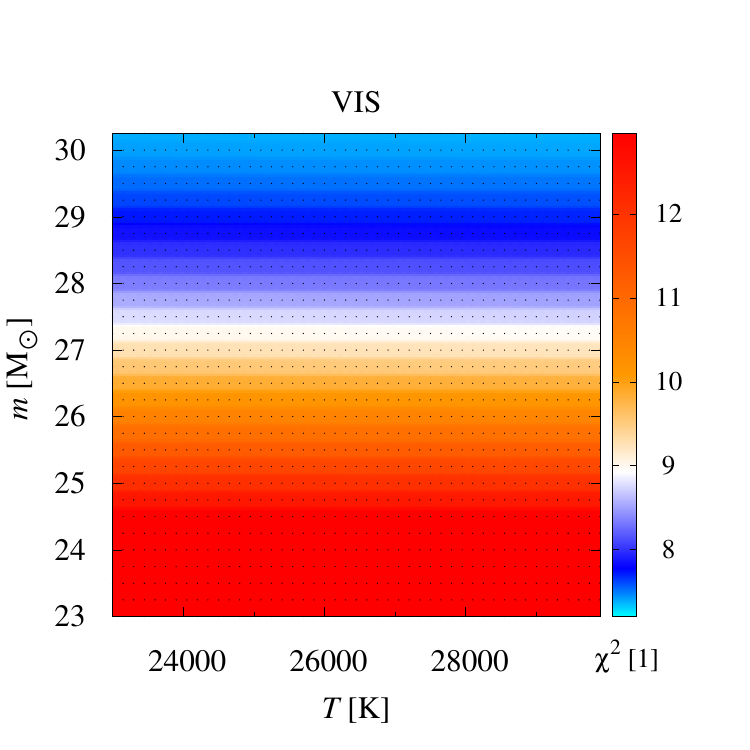} &
\includegraphics[width=0.3\textwidth]{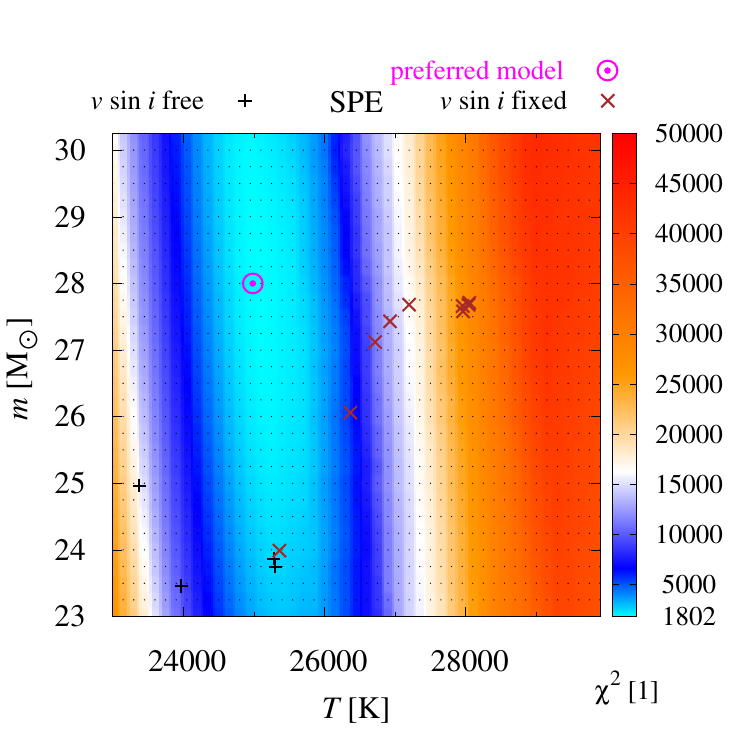} &
\includegraphics[width=0.3\textwidth]{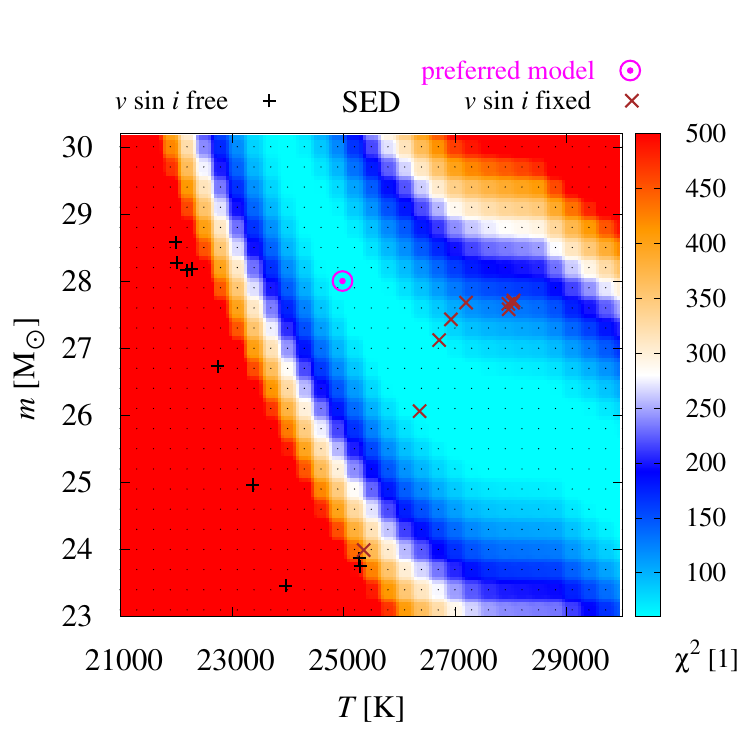} 
\\
\end{tabular}
\caption{
Maps of $\chi^2$ for the $\varepsilon$~Ori model.
We performed 2D mapping of $\chi^2$ across different datasets:
squared visibilities (VIS), 
spectral lines of Balmer series (SPE),
and spectral energy distribution (SED).
We systematically varied two parameters to create a regular grid
($P_\mathrm{rot}$ vs $i$ in the first row,
$T$ vs $m$ in the second row).
The remaining parameters were held fixed.
The colour scale was adjusted as follows:
cyan representing the best fit for a given dataset,
blue, the acceptable fits,
white, the poor fits,
orange, the unacceptable fits,
and red, the forbidden regions.
In the first row,
the fixed parameters are those resulting from the best fit of interferometric data in Fig.~\ref{eps_best_fit_free_vsini}:
$d$, $m$, and $\Omega$; other fixed parameters are
$T=25\,000$\,K,
$\log g=3.011$, and
$\gamma=25.9$\,km\,s$^{-1}$.
The dependent parameters were
$R_\mathrm{equiv}$ and $v \sin i$.
In the second row,
we set $P_\mathrm{rot}$ and $i$
according to the best fit in the first row.
The individual panels, from the upper left, 
are as follows:
I) the VIS dataset shows a preference for a critically rotating star and $i$ around 45\,deg; 
II) the SPE dataset (H lines) shows a preference for a fast-rotating star, $P_\mathrm{rot} = 5$\,d, $i=25$\,deg. 
The correlation between $P_\mathrm{rot}$ and $i$ is due to rotation and $v \sin i$;
III) the SED dataset also allows a critically or fast-rotating star for $i$ close to 20\,deg.
However, it is possible to re-fit $T$, $R$, or $d$ to achieve improved SED fits in other regions of the parameter space;
IV) the VIS dataset shows a flat dependence on temperature and the preferred mass; 
V) the SPE dataset for H lines demonstrates
a weak correlation between $T$ and $m$,
the best fit of $T$ is around 25\,000\,K; and
VI) the SED dataset is strongly correlated between $T$ and $m$
due to the Stefan–Boltzmann law and Eq.\,(\ref{Requiv}).
The crosses indicate the models from Tables\,\ref{vsini_fixed_table} and \ref{vsini_free_table}.
}
\label{maps} 
\end{figure*}

Finally, we computed 2D $\chi^2$ maps
to better understand the mutual relations between datasets
(see Fig.~\ref{maps}).
We always distinguished individual contributions to $\chi^2$,
namely from
interferometry (VIS),
spectroscopy (SPE)
and SED (SED).
In each panel, we were changing only two parameters to get a regular grid,
while other parameters are kept fixed.
For mapping, we chose the period $P_\mathrm{rot}$ from 4\,d to 20\,d with the step of 0.5\,d
versus the inclination $i$ from 0\,deg to 90\,deg with the step of 5\,deg.
Alternatively, we chose
the temperature, $T$, from 21\,000\,K to 30\,000\,K with the step of 500\,K
versus the mass, $m$, from 21\,\Mnom \ to 32\,\Mnom \ with the step of 0.5\,\Mnom.
We recall here that $m$ is always related to $R_\mathrm{equiv}$, as expressed in Eq.~(\ref{Requiv}).

The VIS maps with squared visibility measurements
show a preference for a critically rotating, oblate star,
seen at an oblique angle (40 to 60\,deg).
On the other hand, there is a negligible dependence on temperature.
A higher mass (28 to 30\,\Mnom) is preferred,
but this is certainly correlated with the distance via Eq.~(\ref{theta}).

The SPE maps with Balmer lines demonstrate that
only lower inclinations can appropriately fit
the depth and width of spectral lines.
There is a weak correlation between the mass and temperature, indicating the appropriate temperature is around 25\,000\,K, while the mass is not well-constrained .

\begin{figure}
\centering
\begin{tabular}{@{}c@{}}
\includegraphics[width=0.33\textwidth]{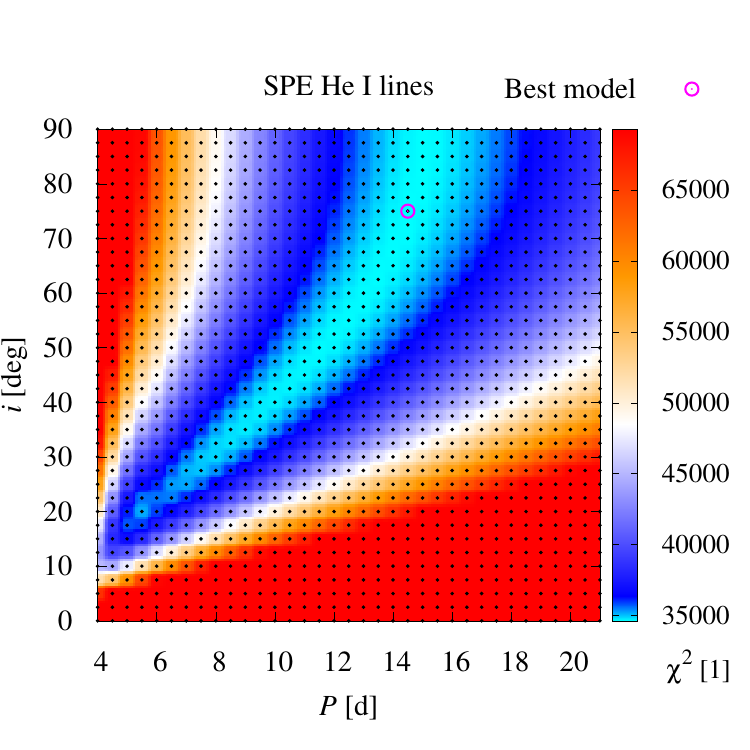} \\[-0.6cm]
\includegraphics[width=0.33\textwidth]{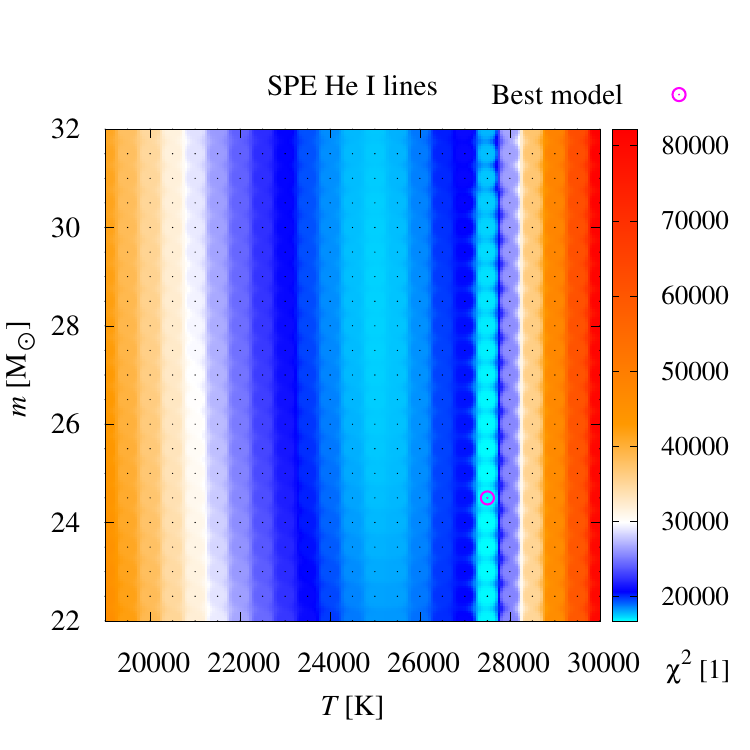}
\end{tabular}
\caption{
Same as Fig.~\ref{maps}, but for the spectroscopic dataset based on \ion{He}{I} lines.
Top:
SED dataset for \ion{He}{I} lines exhibits results very similar 
to spectroscopy based on Balmer lines.
The higher $P_\mathrm{rot}$ and $i$ values are excluded because of other datasets.
Bottom:
SED dataset for \ion{He}{I} lines,
indicating two possible solutions at $T=25\,000$\,K or $27\,500$\,K,
due to systematic offsets in some lines.
The model with $T=25\,000$\,K fits  the depths of \ion{He}{I} lines better.
The map also shows no preference for mass.
}
\label{maps_HeI}
\end{figure}

The SPE maps based on \ion{He}{I} lines
(Fig.~\ref{maps_HeI})
are quite similar.
Although there is a correlation of good fits
between $i$ and $P_\mathrm{rot}$,
we see a possible solution for a fast-rotating star,
with a similar inclination to that derived from the Balmer lines. 
Again, the mass and temperature are weakly correlated,
indicating the temperature 25\,000\,K.
We note that the fits around 28\,000\,K resulted in poorer line depths. The SED maps are, of course, strongly influenced
by the fixed temperature and mass,
but they do show good solutions for a critically rotating star,
with a similar inclination as for spectra.
The mass and temperature are strongly correlated.

We conclude that the inclination is the only uncertain parameter,
for which the results differ according to
interferometry (40 to 60\,deg) versus spectroscopy (10 to 30\,deg).
Other parameters correspond to all three kinds of observations.
As a compromise, we prefer the value around 40\,deg,
which partly explains both the interferometry and the spectroscopy.
The resulting parameters of $\varepsilon$~Ori are summarised
in Table~\ref{resulting_parameters}.

\paragraph{Distance.}

Finally, we tested distances of 350\,pc and 420\,pc (i.e. lower and upper limits from Sect.~\ref{parallax}).
We adjusted parameters of $\varepsilon$~Ori
using the constraints, Eqs.\,(\ref{Prot})--(\ref{theta}),
in particular,
$m$, $P_{\rm rot}$, $R_{\rm equiv}$, and $T_{\rm eff}$,
so that these alternative models are also close to
the respective datasets (VIS, SPE, and SED).
According to $\chi^2$ maps similar to Fig.~\ref{maps},
the first row ($P_{\rm rot}$ vs $i$) remains essentially unchanged,
which would correspond to a poorly constrained distance.
However, the second row ($T_{\rm eff}$, $m$) shows
that these alternative models exhibit even larger tension
between individual datasets (see Figs.\,\ref{maps350} and \ref{maps420}).
We thus conclude that there is no improvement
and the nominal distance of 384\,pc is still the preferred one.

\begin{figure}
\centering
\includegraphics[width=0.41\textwidth]{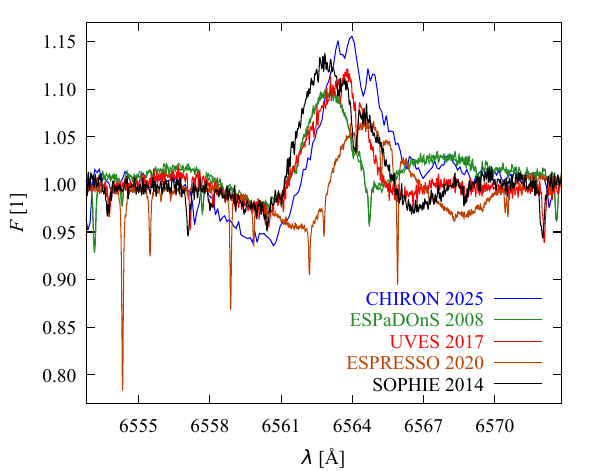}
\caption{
Spectroscopy of $\varepsilon$~Ori showing the H$_\alpha$ line profile
that is highly variable and can be in several morphologies, like
P Cygni profile,
inverse P Cygni profile,
double emission, or
pure emission \citep{Thompson_2013}.
It implies an intense wind (e.g. \citealt{Puebla2016})
or a decretion disk fed by mass loss from the star.
This variable line was not used in our modelling. 
}
\label{H_alpha} 
\end{figure}

\begin{figure}
\centering
\includegraphics[width=0.37\textwidth]{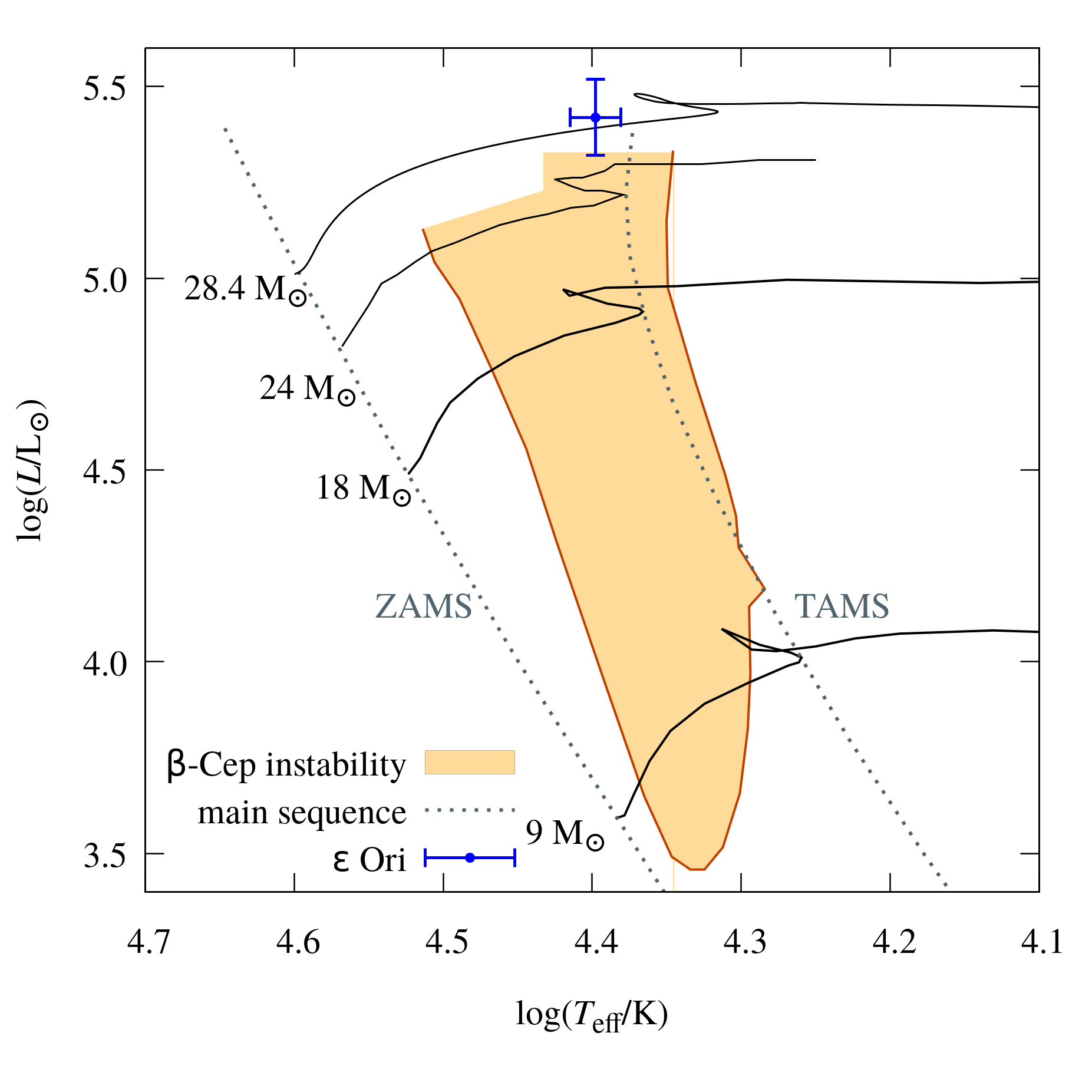}
\caption{
HRD showing the position of $\varepsilon$~Ori
(blue)
in comparison to the $\beta$-Cep instability region
($\ell = 0$; orange), several evolutionary tracks
(black)
from \citet{Paxton_2015ApJS..220...15P},
and evolutionary track for the mass of 
$\varepsilon$~Ori computed by us using the MESA code.
Note: the instability was computed only up to 24\,\Mnom,
but the region most likely continues upwards,
and the instability also occurs for the luminosity class Ia
($\log L/\Lnom \ge 5.3$).}
\label{HRD}
\end{figure}

\begin{table}
\caption{
Resulting parameters of $\varepsilon$~Ori 
based on a compromise among interferometry, spectroscopy, and SED.
}
\label{resulting_parameters}
\centering
\small
\renewcommand{\arraystretch}{1.2}
\begin{tabular*}{\hsize}{l@{\extracolsep{\fill}}cccrl@{\hspace{0.2cm}}}
\noalign{\smallskip}\hline\hline\noalign{\smallskip}
Parameter           & {Value}    & {Uncertainty} & Unit          & Note       \\
\noalign{\smallskip}\hline\noalign{\smallskip}
$m$                  & 28.42         & 2.0                & \Mnom         & fitted  \\
$P_\mathrm{rot}$     & 4.27          & ${}^{+1.0}_{-0.0}$ & d             & fitted  \\
$i$                  & 45            & 20                 & deg           & fitted  \\
$\Omega$             & 301.6         & 30                 & deg           & fitted  \\
$d$                  & 384           & 8                  & pc            & fixed   \\ 
$T_{\mathrm{eff}}$   & 24985         & 1000               & K             & fixed   \\
$\log g$             & 3.011         & 0.1                & cgs           & fixed   \\
$\beta$              & 1.0           & 0.1                & 1             & fixed   \\
$\gamma$             & 25.0          & 0.9                & km\,s$^{-1}$  & derived \\
$\gamma_{\rm LSR}$   & 8.2           & 0.7                & km\,s$^{-1}$  & derived \\
$R_{\rm equiv}$      & 27.36         & 1.5                & \Rnom         & derived \\
$R_{\rm pole}$       & 22.29         & 1.2                & \Rnom         & derived \\
$R_{\rm equ}$        & 33.61         & 1.8                & \Rnom         & derived \\
$\theta_{\rm equiv}$ & 0.667         & 0.035              & mas           & derived \\
$\theta_{\rm pole}$  & 0.540         & 0.064              & mas           & derived \\
$\theta_{\rm equ}$   & 0.814         & 0.073              & mas           & derived \\
$L$                  & 271\,000      & 38\,000            & \Lnom         & derived \\
$v\sin i$            & 220           & ${}^{+40}_{-100}$  & km\,s$^{-1}$  & derived \\
$v$                  & 326           & ${}^{+50}_{-120}$  & km\,s$^{-1}$  & derived \\
$f$                  & 1.47          & 0.35               & rad/d         & derived \\
$F_\mathrm{obs}$     & $57.8 \cdot 10^{-9}$ & $4\cdot 10^{-9}$ & W\,m$^{-2}$   & derived \\
\noalign{\smallskip}\hline\noalign{\smallskip}
\end{tabular*}
\tablefoot{
$m$~denotes the mass;
$P_\mathrm{rot}$, the rotation period;
$i$, the inclination of the stellar rotation axis with respect to the sky;
$\Omega$, the longitude of the ascending node of the star (i.e. equator);
$d$, the distance;
$T_{\mathrm{eff}}$, the effective temperature;
$\log g$, the logarithm of surface gravity;
$\beta$, bolometric gravity brightening exponent;
$\gamma$, the systemic velocity;
$\gamma_{\rm LSR}$, the systemic velocity with respect to the local standard of rest;
$R_{\rm equiv}$, the equivalent radius;
$R_{\rm pole}$, the pole radius;
$R_{\rm equ}$, the equatorial radius;
$\theta_{\rm equiv}$, the equivalent angular diameter;
$\theta_{\rm pole}$, the polar angular diameter;
$\theta_{\rm equ}$, the equatorial angular diameter;
$L$, the luminosity;
$v_\mathrm{rot}\sin i$, the projected rotation velocity;
$v_\mathrm{rot}$, the rotation velocity; and
$f$, the rotation frequency.
}
\end{table}

\section{Discussion}

Our interferometry of $\varepsilon$~Ori can be compared to
\citet{Abeysekara_2020NatAs...4.1164A},
who performed intensity interferometry in B~band.
Their limb-darkened angular diameter,
$\theta_\mathrm{LD} = (0.660\pm 0.018)\,{\rm mas}$,
is in perfect agreement with our measurements in H~band
(Table~\ref{resulting_parameters}).
Moreover, for their baseline T1-T4
\citep[Fig.~2, $u$-direction, east-west]{Abeysekara_2020NatAs...4.1164A},
 approximately corresponding to our baseline A0-J6,
their observations of $V^2$ are above their spherical model,
indicating a shorter angular size in that direction,
which agrees with our non-spherical shape.
Our model is also consistent with seminal papers 
\citet{Hanbury-Brown1974MNRAS.167..121H},
even though their value of $\theta = (0.69 \pm 0.04)\,{\rm mas}$
was determined using only two baselines,
and \citet{Code1976ApJ...203..417C},
reporting the bolometric flux of
$F_\mathrm{obs} = (60.2 \pm 5.4) \cdot 10^{-9}$\,W\,m$^{-2}$
and the effective temperature of 
$T = (24\,820 \pm 920)$\,K. However, as discussed in Sect.~\ref{sec3},
the observed closure phases are high (up to 4\,deg) 
compared to our model.
This raises the question of what might have been missing in our model.

\paragraph{Possible binarity.}
Such a modest amplitude in closure phases
indicates a flux asymmetry with low contrast,
up to $4^\circ = 0.07\,\mathrm{rad} = 7$\%,
in the Johnson:H passband.
For this estimate, we assumed a maximally asymmetric object
(i.e. a binary)
and used the following closure phase approximation,
$\arg T_3 \approx F_2/F_1\,\sin\left(2\pi\frac{B\phi}{\lambda}\right)$,
where $F_1$, $F_2$ denote the fluxes of the two stars
and $\phi$ the angular scale.
Based on this, we estimated the mass of the companion,
assuming that it is in the main sequence,
up to ${\sim}12\,\Mnom$.
To explain the observed dependence on $B/\lambda$,
the angular scale should be of the order of 0.5\,mas,
corresponding to ${\sim}40\,\Rnom$.
From this, we estimated the orbital period,
$P = 2\pi\sqrt{a^3/(GM)} \approx 6\,{\rm d}$,
assuming the total mass ${\sim}30$\,\Mnom;  namely, of the same order as the primary.
The RV amplitudes for the primary and secondary are
of the order of a hundred ${\rm km}\,{\rm s}^{-1}$,
but these could be suppressed by orbit orientation
(e.g. for $i = 5^\circ$, $15\,{\rm km}\,{\rm s}^{-1}$).
In the dataset of \citet{Thompson_2013},
there are RV variations of this order observed
in the He I 5876\,\AA \ line.
However, they are not persistent, and they cannot be phased
with a fixed period.
Only for one season,
\citet{Thompson_2013} reported a period of about 5\,d.
In the context of our model, this could correspond
to the rotation period of the primary
or (even more likely) to the variability of circumstellar material. More information on RV changes are given in Appendix \ref{radial_velocities}.

\paragraph{Wind.}
Another asymmetry could be related to wind \citep{Puebla2016},
if the material expelled from $\varepsilon$~Ori
is not evenly distributed,
but concentrated in relatively dense clumps.
This phenomenon is especially relevant for hot, massive stars such as
B supergiants,
O-type stars, or
Wolf-Rayet (WR) stars
\citep{Puls2008A&ARv..16..209P,Krticka_2021A&A...647A..28K}.
It is only when clumps are locally optically thick in the Johnson:H passband that
they contribute to the respective continuum flux
and the closure phase signal observed by PIONIER.
Otherwise, wind is observed in the H$_\alpha$ line
(Fig.~\ref{H_alpha}),
which is highly variable and exhibits a range of morphologies,
from a classical P Cygni profile, 
an inverse P Cygni profile,
double emission,
to pure emission
(see also \citealt{Thompson_2013}, Fig.~1).
Because of the variability in terms of H$_\alpha$ as well as the intense stellar wind and associated mass loss, we should be cautious when considering $\varepsilon$~Ori as a standard for B-type stars \citep{Negueruela2024A&A...690A.176N}.

\paragraph{Disk.}
An additional asymmetry might be produced by an unresolved disk,
fed by mass loss from the star
and partially eclipsed by the star.
If this is the source of the double emission sometimes observed 
in the wings of H$_\alpha$
(Fig.~\ref{H_alpha}),
the redshift and the blueshift,
$\Delta\lambda \simeq 6\,\AA$,
are interpreted as rotation,
$\Delta v = \Delta\lambda/\lambda\,c = 274\,{\rm km}\,{\rm s}^{-1}$.
This is surprisingly similar to the projected velocity $v\sin i$
in our models of $\varepsilon$~Ori.
Finally, if the star is close to critical rotation,
as indicated by the PIONIER observations,
we naturally expect an ongoing outflow from the equator.

\paragraph{Pulsations.}
Moreover,
$\varepsilon$~Ori is close to the $\beta$~Cep instability region
\citep{Paxton_2015ApJS..220...15P}
for low-order modes, $\ell = 0, 1, 2$
(see Fig.~\ref{HRD}).
Photometric observations of $\varepsilon$~Ori
seem to be compatible with
typical periods ranging from 0.1 to 0.6\,d,
and the amplitudes $\sim$0.1\,mag
for $\beta$~Cep variables
\citep{Lesh_1978ARA&A..16..215L}.
Such pulsations can enhance mass loss,
especially if the star is rotating near its critical velocity.
On the other hand, the periodograms for BRITE light curves
or for RVs do not exhibit distinct peaks
\citep{Krticka2018A&A...617A.121K,Thompson_2013},
so their interpretation is
stochastic, episodic activity due to mass loss.

\paragraph{Stellar evolution.}
If we, only for the moment, assumed a standard stellar evolution model
for a 28.4\,\Mnom, rotating star,
it would take ${\sim}7.3\,{\rm My}$
to evolve from the zero-age main sequence (ZAMS) to B0Ia
\citep{Ekstrom_2012A&A...537A.146E,Paxton_2013ApJS..208....4P}.
This is about $25\%$ longer than for a non-rotating star
due to enhanced mixing and homogenised composition.
For comparison,
there are two groups of
low-mass, young stellar objects (YSOs)
observed in the vicinity of $\varepsilon$~Ori,
denoted \enquote*{Orion~C} and \enquote*{Orion~D}
\citep{Kounkel_2018AJ....156...84K}.
Their ages span from approximately 0.3 to 5.5\,My.
They were classified as Class II or III,
as some are still accreting and associated with translucent gas clouds
\citep{Briceno_2005AJ....129..907B}.
This implies either that the stellar evolution for $\varepsilon$~Ori was non-standard
(e.g. with extreme mass loss)
or the formation of low- and high-mass stars was not
eodem loco et tempore 
Additionally,
both the distance and the systemic velocity of $\varepsilon$~Ori
(384\,pc and $8\,{\rm km}\,{\rm s}^{-1}$,
measured with respect to the local standard of rest)
seem to be in the middle, namely, between Orion~C and Orion~D groups
(\citealt{Kounkel_2018AJ....156...84K}, Fig.~11).
This relation might suggest that high-mass stars form
either a bit earlier than low-mass stars or elsewhere
\citep{Sanhueza_2017ApJ...841...97S,Maud_2018A&A...620A..31M}.

\newpage

\section{Conclusions}

We obtained and calibrated 
VLTI/GRAVITY and PIONIER interferometric data for the brightest stars
in  Orion's Belt,
our closest star-forming region.
In this first paper, 
we modelled the supergiant $\varepsilon$~Ori based on
VLTI/PIONIER visibility data, 
CFHT and CTIO spectra, and
absolute fluxes. 
While the models based on interferometry indicate an oblate,
critically rotating star,
the models based on spectroscopy indicate a fast-rotating, 
less oblate star.
The fast rotation might imply that $\varepsilon$~Ori is a merger,
for instance, similar to blue supergiants from \cite{Menon2024ApJ...963L..42M}.

We also discussed the possibility of binarity,
based on the asymmetry visible in closure phase data.
However, due to the lack of consistent radial velocity variations,
we excluded this binary model.
We rather attributed the asymmetry to the clumped stellar wind
or possibly to a decretion disk.
To better constrain its properties,
it is necessary to observe $\varepsilon$~Ori
with a spectro-interferometer such as CHARA/SPICA
\citep{Mourard_2024SPIE13095E..03M},
which is capable of scanning across the H$_\alpha$ line. In a forthcoming second paper, we will focus on the
multiple stellar systems in  Orion's Belt.

\leavevmode\vskip\baselineskip

\begin{acknowledgements}
A.O. was supported by GA UK grant no. 113224 of the Grant Agency of Charles University.
M.B. was supported by GA ČR grant no. 25-16507S of the Czech Science Foundation.
We thank other observers at the Hvar Observatory, 
Hrvoje Bo\v{z}i\'c,
Domagoj Ru\v{z}djak, and
Davor Sudar
for securing data for $\varepsilon$~Ori.
Thanks are due to an anonymous referee whose constructive feedback helped to improve the paper.
\end{acknowledgements}

\bibliographystyle{aa}
\bibliography{Orion_belt}

\begin{thebibliography}{93}
\expandafter\ifx\csname natexlab\endcsname\relax\def\natexlab#1{#1}\fi

\bibitem[{Abbott {et~al.}(2021)Abbott, Abbott, Abraham, Acernese, Ackley,
  Adams, Adams, Adhikari, Adya, Affeldt, Agathos, Agatsuma, Aggarwal, Aguiar,
  Aiello, Ain, Ajith, Allen, Allocca, Altin, Amato, Anand, Ananyeva, Anderson,
  Anderson, Angelova, Ansoldi, Antelis, Antier, Appert, Arai, Araya, Areeda,
  Arène, Arnaud, Aronson, Arun, Asali, Ascenzi, Ashton, Aston, Astone, Aubin,
  Aufmuth, AultONeal, Austin, Avendano, Babak, Badaracco, Bader, Bae, Baer,
  Bagnasco, Baird, Ball, Ballardin, Ballmer, Bals, Balsamo, Baltus, Banagiri,
  Bankar, Bankar, Barayoga, Barbieri, Barish, Barker, Barneo, Barnum, Barone,
  Barr, Barsotti, Barsuglia, Barta, Bartlett, Bartos, Bassiri, Basti, Bawaj,
  Bayley, Bazzan, Becher, Bécsy, Bedakihale, Bejger, Belahcene, Beniwal,
  Benjamin, Bennett, Bentley, Bergamin, Berger, Bergmann, Bernuzzi, Berry,
  Bersanetti, Bertolini, Betzwieser, Bhandare, Bhandari, Bhattacharjee, Bidler,
  Bilenko, Billingsley, Birney, Birnholtz, Biscans, Bischi, Biscoveanu, Bisht,
  Bitossi, Bizouard, Blackburn, Blackman, Blair, Blair, Blair, Blanch, Bobba,
  Bode, Boer, Boetzel, Bogaert, Boldrini, Bondu, Bonilla, Bonnand, Booker,
  Boom, Bork, Boschi, Bose, Bossilkov, Boudart, Bouffanais, Bozzi, Bradaschia,
  Brady, Bramley, Branchesi, Brau, Breschi, Briant, Briggs, Brighenti, Brillet,
  Brinkmann, Brockill, Brooks, Brooks, Brown, Brunett, Bruno, Bruntz, Buikema,
  Bulik, Bulten, Buonanno, Buscicchio, Buskulic, Byer, Cabero, Cadonati,
  Caesar, Cagnoli, Cahillane, Bustillo, Callaghan, Callister, Calloni, Camp,
  Canepa, Cannon, Cao, Cao, Carapella, Carbognani, Carney, Carpinelli, Carullo,
  Carver, Diaz, Casentini, Caudill, Cavaglià, Cavalier, Cavalieri, Cella,
  Cerdá-Durán, Cesarini, Chaibi, Chakravarti, Chan, Chan, Chandra, Chanial,
  Chao, Charlton, Chase, Chassande-Mottin, Chatterjee, Chattopadhyay,
  Chaturvedi, Chatziioannou, Chen, Chen, Chen, Chen, Cheng, Cheong, Chia,
  Chiadini, Chierici, Chincarini, Chiummo, Cho, Cho, Cho, Choate, Christensen,
  Chu, Chua, Chung, Chung, Ciani, Ciecielag, Cieślar, Cifaldi, Ciobanu,
  Ciolfi, Cipriano, Cirone, Clara, Clark, Clark, Clarke, Clearwater, Clesse,
  Cleva, Coccia, Cohadon, Cohen, Colleoni, Collette, Collins, Colpi,
  Constancio, Conti, Cooper, Corban, Corbitt, Cordero-Carrión, Corezzi,
  Corley, Cornish, Corre, Corsi, Cortese, Costa, Cotesta, Coughlin, Coughlin,
  Coulon, Countryman, Couvares, Covas, Coward, Cowart, Coyne, Coyne, Creighton,
  Creighton, Croquette, Crowder, Cudell, Cullen, Cumming, Cummings, Cunningham,
  Cuoco, Curylo, Canton, Dálya, Dana, DaneshgaranBajastani, D’Angelo,
  Danilishin, D’Antonio, Danzmann, Darsow-Fromm, Dasgupta, Datrier, Dattilo,
  Dave, Davier, Davies, Davis, Daw, Dean, DeBra, Deenadayalan, Degallaix,
  Laurentis, Deléglise, Favero, Lillo, Lillo, Pozzo, DeMarchi, Matteis,
  D’Emilio, Demos, Denker, Dent, Depasse, Pietri, Rosa, Rossi, DeSalvo,
  de~Varona, Dhurandhar, Díaz, Diaz-Ortiz, Didio, Dietrich, Fiore, DiFronzo,
  Giorgio, Giovanni, Giovanni, Girolamo, Lieto, Ding, Pace, Palma, Renzo,
  Divakarla, Dmitriev, Doctor, D’Onofrio, Donovan, Dooley, Doravari,
  Dorrington, Downes, Drago, Driggers, Du, Ducoin, Dupej, Durante, D’Urso,
  Duverne, Dwyer, Easter, Eddolls, Edelman, Edo, Edy, Effler, Eichholz,
  Eikenberry, Eisenmann, Eisenstein, Ejlli, Errico, Essick, Estellés, Estevez,
  Etienne, Etzel, Evans, Evans, Ewing, Fafone, Fair, Fairhurst, Fan, Farah,
  Farinon, Farr, Farr, Fauchon-Jones, Favata, Fays, Fazio, Feicht, Fejer, Feng,
  Fenyvesi, Ferguson, Fernandez-Galiana, Ferrante, Ferreira, Fidecaro, Figura,
  Fiori, Fiorucci, Fishbach, Fisher, Fishner, Fittipaldi, Fitz-Axen, Fiumara,
  Flaminio, Floden, Flynn, Fong, Font, Forsyth, Fournier, Frasca, Frasconi,
  Frei, Freise, Frey, Frey, Fritschel, Frolov, Fronzé, Fulda, Fyffe, Gabbard,
  Gadre, Gaebel, Gair, Gais, Galaudage, Gamba, Ganapathy, Ganguly, Gaonkar,
  Garaventa, García-Quirós, Garufi, Gateley, Gaudio, Gayathri, Gemme, Gennai,
  George, George, Gergely, Ghonge, Ghosh, Ghosh, Ghosh, Giacomazzo, Giacoppo,
  Giaime, Giardina, Gibson, Gier, Gill, Giri, Glanzer, Gleckl, Godwin, Goetz,
  Goetz, Gohlke, Goncharov, González, Gopakumar, Gossan, Gosselin, Gouaty,
  Grace, Grado, Granata, Granata, Grant, Gras, Grassia, Gray, Gray, Greco,
  Green, Green, Gretarsson, Griggs, Grignani, Grimaldi, Grimes, Grimm, Grote,
  Grunewald, Gruning, Guerrero, Guidi, Guimaraes, Guixé, Gulati, Guo, Gupta,
  Gupta, Gupta, Gustafson, Gustafson, Guzman, Haegel, Halim, Hall, Hamilton,
  Hammond, Haney, Hanke, Hanks, Hanna, Hannuksela, Hannuksela, Hansen, Hansen,
  Hanson, Harder, Hardwick, Haris, Harms, Harry, Harry, Hartwig, Hasskew,
  Haster, Haughian, Hayes, Healy, Heidmann, Heintze, Heinze, Heinzel, Heitmann,
  Hellman, Hello, Helmling-Cornell, Hemming, Hendry, Heng, Hennes, Hennig,
  Hennig, Vivanco, Heurs, Hild, Hill, Hines, Hochheim, Hofgard, Hofman,
  Hohmann, Holgado, Holland, Hollows, Holmes, Holt, Holz, Hopkins, Horst,
  Hough, Howell, Hoy, Hoyland, Huang, Hübner, Huddart, Huerta, Hughey, Hui,
  Husa, Huttner, Hutzler, Huxford, Huynh-Dinh, Idzkowski, Iess, Imperato,
  Inchauspe, Ingram, Intini, Isi, Iyer, JaberianHamedan, Jacqmin, Jadhav,
  Jadhav, James, Jani, Janssens, Janthalur, Jaranowski, Jariwala, Jaume,
  Jenkins, Jeunon, Jiang, Johns, Jones, Jones, Jones, Jones, Jones, Jonker, Ju,
  Junker, Kalaghatgi, Kalogera, Kamai, Kandhasamy, Kang, Kanner, Kapadia,
  Kapasi, Karathanasis, Karki, Kashyap, Kasprzack, Kastaun, Katsanevas,
  Katsavounidis, Katzman, Kawabe, Kéfélian, Keitel, Key, Khadka, Khalili,
  Khan, Khan, Khazanov, Khetan, Khursheed, Kijbunchoo, Kim, Kim, Kim, Kim, Kim,
  Kim, Kimball, King, Kinley-Hanlon, Kirchhoff, Kissel, Kleybolte, Klimenko,
  Knowles, Knyazev, Koch, Koehlenbeck, Koekoek, Koley, Kolstein, Komori,
  Kondrashov, Kontos, Koper, Korobko, Korth, Kovalam, Kozak, Krämer, Kringel,
  Krishnendu, Królak, Kuehn, Kumar, Kumar, Kumar, Kumar, Kuns, Kwang, Lackey,
  Laghi, Lalande, Lam, Lamberts, Landry, Lane, Lang, Lange, Lantz, Lanza, Rosa,
  Lartaux-Vollard, Lasky, Laxen, Lazzarini, Lazzaro, Leaci, Leavey, Lecoeuche,
  Lee, Lee, Lee, Lee, Lehmann, Leon, Leroy, Letendre, Levin, Li, Li, Li, Li,
  Li, Linde, Linker, Linley, Littenberg, Liu, Liu, Llorens-Monteagudo, Lo,
  Lockwood, London, Longo, Lorenzini, Loriette, Lormand, Losurdo, Lough,
  Lousto, Lovelace, Lück, Lumaca, Lundgren, Ma, Macas, MacInnis, Macleod,
  MacMillan, Macquet, Hernandez, Magaña-Sandoval, Magazzù, Magee, Majorana,
  Maksimovic, Maliakal, Malik, Man, Mandic, Mangano, Mansell, Manske,
  Mantovani, Mapelli, Marchesoni, Marion, Márka, Márka, Markakis, Markosyan,
  Markowitz, Maros, Marquina, Marsat, Martelli, Martin, Martin, Martinez,
  Martinez, Martynov, Masalehdan, Mason, Massera, Masserot, Massinger,
  Masso-Reid, Mastrogiovanni, Matas, Mateu-Lucena, Matichard, Matiushechkina,
  Mavalvala, Maynard, McCann, McCarthy, McClelland, McCormick, McCuller,
  McGuire, McIsaac, McIver, McManus, McRae, McWilliams, Meacher, Meadors,
  Mehmet, Mehta, Melatos, Melchor, Mendell, Menendez-Vazquez, Mercer, Mereni,
  Merfeld, Merilh, Merritt, Merzougui, Meshkov, Messenger, Messick, Metzdorff,
  Meyers, Meylahn, Mhaske, Miani, Miao, Michaloliakos, Michel, Middleton,
  Milano, Miller, Miller, Millhouse, Mills, Milotti, Milovich-Goff, Minazzoli,
  Minenkov, Mir, Mishkin, Mishra, Mistry, Mitra, Mitrofanov, Mitselmakher,
  Mittleman, Mo, Mogushi, Mohapatra, Mohite, Molina, Molina-Ruiz, Mondin,
  Montani, Moore, Moraru, Morawski, Moreno, Morisaki, Mours, Mow-Lowry, Mozzon,
  Muciaccia, Mukherjee, Mukherjee, Mukherjee, Mukherjee, Mukund, Mullavey,
  Munch, Muñiz, Murray, Nadji, Nagar, Nardecchia, Naticchioni, Nayak, Neil,
  Neilson, Nelemans, Nelson, Nery, Neunzert, Ng, Ng, Nguyen, Nguyen, Nguyen,
  Nichols, Nissanke, Nocera, Noh, North, Nothard, Nuttall, Oberling, O’Brien,
  O’Dell, Oganesyan, Ogin, Oh, Oh, Ohme, Ohta, Okada, Olivetto, Oppermann,
  Oram, O’Reilly, Ormiston, Ormsby, Ortega, O’Shaughnessy, Ossokine,
  Osthelder, Ottaway, Overmier, Owen, Pace, Pagano, Page, Pagliaroli, Pai, Pai,
  Palamos, Palashov, Palomba, Pan, Panda, Pang, Pankow, Pannarale, Pant,
  Paoletti, Paoli, Paolone, Parker, Pascucci, Pasqualetti, Passaquieti,
  Passuello, Patel, Patricelli, Payne, Pechsiri, Pedraza, Pegoraro, Pele, Penn,
  Perego, Perez, Périgois, Perreca, Perriès, Petermann, Petterson, Pfeiffer,
  Pham, Phukon, Piccinni, Pichot, Piendibene, Piergiovanni, Pierini, Pierro,
  Pillant, Pilo, Pinard, Pinto, Piotrzkowski, Pirello, Pitkin, Placidi,
  Plastino, Pluchar, Poggiani, Polini, Pong, Ponrathnam, Popolizio, Porter,
  Poverman, Powell, Pracchia, Prajapati, Prasai, Prasanna, Pratten, Prestegard,
  Principe, Prodi, Prokhorov, Prosposito, Puecher, Punturo, Puosi, Puppo,
  Pürrer, Qi, Quetschke, Quinonez, Quitzow-James, Raab, Raaijmakers, Radkins,
  Radulesco, Raffai, Rafferty, Rail, Raja, Rajan, Rajbhandari, Rakhmanov,
  Ramirez, Ramirez, Ramos-Buades, Rana, Rao, Rapagnani, Rapol, Ratto, Raymond,
  Razzano, Read, Regimbau, Rei, Reid, Reitze, Rettegno, Ricci, Richardson,
  Richardson, Richardson, Ricker, Riemenschneider, Riles, Rizzo, Robertson,
  Robinet, Rocchi, Rocha, Rodriguez, Rodriguez-Soto, Rolland, Rollins, Roma,
  Romanelli, Romano, Romel, Romero, Romero-Shaw, Romie, Ronchini, Rose, Rose,
  Rose, Rosell, Rosińska, Rosofsky, Ross, Rowan, Rowlinson, Roy, Roy, Ruggi,
  Ryan, Sachdev, Sadecki, Sakellariadou, Salafia, Salconi, Saleem, Samajdar,
  Sanchez, Sanchez, Sanchez, Sanchis-Gual, Sanders, Santiago, Santos,
  Saravanan, Sarin, Sassolas, Sathyaprakash, Sauter, Savage, Savant, Sawant,
  Sayah, Schaetzl, Schale, Scheel, Scheuer, Schindler-Tyka, Schmidt, Schnabel,
  Schofield, Schönbeck, Schreiber, Schulte, Schutz, Schwarm, Schwartz, Scott,
  Scott, Seglar-Arroyo, Seidel, Sellers, Sengupta, Sennett, Sentenac, Sequino,
  Sergeev, Setyawati, Shaffer, Shahriar, Sharifi, Sharma, Sharma, Shawhan,
  Shen, Shikauchi, Shink, Shoemaker, Shoemaker, Shukla, ShyamSundar,
  Sieniawska, Sigg, Singer, Singh, Singh, Singha, Singhal, Sintes, Sipala,
  Skliris, Slagmolen, Slaven-Blair, Smetana, Smith, Smith, Somala, Son, Soni,
  Sorazu, Sordini, Sorrentino, Sorrentino, Soulard, Souradeep, Sowell, Spencer,
  Spera, Srivastava, Srivastava, Staats, Stachie, Steer, Steinke, Steinlechner,
  Steinlechner, Steinmeyer, Stevenson, Stolle-McAllister, Stops, Stover,
  Strain, Stratta, Strunk, Sturani, Stuver, Südbeck, Sudhagar, Sudhir, Suh,
  Summerscales, Sun, Sun, Sunil, Sur, Suresh, Sutton, Swinkels, Szczepańczyk,
  Tacca, Tait, Talbot, Tanasijczuk, Tanner, Tao, Tapia, Martin, Tasson, Taylor,
  Tenorio, Terkowski, Thirugnanasambandam, Thomas, Thomas, Thomas, Thompson,
  Thondapu, Thorne, Thrane, Tiwari, Tiwari, Tiwari, Toland, Tolley, Tonelli,
  Tornasi, Torres-Forné, Torrie, e~Melo, Töyrä, Tran, Trapananti, Travasso,
  Traylor, Tringali, Tripathee, Trovato, Trudeau, Tsai, Tsang, Tse, Tso,
  Tsukada, Tsuna, Tsutsui, Turconi, Ubhi, Udall, Ueno, Ugolini, Unnikrishnan,
  Urban, Usman, Utina, Vahlbruch, Vajente, Vajpeyi, Valdes, Valentini, Valsan,
  van Bakel, van Beuzekom, van~den Brand, Broeck, Vander-Hyde, van~der Schaaf,
  van Heijningen, Vardaro, Vargas, Varma, Vass, Vasúth, Vecchio, Vedovato,
  Veitch, Veitch, Venkateswara, Venneberg, Venugopalan, Verkindt, Verma, Veske,
  Vetrano, Viceré, Viets, Villa-Ortega, Vinet, Vitale, Vo, Vocca, Vorvick,
  Vyatchanin, Wade, Wade, Wade, Walet, Walker, Wallace, Wallace, Walsh, Wang,
  Wang, Wang, Wang, Ward, Warner, Was, Washington, Watchi, Weaver, Wei,
  Weinert, Weinstein, Weiss, Wellmann, Wen, Weßels, Westhouse, Wette, Whelan,
  White, White, Whiting, Whittle, Wilken, Williams, Williams, Williamson,
  Willis, Willke, Wilson, Wimmer, Winkler, Wipf, Woan, Woehler, Wofford, Wong,
  Wrangel, Wright, Wu, Wysocki, Xiao, Yamamoto, Yang, Yang, Yang, Yap, Yeeles,
  Yoon, Yu, Yu, Yuen, Zadrożny, Zanolin, Zelenova, Zendri, Zevin, Zhang,
  Zhang, Zhang, Zhang, Zhao, Zhao, Zhou, Zhou, Zhu, Zimmerman, Zucker, Zweizig,
  Collaboration, \& the Virgo~Collaboration}]{Abbott_2021}
Abbott, R., Abbott, T.~D., Abraham, S., {et~al.} 2021, \apjl, 913, L7

\bibitem[{{Abeysekara} {et~al.}(2020){Abeysekara}, {Benbow}, {Brill},
  {Buckley}, {Christiansen}, {Chromey}, {Daniel}, {Davis}, {Falcone}, {Feng},
  {Finley}, {Fortson}, {Furniss}, {Gent}, {Giuri}, {Gueta}, {Hanna}, {Hassan},
  {Hervet}, {Holder}, {Hughes}, {Humensky}, {Kaaret}, {Kertzman}, {Kieda},
  {Krennrich}, {Kumar}, {LeBohec}, {Lin}, {Lundy}, {Maier}, {Matthews},
  {Moriarty}, {Mukherjee}, {Nievas-Rosillo}, {O'Brien}, {Ong}, {Otte},
  {Pfrang}, {Pohl}, {Prado}, {Pueschel}, {Quinn}, {Ragan}, {Reynolds},
  {Ribeiro}, {Richards}, {Roache}, {Ryan}, {Santander}, {Sembroski}, {Wakely},
  {Weinstein}, {Wilcox}, {Williams}, \&
  {Williamson}}]{Abeysekara_2020NatAs...4.1164A}
{Abeysekara}, A.~U., {Benbow}, W., {Brill}, A., {et~al.} 2020, Nature
  Astronomy, 4, 1164

\bibitem[{{Allen} {et~al.}(2014){Allen}, {Ochsenbein}, {Derriere}, {Boch},
  {Fernique}, \& {Landais}}]{2014ASPC..485..219A}
{Allen}, M.~G., {Ochsenbein}, F., {Derriere}, S., {et~al.} 2014, in
  Astronomical Society of the Pacific Conference Series, Vol. 485, Astronomical
  Data Analysis Software and Systems XXIII, ed. N.~{Manset} \& P.~{Forshay},
  219

\bibitem[{{Almeida} {et~al.}(2015){Almeida}, {Sana}, {de Mink}, {Tramper},
  {Soszy{\'n}ski}, {Langer}, {Barb{\'a}}, {Cantiello}, {Damineli}, {de Koter},
  {Garcia}, {Gr{\"a}fener}, {Herrero}, {Howarth}, {Ma{\'\i}z Apell{\'a}niz},
  {Norman}, {Ram{\'\i}rez-Agudelo}, \& {Vink}}]{Almeida2015ApJ...812..102A}
{Almeida}, L.~A., {Sana}, H., {de Mink}, S.~E., {et~al.} 2015, \apj, 812, 102

\bibitem[{{Anderson} \& {Francis}(2012)}]{Anderson_2012AstL...38..331A}
{Anderson}, E. \& {Francis}, C. 2012, Astronomy Letters, 38, 331

\bibitem[{{Atri} {et~al.}(2019){Atri}, {Miller-Jones}, {Bahramian}, {Plotkin},
  {Jonker}, {Nelemans}, {Maccarone}, {Sivakoff}, {Deller}, {Chaty}, {Torres},
  {Horiuchi}, {McCallum}, {Natusch}, {Phillips}, {Stevens}, \&
  {Weston}}]{Atri2019MNRAS.489.3116A}
{Atri}, P., {Miller-Jones}, J.~C.~A., {Bahramian}, A., {et~al.} 2019, \mnras,
  489, 3116

\bibitem[{{Baumann} {et~al.}(2022){Baumann}, {Boch}, {Pineau}, {Fernique},
  {Bot}, \& {Allen}}]{Aladin2022ASPC..532....7B}
{Baumann}, M., {Boch}, T., {Pineau}, F.-X., {et~al.} 2022, in Astronomical
  Society of the Pacific Conference Series, Vol. 532, Astronomical Data
  Analysis Software and Systems XXX, ed. J.~E. {Ruiz}, F.~{Pierfedereci}, \&
  P.~{Teuben}, 7

\bibitem[{{Bessell}(2000)}]{Bessell2000eaa..bookE1939B}
{Bessell}, M. 2000, in Encyclopedia of Astronomy and Astrophysics, ed.
  P.~{Murdin}, 1939

\bibitem[{{Blanco} {et~al.}(1970){Blanco}, {Demers}, \&
  {Douglass}}]{Blanco1970pcmc.book.....B}
{Blanco}, M., {Demers}, S., \& {Douglass}, G.~G. 1970, {Photoelectric catalogue
  - Magnitudes and colors or stars in the U, B, V and Uc, B, V systems}

\bibitem[{Bo\v{z}i\'c \& Harmanec(2023)}]{Hvar}
Bo\v{z}i\'c, H. \& Harmanec, P. 2023, Highlights of a half century of the UBV
  photometry at Hvar

\bibitem[{{Brice{\~n}o} {et~al.}(2005){Brice{\~n}o}, {Calvet}, {Hern{\'a}ndez},
  {Vivas}, {Hartmann}, {Downes}, \& {Berlind}}]{Briceno_2005AJ....129..907B}
{Brice{\~n}o}, C., {Calvet}, N., {Hern{\'a}ndez}, J., {et~al.} 2005, \aj, 129,
  907

\bibitem[{{Bro{\v{z}}}(2017)}]{Broz_2017ApJS..230...19B}
{Bro{\v{z}}}, M. 2017, \apjs, 230, 19

\bibitem[{{Brož} {et~al.}(2025{\natexlab{a}}){Brož}, {Prša}, {Conroy}, \&
  {Abdul-Masih}}]{Broz_2025b}
{Brož}, M., {Prša}, A., {Conroy}, K.~E., \& {Abdul-Masih}, M.
  2025{\natexlab{a}}, \url{https://arxiv.org/abs/2506.20868}
  [\eprint[arXiv]{2506.20868}]

\bibitem[{{Brož} {et~al.}(2025{\natexlab{b}}){Brož}, {Prša}, {Conroy},
  {Oplištilová}, \& {Horvat}}]{Broz_2025a}
{Brož}, M., {Prša}, A., {Conroy}, K.~E., {Oplištilová}, A., \& {Horvat}, M.
  2025{\natexlab{b}}, \url{https://arxiv.org/abs/2506.20865}
  [\eprint[arXiv]{2506.20865}]

\bibitem[{{Castelli} \& {Kurucz}(2003)}]{Castelli2003IAUS..210P.A20C}
{Castelli}, F. \& {Kurucz}, R.~L. 2003, in IAU Symposium, Vol. 210, Modelling
  of Stellar Atmospheres, ed. N.~{Piskunov}, W.~W. {Weiss}, \& D.~F. {Gray},
  A20

\bibitem[{{Clayton}(1983)}]{Clayton1983psen.book.....C}
{Clayton}, D.~D. 1983, {Principles of stellar evolution and nucleosynthesis}

\bibitem[{{Code} {et~al.}(1976){Code}, {Davis}, {Bless}, \&
  {Brown}}]{Code1976ApJ...203..417C}
{Code}, A.~D., {Davis}, J., {Bless}, R.~C., \& {Brown}, R.~H. 1976, \apj, 203,
  417

\bibitem[{{Conroy} {et~al.}(2020){Conroy}, {Kochoska}, {Hey}, {Pablo},
  {Hambleton}, {Jones}, {Giammarco}, {Abdul-Masih}, \&
  {Pr{\v{s}}a}}]{Conroy_2020ApJS..250...34C}
{Conroy}, K.~E., {Kochoska}, A., {Hey}, D., {et~al.} 2020, \apjs, 250, 34

\bibitem[{{Corcoran} {et~al.}(2015){Corcoran}, {Nichols}, {Pablo}, {Shenar},
  {Pollock}, {Waldron}, {Moffat}, {Richardson}, {Russell}, {Hamaguchi},
  {Huenemoerder}, {Oskinova}, {Hamann}, {Naz{\'e}}, {Ignace}, {Evans}, {Lomax},
  {Hoffman}, {Gayley}, {Owocki}, {Leutenegger}, {Gull}, {Hole}, {Lauer}, \&
  {Iping}}]{corcoran2015}
{Corcoran}, M.~F., {Nichols}, J.~S., {Pablo}, H., {et~al.} 2015, \apj, 809, 132

\bibitem[{{Cutri} {et~al.}(2012){Cutri}, L., \&
  {Conrow}}]{Cutri_2012yCat.2311....0C}
{Cutri}, R.~M., L., W.~E., \& {Conrow}, T. 2012, VizieR Online Data Catalog,
  II/311

\bibitem[{{Cutri} {et~al.}(2003){Cutri}, {Skrutskie}, {van Dyk}, {Beichman},
  {Carpenter}, {Chester}, {Cambresy}, {Evans}, {Fowler}, {Gizis}, {Howard},
  {Huchra}, {Jarrett}, {Kopan}, {Kirkpatrick}, {Light}, {Marsh}, {McCallon},
  {Schneider}, {Stiening}, {Sykes}, {Weinberg}, {Wheaton}, {Wheelock}, \&
  {Zacarias}}]{Cutri_2003yCat.2246....0C}
{Cutri}, R.~M., {Skrutskie}, M.~F., {van Dyk}, S., {et~al.} 2003, VizieR Online
  Data Catalog, II/246

\bibitem[{{Ducati}(2002{\natexlab{a}})}]{2002yCat.2237....0D}
{Ducati}, J.~R. 2002{\natexlab{a}}, {VizieR Online Data Catalog: Catalogue of
  Stellar Photometry in Johnson's 11-color system.}, CDS/ADC Collection of
  Electronic Catalogues, 2237, 0 (2002)

\bibitem[{{Ducati}(2002{\natexlab{b}})}]{Ducati_2002yCat.2237....0D}
{Ducati}, J.~R. 2002{\natexlab{b}}, VizieR Online Data Catalog

\bibitem[{{Egan} {et~al.}(2003){Egan}, {Price}, {Kraemer}, {Mizuno}, {Carey},
  {Wright}, {Engelke}, {Cohen}, \& {Gugliotti}}]{Egan_2003yCat.5114....0E}
{Egan}, M.~P., {Price}, S.~D., {Kraemer}, K.~E., {et~al.} 2003, VizieR Online
  Data Catalog, V/114

\bibitem[{{Eisenhauer} {et~al.}(2011){Eisenhauer}, {Perrin}, {Brandner},
  {Straubmeier}, {Perraut}, {Amorim}, {Sch{\"o}ller}, {Gillessen}, {Kervella},
  {Benisty}, {Araujo-Hauck}, {Jocou}, {Lima}, {Jakob}, {Haug}, {Cl{\'e}net},
  {Henning}, {Eckart}, {Berger}, {Garcia}, {Abuter}, {Kellner}, {Paumard},
  {Hippler}, {Fischer}, {Moulin}, {Villate}, {Avila}, {Gr{\"a}ter}, {Lacour},
  {Huber}, {Wiest}, {Nolot}, {Carvas}, {Dorn}, {Pfuhl}, {Gendron}, {Kendrew},
  {Yazici}, {Anton}, {Jung}, {Thiel}, {Choquet}, {Klein}, {Teixeira}, {Gitton},
  {Moch}, {Vincent}, {Kudryavtseva}, {Str{\"o}bele}, {Sturm}, {F{\'e}dou},
  {Lenzen}, {Jolley}, {Kister}, {Lapeyr{\`e}re}, {Naranjo}, {Lucuix},
  {Hofmann}, {Chapron}, {Neumann}, {Mehrgan}, {Hans}, {Rousset}, {Ramos},
  {Suarez}, {Lederer}, {Reess}, {Rohloff}, {Haguenauer}, {Bartko}, {Sevin},
  {Wagner}, {Lizon}, {Rabien}, {Collin}, {Finger}, {Davies}, {Rouan},
  {Wittkowski}, {Dodds-Eden}, {Ziegler}, {Cassaing}, {Bonnet}, {Casali},
  {Genzel}, \& {Lena}}]{GRAVITY2011Msngr.143...16E}
{Eisenhauer}, F., {Perrin}, G., {Brandner}, W., {et~al.} 2011, The Messenger,
  143, 16

\bibitem[{{Ekstr{\"o}m} {et~al.}(2012){Ekstr{\"o}m}, {Georgy}, {Eggenberger},
  {Meynet}, {Mowlavi}, {Wyttenbach}, {Granada}, {Decressin}, {Hirschi},
  {Frischknecht}, {Charbonnel}, \& {Maeder}}]{Ekstrom_2012A&A...537A.146E}
{Ekstr{\"o}m}, S., {Georgy}, C., {Eggenberger}, P., {et~al.} 2012, \aap, 537,
  A146

\bibitem[{{Fan} {et~al.}(2017){Fan}, {Welty}, {York}, {Sonnentrucker},
  {Dahlstrom}, {Baskes}, {Friedman}, {Hobbs}, {Jiang}, {Rachford}, {Snow},
  {Sherman}, \& {Zhao}}]{Fan2017ApJ...850..194F}
{Fan}, H., {Welty}, D.~E., {York}, D.~G., {et~al.} 2017, \apj, 850, 194

\bibitem[{{Freudling} {et~al.}(2013){Freudling}, {Romaniello}, {Bramich},
  {Ballester}, {Forchi}, {Garc{\'{\i}}a-Dabl{\'o}}, {Moehler}, \&
  {Neeser}}]{Freudling2013A&A...559A..96F}
{Freudling}, W., {Romaniello}, M., {Bramich}, D.~M., {et~al.} 2013, \aap, 559,
  A96

\bibitem[{{Friedman} {et~al.}(2011){Friedman}, {York}, {McCall}, {Dahlstrom},
  {Sonnentrucker}, {Welty}, {Drosback}, {Hobbs}, {Rachford}, \&
  {Snow}}]{Friedman2011ApJ...727...33F}
{Friedman}, S.~D., {York}, D.~G., {McCall}, B.~J., {et~al.} 2011, \apj, 727, 33

\bibitem[{{Gaia Collaboration}(2020)}]{Gaia_2020yCat.1350....0G}
{Gaia Collaboration}. 2020, VizieR Online Data Catalog, I/350

\bibitem[{{Gaia Collaboration} {et~al.}(2021){Gaia Collaboration}, {Brown},
  {Vallenari}, {Prusti}, {de Bruijne}, {Babusiaux}, {Biermann}, {Creevey},
  {Evans}, {Eyer}, {Hutton}, {Jansen}, {Jordi}, {Klioner}, {Lammers},
  {Lindegren}, {Luri}, {Mignard}, {Panem}, {Pourbaix}, {Randich}, {Sartoretti},
  {Soubiran}, {Walton}, {Arenou}, {Bailer-Jones}, {Bastian}, {Cropper},
  {Drimmel}, {Katz}, {Lattanzi}, {van Leeuwen}, {Bakker}, {Cacciari},
  {Casta{\~n}eda}, {De Angeli}, {Ducourant}, {Fabricius}, {Fouesneau},
  {Fr{\'e}mat}, {Guerra}, {Guerrier}, {Guiraud}, {Jean-Antoine Piccolo},
  {Masana}, {Messineo}, {Mowlavi}, {Nicolas}, {Nienartowicz}, {Pailler},
  {Panuzzo}, {Riclet}, {Roux}, {Seabroke}, {Sordo}, {Tanga}, {Th{\'e}venin},
  {Gracia-Abril}, {Portell}, {Teyssier}, {Altmann}, {Andrae}, {Bellas-Velidis},
  {Benson}, {Berthier}, {Blomme}, {Brugaletta}, {Burgess}, {Busso}, {Carry},
  {Cellino}, {Cheek}, {Clementini}, {Damerdji}, {Davidson}, {Delchambre},
  {Dell'Oro}, {Fern{\'a}ndez-Hern{\'a}ndez}, {Galluccio}, {Garc{\'\i}a-Lario},
  {Garcia-Reinaldos}, {Gonz{\'a}lez-N{\'u}{\~n}ez}, {Gosset}, {Haigron},
  {Halbwachs}, {Hambly}, {Harrison}, {Hatzidimitriou}, {Heiter},
  {Hern{\'a}ndez}, {Hestroffer}, {Hodgkin}, {Holl}, {Jan{\ss}en}, {Jevardat de
  Fombelle}, {Jordan}, {Krone-Martins}, {Lanzafame}, {L{\"o}ffler}, {Lorca},
  {Manteiga}, {Marchal}, {Marrese}, {Moitinho}, {Mora}, {Muinonen}, {Osborne},
  {Pancino}, {Pauwels}, {Petit}, {Recio-Blanco}, {Richards}, {Riello},
  {Rimoldini}, {Robin}, {Roegiers}, {Rybizki}, {Sarro}, {Siopis}, {Smith},
  {Sozzetti}, {Ulla}, {Utrilla}, {van Leeuwen}, {van Reeven}, {Abbas}, {Abreu
  Aramburu}, {Accart}, {Aerts}, {Aguado}, {Ajaj}, {Altavilla}, {{\'A}lvarez},
  {{\'A}lvarez Cid-Fuentes}, {Alves}, {Anderson}, {Anglada Varela}, {Antoja},
  {Audard}, {Baines}, {Baker}, {Balaguer-N{\'u}{\~n}ez}, {Balbinot}, {Balog},
  {Barache}, {Barbato}, {Barros}, {Barstow}, {Bartolom{\'e}}, {Bassilana},
  {Bauchet}, {Baudesson-Stella}, {Becciani}, {Bellazzini}, {Bernet}, {Bertone},
  {Bianchi}, {Blanco-Cuaresma}, {Boch}, {Bombrun}, {Bossini}, {Bouquillon},
  {Bragaglia}, {Bramante}, {Breedt}, {Bressan}, {Brouillet}, {Bucciarelli},
  {Burlacu}, {Busonero}, {Butkevich}, {Buzzi}, {Caffau}, {Cancelliere},
  {C{\'a}novas}, {Cantat-Gaudin}, {Carballo}, {Carlucci}, {Carnerero},
  {Carrasco}, {Casamiquela}, {Castellani}, {Castro-Ginard}, {Castro Sampol},
  {Chaoul}, {Charlot}, {Chemin}, {Chiavassa}, {Cioni}, {Comoretto}, {Cooper},
  {Cornez}, {Cowell}, {Crifo}, {Crosta}, {Crowley}, {Dafonte}, {Dapergolas},
  {David}, {David}, {de Laverny}, {De Luise}, {De March}, {De Ridder}, {de
  Souza}, {de Teodoro}, {de Torres}, {del Peloso}, {del Pozo}, {Delbo},
  {Delgado}, {Delgado}, {Delisle}, {Di Matteo}, {Diakite}, {Diener},
  {Distefano}, {Dolding}, {Eappachen}, {Edvardsson}, {Enke}, {Esquej}, {Fabre},
  {Fabrizio}, {Faigler}, {Fedorets}, {Fernique}, {Fienga}, {Figueras},
  {Fouron}, {Fragkoudi}, {Fraile}, {Franke}, {Gai}, {Garabato},
  {Garcia-Gutierrez}, {Garc{\'\i}a-Torres}, {Garofalo}, {Gavras}, {Gerlach},
  {Geyer}, {Giacobbe}, {Gilmore}, {Girona}, {Giuffrida}, {Gomel}, {Gomez},
  {Gonzalez-Santamaria}, {Gonz{\'a}lez-Vidal}, {Granvik},
  {Guti{\'e}rrez-S{\'a}nchez}, {Guy}, {Hauser}, {Haywood}, {Helmi}, {Hidalgo},
  {Hilger}, {H{\l}adczuk}, {Hobbs}, {Holland}, {Huckle}, {Jasniewicz},
  {Jonker}, {Juaristi Campillo}, {Julbe}, {Karbevska}, {Kervella}, {Khanna},
  {Kochoska}, {Kontizas}, {Kordopatis}, {Korn}, {Kostrzewa-Rutkowska},
  {Kruszy{\'n}ska}, {Lambert}, {Lanza}, {Lasne}, {Le Campion}, {Le Fustec},
  {Lebreton}, {Lebzelter}, {Leccia}, {Leclerc}, {Lecoeur-Taibi}, {Liao},
  {Licata}, {Lindstr{\o}m}, {Lister}, {Livanou}, {Lobel}, {Madrero Pardo},
  {Managau}, {Mann}, {Marchant}, {Marconi}, {Marcos Santos}, {Marinoni},
  {Marocco}, {Marshall}, {Martin Polo}, {Mart{\'\i}n-Fleitas}, {Masip},
  {Massari}, {Mastrobuono-Battisti}, {Mazeh}, {McMillan}, {Messina},
  {Michalik}, {Millar}, {Mints}, {Molina}, {Molinaro}, {Moln{\'a}r},
  {Montegriffo}, {Mor}, {Morbidelli}, {Morel}, {Morris}, {Mulone}, {Munoz},
  {Muraveva}, {Murphy}, {Musella}, {Noval}, {Ord{\'e}novic}, {Orr{\`u}},
  {Osinde}, {Pagani}, {Pagano}, {Palaversa}, {Palicio}, {Panahi}, {Pawlak},
  {Pe{\~n}alosa Esteller}, {Penttil{\"a}}, {Piersimoni}, {Pineau}, {Plachy},
  {Plum}, {Poggio}, {Poretti}, {Poujoulet}, {Pr{\v{s}}a}, {Pulone}, {Racero},
  {Ragaini}, {Rainer}, {Raiteri}, {Rambaux}, {Ramos}, {Ramos-Lerate}, {Re
  Fiorentin}, {Regibo}, {Reyl{\'e}}, {Ripepi}, {Riva}, {Rixon}, {Robichon},
  {Robin}, {Roelens}, {Rohrbasser}, {Romero-G{\'o}mez}, {Rowell}, {Royer},
  {Rybicki}, {Sadowski}, {Sagrist{\`a} Sell{\'e}s}, {Sahlmann}, {Salgado},
  {Salguero}, {Samaras}, {Sanchez Gimenez}, {Sanna}, {Santove{\~n}a},
  {Sarasso}, {Schultheis}, {Sciacca}, {Segol}, {Segovia}, {S{\'e}gransan},
  {Semeux}, {Shahaf}, {Siddiqui}, {Siebert}, {Siltala}, {Slezak}, {Smart},
  {Solano}, {Solitro}, {Souami}, {Souchay}, {Spagna}, {Spoto}, {Steele},
  {Steidelm{\"u}ller}, {Stephenson}, {S{\"u}veges}, {Szabados}, {Szegedi-Elek},
  {Taris}, {Tauran}, {Taylor}, {Teixeira}, {Thuillot}, {Tonello}, {Torra},
  {Torra}, {Turon}, {Unger}, {Vaillant}, {van Dillen}, {Vanel}, {Vecchiato},
  {Viala}, {Vicente}, {Voutsinas}, {Weiler}, {Wevers}, {Wyrzykowski}, {Yoldas},
  {Yvard}, {Zhao}, {Zorec}, {Zucker}, {Zurbach}, \&
  {Zwitter}}]{Brown2021A&A...649A...1G}
{Gaia Collaboration}, {Brown}, A.~G.~A., {Vallenari}, A., {et~al.} 2021, \aap,
  649, A1

\bibitem[{{Golay}(1974)}]{Golay1974ASSL...41.....G}
{Golay}, M. 1974, {Introduction to astronomical photometry}

\bibitem[{{GRAVITY Collaboration} {et~al.}(2017){GRAVITY Collaboration},
  {Abuter}, {Accardo}, {Amorim}, {Anugu}, {{\'A}vila}, {Azouaoui}, {Benisty},
  {Berger}, {Blind}, {Bonnet}, {Bourget}, {Brandner}, {Brast}, {Buron},
  {Burtscher}, {Cassaing}, {Chapron}, {Choquet}, {Cl{\'e}net}, {Collin},
  {Coud{\'e} Du Foresto}, {de Wit}, {de Zeeuw}, {Deen},
  {Delplancke-Str{\"o}bele}, {Dembet}, {Derie}, {Dexter}, {Duvert}, {Ebert},
  {Eckart}, {Eisenhauer}, {Esselborn}, {F{\'e}dou}, {Finger}, {Garcia}, {Garcia
  Dabo}, {Garcia Lopez}, {Gendron}, {Genzel}, {Gillessen}, {Gonte}, {Gordo},
  {Grould}, {Gr{\"o}zinger}, {Guieu}, {Haguenauer}, {Hans}, {Haubois}, {Haug},
  {Haussmann}, {Henning}, {Hippler}, {Horrobin}, {Huber}, {Hubert}, {Hubin},
  {Hummel}, {Jakob}, {Janssen}, {Jochum}, {Jocou}, {Kaufer}, {Kellner},
  {Kendrew}, {Kern}, {Kervella}, {Kiekebusch}, {Klein}, {Kok}, {Kolb}, {Kulas},
  {Lacour}, {Lapeyr{\`e}re}, {Lazareff}, {Le Bouquin}, {L{\`e}na}, {Lenzen},
  {L{\'e}v{\^e}que}, {Lippa}, {Magnard}, {Mehrgan}, {Mellein}, {M{\'e}rand},
  {Moreno-Ventas}, {Moulin}, {M{\"u}ller}, {M{\"u}ller}, {Neumann}, {Oberti},
  {Ott}, {Pallanca}, {Panduro}, {Pasquini}, {Paumard}, {Percheron}, {Perraut},
  {Perrin}, {Pfl{\"u}ger}, {Pfuhl}, {Phan Duc}, {Plewa}, {Popovic}, {Rabien},
  {Ram{\'\i}rez}, {Ramos}, {Rau}, {Riquelme}, {Rohloff}, {Rousset},
  {Sanchez-Bermudez}, {Scheithauer}, {Sch{\"o}ller}, {Schuhler}, {Spyromilio},
  {Straubmeier}, {Sturm}, {Suarez}, {Tristram}, {Ventura}, {Vincent},
  {Waisberg}, {Wank}, {Weber}, {Wieprecht}, {Wiest}, {Wiezorrek}, {Wittkowski},
  {Woillez}, {Wolff}, {Yazici}, {Ziegler}, \&
  {Zins}}]{GRAVITY2017A&A...602A..94G}
{GRAVITY Collaboration}, {Abuter}, R., {Accardo}, M., {et~al.} 2017, \aap, 602,
  A94

\bibitem[{{Green} {et~al.}(2019){Green}, {Schlafly}, {Zucker}, {Speagle}, \&
  {Finkbeiner}}]{Green_2019ApJ...887...93G}
{Green}, G.~M., {Schlafly}, E., {Zucker}, C., {Speagle}, J.~S., \&
  {Finkbeiner}, D. 2019, \apj, 887, 93

\bibitem[{{Guinan} {et~al.}(2012){Guinan}, {Mayer}, {Harmanec},
  {Bo{\v{z}}i{\'c}}, {Bro{\v{z}}}, {Nemravov{\'a}}, {Engle}, {{\v{S}}lechta},
  {Zasche}, {Wolf}, {Kor{\v{c}}{\'a}kov{\'a}}, \&
  {Johnston}}]{Guinan2012A&A...546A.123G}
{Guinan}, E.~F., {Mayer}, P., {Harmanec}, P., {et~al.} 2012, \aap, 546, A123

\bibitem[{{Hainich} {et~al.}(2014){Hainich}, {R{\"u}hling}, {Todt}, {Oskinova},
  {Liermann}, {Gr{\"a}fener}, {Foellmi}, {Schnurr}, \&
  {Hamann}}]{Hainich2014A&A...565A..27H}
{Hainich}, R., {R{\"u}hling}, U., {Todt}, H., {et~al.} 2014, \aap, 565, A27

\bibitem[{{Hanbury-Brown} {et~al.}(1974){Hanbury-Brown}, {Davis}, \&
  {Allen}}]{Hanbury-Brown1974MNRAS.167..121H}
{Hanbury-Brown}, R., {Davis}, J., \& {Allen}, L.~R. 1974, \mnras, 167, 121

\bibitem[{{Horvat} {et~al.}(2018){Horvat}, {Conroy}, {Pablo}, {Hambleton},
  {Kochoska}, {Giammarco}, \& {Pr{\v{s}}a}}]{Horvat_2018ApJS..237...26H}
{Horvat}, M., {Conroy}, K.~E., {Pablo}, H., {et~al.} 2018, \apjs, 237, 26

\bibitem[{{Houk} \& {Swift}(1999)}]{Houk1999MSS...C05....0H}
{Houk}, N. \& {Swift}, C. 1999, Michigan Spectral Survey, 5, 0

\bibitem[{{Hummel} {et~al.}(2013){Hummel}, {Rivinius}, {Nieva}, {Stahl}, {van
  Belle}, \& {Zavala}}]{Hummel2013}
{Hummel}, C.~A., {Rivinius}, T., {Nieva}, M.~F., {et~al.} 2013, \aap, 554, A52

\bibitem[{{Hummel} {et~al.}(2000){Hummel}, {White}, {Elias}, {Hajian}, \&
  {Nordgren}}]{Hummel2000ApJ...540L..91H}
{Hummel}, C.~A., {White}, N.~M., {Elias}, N.~M., I., {Hajian}, A.~R., \&
  {Nordgren}, T.~E. 2000, \apjl, 540, L91

\bibitem[{{Ishihara} {et~al.}(2010){Ishihara}, {Onaka}, {Kataza}, {Salama},
  {Alfageme}, {Cassatella}, {Cox}, {Garc{\'\i}a-Lario}, {Stephenson}, {Cohen},
  {Fujishiro}, {Fujiwara}, {Hasegawa}, {Ita}, {Kim}, {Matsuhara}, {Murakami},
  {M{\"u}ller}, {Nakagawa}, {Ohyama}, {Oyabu}, {Pyo}, {Sakon}, {Shibai},
  {Takita}, {Tanab{\'e}}, {Uemizu}, {Ueno}, {Usui}, {Wada}, {Watarai},
  {Yamamura}, \& {Yamauchi}}]{Ishihara_2010A&A...514A...1I}
{Ishihara}, D., {Onaka}, T., {Kataza}, H., {et~al.} 2010, \aap, 514, A1

\bibitem[{{Johnson} \& {Morgan}(1953)}]{Johnson1953ApJ...117..313J}
{Johnson}, H.~L. \& {Morgan}, W.~W. 1953, \apj, 117, 313

\bibitem[{{Jones} {et~al.}(2020){Jones}, {Conroy}, {Horvat}, {Giammarco},
  {Kochoska}, {Pablo}, {Brown}, {Sowicka}, \&
  {Pr{\v{s}}a}}]{Jones_2020ApJS..247...63J}
{Jones}, D., {Conroy}, K.~E., {Horvat}, M., {et~al.} 2020, \apjs, 247, 63

\bibitem[{{Kalari} {et~al.}(2022){Kalari}, {Horch}, {Salinas}, {Vink},
  {Andersen}, {Bestenlehner}, \& {Rubio}}]{Kalari2022ApJ...935..162K}
{Kalari}, V.~M., {Horch}, E.~P., {Salinas}, R., {et~al.} 2022, \apj, 935, 162

\bibitem[{{Keller} {et~al.}(2007){Keller}, {Schmidt}, {Bessell}, {Conroy},
  {Francis}, {Granlund}, {Kowald}, {Oates}, {Martin-Jones}, {Preston},
  {Tisserand}, {Vaccarella}, \& {Waterson}}]{Keller2007PASA...24....1K}
{Keller}, S.~C., {Schmidt}, B.~P., {Bessell}, M.~S., {et~al.} 2007, \pasa, 24,
  1

\bibitem[{{Kounkel} {et~al.}(2018){Kounkel}, {Covey}, {Su{\'a}rez},
  {Rom{\'a}n-Z{\'u}{\~n}iga}, {Hernandez}, {Stassun}, {Jaehnig}, {Feigelson},
  {Pe{\~n}a Ram{\'\i}rez}, {Roman-Lopes}, {Da Rio}, {Stringfellow}, {Kim},
  {Borissova}, {Fern{\'a}ndez-Trincado}, {Burgasser},
  {Garc{\'\i}a-Hern{\'a}ndez}, {Zamora}, {Pan}, \&
  {Nitschelm}}]{Kounkel_2018AJ....156...84K}
{Kounkel}, M., {Covey}, K., {Su{\'a}rez}, G., {et~al.} 2018, \aj, 156, 84

\bibitem[{{Krti{\v{c}}ka} \& {Feldmeier}(2018)}]{Krticka2018A&A...617A.121K}
{Krti{\v{c}}ka}, J. \& {Feldmeier}, A. 2018, \aap, 617, A121

\bibitem[{{Krti{\v{c}}ka} {et~al.}(2021){Krti{\v{c}}ka}, {Kub{\'a}t}, \&
  {Krti{\v{c}}kov{\'a}}}]{Krticka_2021A&A...647A..28K}
{Krti{\v{c}}ka}, J., {Kub{\'a}t}, J., \& {Krti{\v{c}}kov{\'a}}, I. 2021, \aap,
  647, A28

\bibitem[{{Kuhn} {et~al.}(2010){Kuhn}, {Getman}, {Feigelson}, {Reipurth},
  {Rodney}, \& {Garmire}}]{Kuhn2010ApJ...725.2485K}
{Kuhn}, M.~A., {Getman}, K.~V., {Feigelson}, E.~D., {et~al.} 2010, \apj, 725,
  2485

\bibitem[{{Lanz} \& {Huben\'y}(2003)}]{Lanz_2003ApJS..146..417L}
{Lanz}, T. \& {Huben\'y}, I. 2003, \apjs, 146, 417

\bibitem[{{Lanz} \& {Hubený}(2007)}]{Lanz_2007ApJS..169...83L}
{Lanz}, T. \& {Hubený}, I. 2007, \apjs, 169, 83

\bibitem[{{Le Bouquin} {et~al.}(2011){Le Bouquin}, {Berger}, {Lazareff},
  {Zins}, {Haguenauer}, {Jocou}, {Kern}, {Millan-Gabet}, {Traub}, {Absil},
  {Augereau}, {Benisty}, {Blind}, {Bonfils}, {Bourget}, {Delboulbe},
  {Feautrier}, {Germain}, {Gitton}, {Gillier}, {Kiekebusch}, {Kluska},
  {Knudstrup}, {Labeye}, {Lizon}, {Monin}, {Magnard}, {Malbet}, {Maurel},
  {M{\'e}nard}, {Micallef}, {Michaud}, {Montagnier}, {Morel}, {Moulin},
  {Perraut}, {Popovic}, {Rabou}, {Rochat}, {Rojas}, {Roussel}, {Roux},
  {Stadler}, {Stefl}, {Tatulli}, \& {Ventura}}]{Bouquin2011A&A...535A..67L}
{Le Bouquin}, J.~B., {Berger}, J.~P., {Lazareff}, B., {et~al.} 2011, \aap, 535,
  A67

\bibitem[{{Lesh} \& {Aizenman}(1978)}]{Lesh_1978ARA&A..16..215L}
{Lesh}, J.~R. \& {Aizenman}, M.~L. 1978, \araa, 16, 215

\bibitem[{{Ma{\'\i}z Apell{\'a}niz} {et~al.}(2019){Ma{\'\i}z Apell{\'a}niz},
  {Trigueros P{\'a}ez}, {Negueruela}, {Barb{\'a}}, {Sim{\'o}n-D{\'\i}az},
  {Lorenzo}, {Sota}, {Gamen}, {Fari{\~n}a}, {Salas}, {Caballero}, {Morrell},
  {Pellerin}, {Alfaro}, {Herrero}, {Arias}, \&
  {Marco}}]{Apellaniz2019A&A...626A..20M}
{Ma{\'\i}z Apell{\'a}niz}, J., {Trigueros P{\'a}ez}, E., {Negueruela}, I.,
  {et~al.} 2019, \aap, 626, A20

\bibitem[{{Marchant} {et~al.}(2021){Marchant}, {Pappas}, {Gallegos-Garcia},
  {Berry}, {Taam}, {Kalogera}, \&
  {Podsiadlowski}}]{Marchant2021A&A...650A.107M}
{Marchant}, P., {Pappas}, K. M.~W., {Gallegos-Garcia}, M., {et~al.} 2021, \aap,
  650, A107

\bibitem[{{Maud} {et~al.}(2018){Maud}, {Cesaroni}, {Kumar}, {van der Tak},
  {Allen}, {Hoare}, {Klaassen}, {Harsono}, {Hogerheijde}, {S{\'a}nchez-Monge},
  {Schilke}, {Ahmadi}, {Beltr{\'a}n}, {Beuther}, {Csengeri}, {Etoka}, {Fuller},
  {Galv{\'a}n-Madrid}, {Goddi}, {Henning}, {Johnston}, {Kuiper}, {Lumsden},
  {Moscadelli}, {Mottram}, {Peters}, {Rivilla}, {Testi}, {Vig}, {de Wit}, \&
  {Zinnecker}}]{Maud_2018A&A...620A..31M}
{Maud}, L.~T., {Cesaroni}, R., {Kumar}, M.~S.~N., {et~al.} 2018, \aap, 620, A31

\bibitem[{{Menon} {et~al.}(2024){Menon}, {Ercolino}, {Urbaneja}, {Lennon},
  {Herrero}, {Hirai}, {Langer}, {Schootemeijer}, {Chatzopoulos}, {Frank}, \&
  {Shiber}}]{Menon2024ApJ...963L..42M}
{Menon}, A., {Ercolino}, A., {Urbaneja}, M.~A., {et~al.} 2024, \apjl, 963, L42

\bibitem[{{Mermilliod}(1994)}]{Mermilliod1994BICDS..45....3M}
{Mermilliod}, J.~C. 1994, Bulletin d'Information du Centre de Donnees
  Stellaires, 45, 3

\bibitem[{{Mourard} {et~al.}(2024){Mourard}, {Meilland}, {Iba{\~n}ez Bustos},
  {Jonak}, {Berio}, {Dejonghe}, {Lecron}, {Morand}, {Salabert}, {Allouche},
  {Anugu}, {Bosio}, {Bourges}, {Creevey}, {Deheuvels}, {Domiciano de Souza},
  {Ebrahimkutty}, {Gies}, {Kubiak}, {Ligi}, {Ligon}, {Mella}, {Nardetto},
  {Perraut}, {Pitiot}, {Rousseau}, {Vrard}, {Schaefer}, {Spang}, {Turner},
  {Wittkowski}, \& {Zumbo}}]{Mourard_2024SPIE13095E..03M}
{Mourard}, D., {Meilland}, A., {Iba{\~n}ez Bustos}, R., {et~al.} 2024, in
  Society of Photo-Optical Instrumentation Engineers (SPIE) Conference Series,
  Vol. 13095, Optical and Infrared Interferometry and Imaging IX, ed.
  J.~{Kammerer}, S.~{Sallum}, \& J.~{Sanchez-Bermudez}, 1309503

\bibitem[{{Negueruela} {et~al.}(2024){Negueruela}, {Sim{\'o}n-D{\'\i}az}, {de
  Burgos}, {Casasbuenas}, \& {Beck}}]{Negueruela2024A&A...690A.176N}
{Negueruela}, I., {Sim{\'o}n-D{\'\i}az}, S., {de Burgos}, A., {Casasbuenas},
  A., \& {Beck}, P.~G. 2024, \aap, 690, A176

\bibitem[{Nelder \& Mead(1965)}]{Nelder_1965}
Nelder, J.~A. \& Mead, R. 1965, The Computer Journal, 7, 308

\bibitem[{{Neugebauer} {et~al.}(1984){Neugebauer}, {Habing}, {van Duinen},
  {Aumann}, {Baud}, {Beichman}, {Beintema}, {Boggess}, {Clegg}, {de Jong},
  {Emerson}, {Gautier}, {Gillett}, {Harris}, {Hauser}, {Houck}, {Jennings},
  {Low}, {Marsden}, {Miley}, {Olnon}, {Pottasch}, {Raimond}, {Rowan-Robinson},
  {Soifer}, {Walker}, {Wesselius}, \& {Young}}]{Neugebauer_1984ApJ...278L...1N}
{Neugebauer}, G., {Habing}, H.~J., {van Duinen}, R., {et~al.} 1984, \apjl, 278,
  L1

\bibitem[{{Nichols} {et~al.}(2015){Nichols}, {Huenemoerder}, {Corcoran},
  {Waldron}, {Naz{\'e}}, {Pollock}, {Moffat}, {Lauer}, {Shenar}, {Russell},
  {Richardson}, {Pablo}, {Evans}, {Hamaguchi}, {Gull}, {Hamann}, {Oskinova},
  {Ignace}, {Hoffman}, {Hole}, \& {Lomax}}]{nichols2015}
{Nichols}, J., {Huenemoerder}, D.~P., {Corcoran}, M.~F., {et~al.} 2015, \apj,
  809, 133

\bibitem[{{Opli{\v{s}}tilov{\'a}} {et~al.}(2023){Opli{\v{s}}tilov{\'a}},
  {Mayer}, {Harmanec}, {Bro{\v{z}}}, {Pigulski}, {Bo{\v{z}}i{\'c}}, {Zasche},
  {{\v{S}}lechta}, {Pablo}, {Ko{\l}aczek-Szyma{\'n}ski}, {Moffat}, {Lovekin},
  {Wade}, {Zwintz}, {Popowicz}, \& {Weiss}}]{Oplistilova2023AA...672A..31O}
{Opli{\v{s}}tilov{\'a}}, A., {Mayer}, P., {Harmanec}, P., {et~al.} 2023, \aap,
  672, A31

\bibitem[{{Pablo} {et~al.}(2015){Pablo}, {Richardson}, {Moffat}, {Corcoran},
  {Shenar}, {Benvenuto}, {Fuller}, {Naz{\'e}}, {Hoffman}, {Miroshnichenko},
  {Ma{\'\i}z Apell{\'a}niz}, {Evans}, {Eversberg}, {Gayley}, {Gull},
  {Hamaguchi}, {Hamann}, {Henrichs}, {Hole}, {Ignace}, {Iping}, {Lauer},
  {Leutenegger}, {Lomax}, {Nichols}, {Oskinova}, {Owocki}, {Pollock},
  {Russell}, {Waldron}, {Buil}, {Garrel}, {Graham}, {Heathcote}, {Lemoult},
  {Li}, {Mauclaire}, {Potter}, {Ribeiro}, {Matthews}, {Cameron}, {Guenther},
  {Kuschnig}, {Rowe}, {Rucinski}, {Sasselov}, \& {Weiss}}]{pablo2015}
{Pablo}, H., {Richardson}, N.~D., {Moffat}, A. F.~J., {et~al.} 2015, \apj, 809,
  134

\bibitem[{{Pauwels} {et~al.}(2023){Pauwels}, {Reggiani}, {Sana}, {Rainot}, \&
  {Kratter}}]{Pauwels2023A&A...678A.172P}
{Pauwels}, T., {Reggiani}, M., {Sana}, H., {Rainot}, A., \& {Kratter}, K. 2023,
  \aap, 678, A172

\bibitem[{{Paxton} {et~al.}(2013){Paxton}, {Cantiello}, {Arras}, {Bildsten},
  {Brown}, {Dotter}, {Mankovich}, {Montgomery}, {Stello}, {Timmes}, \&
  {Townsend}}]{Paxton_2013ApJS..208....4P}
{Paxton}, B., {Cantiello}, M., {Arras}, P., {et~al.} 2013, \apjs, 208, 4

\bibitem[{{Paxton} {et~al.}(2015){Paxton}, {Marchant}, {Schwab}, {Bauer},
  {Bildsten}, {Cantiello}, {Dessart}, {Farmer}, {Hu}, {Langer}, {Townsend},
  {Townsley}, \& {Timmes}}]{Paxton_2015ApJS..220...15P}
{Paxton}, B., {Marchant}, P., {Schwab}, J., {et~al.} 2015, \apjs, 220, 15

\bibitem[{{Perryman} {et~al.}(1997){Perryman}, {Lindegren}, {Kovalevsky},
  {Hoeg}, {Bastian}, {Bernacca}, {Cr{\'e}z{\'e}}, {Donati}, {Grenon},
  {Grewing}, {van Leeuwen}, {van der Marel}, {Mignard}, {Murray}, {Le Poole},
  {Schrijver}, {Turon}, {Arenou}, {Froeschl{\'e}}, \&
  {Petersen}}]{Perryman1997A&A...323L..49P}
{Perryman}, M.~A.~C., {Lindegren}, L., {Kovalevsky}, J., {et~al.} 1997, \aap,
  323, L49

\bibitem[{{Pietrzy{\'n}ski} {et~al.}(2019){Pietrzy{\'n}ski}, {Graczyk},
  {Gallenne}, {Gieren}, {Thompson}, {Pilecki}, {Karczmarek}, {G{\'o}rski},
  {Suchomska}, {Taormina}, {Zgirski}, {Wielg{\'o}rski}, {Ko{\l}aczkowski},
  {Konorski}, {Villanova}, {Nardetto}, {Kervella}, {Bresolin}, {Kudritzki},
  {Storm}, {Smolec}, \& {Narloch}}]{Pietrzynski_2019Natur.567..200P}
{Pietrzy{\'n}ski}, G., {Graczyk}, D., {Gallenne}, A., {et~al.} 2019, \nat, 567,
  200

\bibitem[{Pr{\v{s}}a(2011)}]{Prsa2011PHOEBE}
Pr{\v{s}}a, A. 2011, {PHOEBE Scientific Reference, Version 0.30}, Villanova
  University, Department of Astronomy \& Astrophysics, accessed: 2025-08-08

\bibitem[{{Pr{\v{s}}a} {et~al.}(2016){Pr{\v{s}}a}, {Conroy}, {Horvat}, {Pablo},
  {Kochoska}, {Bloemen}, {Giammarco}, {Hambleton}, \&
  {Degroote}}]{Prsa_2016ApJS..227...29P}
{Pr{\v{s}}a}, A., {Conroy}, K.~E., {Horvat}, M., {et~al.} 2016, \apjs, 227, 29

\bibitem[{{Puebla} {et~al.}(2016{\natexlab{a}}){Puebla}, {Hillier},
  {Zsarg{\'o}}, {Cohen}, \& {Leutenegger}}]{Puebla2016MNRAS.456.2907P}
{Puebla}, R.~E., {Hillier}, D.~J., {Zsarg{\'o}}, J., {Cohen}, D.~H., \&
  {Leutenegger}, M.~A. 2016{\natexlab{a}}, \mnras, 456, 2907

\bibitem[{{Puebla} {et~al.}(2016{\natexlab{b}}){Puebla}, {Hillier},
  {Zsarg{\'o}}, {Cohen}, \& {Leutenegger}}]{Puebla2016}
{Puebla}, R.~E., {Hillier}, D.~J., {Zsarg{\'o}}, J., {Cohen}, D.~H., \&
  {Leutenegger}, M.~A. 2016{\natexlab{b}}, \mnras, 456, 2907

\bibitem[{{Puls} {et~al.}(2008){Puls}, {Vink}, \&
  {Najarro}}]{Puls2008A&ARv..16..209P}
{Puls}, J., {Vink}, J.~S., \& {Najarro}, F. 2008, \aapr, 16, 209

\bibitem[{{Renzo} {et~al.}(2022){Renzo}, {Hendriks}, {van Son}, \&
  {Farmer}}]{Renzo2022RNAAS...6...25R}
{Renzo}, M., {Hendriks}, D.~D., {van Son}, L.~A.~C., \& {Farmer}, R. 2022,
  Research Notes of the American Astronomical Society, 6, 25

\bibitem[{{Repolust} {et~al.}(2005){Repolust}, {Puls}, {Hanson}, {Kudritzki},
  \& {Mokiem}}]{2005A&A...440..261R}
{Repolust}, T., {Puls}, J., {Hanson}, M.~M., {Kudritzki}, R.~P., \& {Mokiem},
  M.~R. 2005, \aap, 440, 261

\bibitem[{{Sana} {et~al.}(2014){Sana}, {Le Bouquin}, {Lacour}, {Berger},
  {Duvert}, {Gauchet}, {Norris}, {Olofsson}, {Pickel}, {Zins}, {Absil}, {de
  Koter}, {Kratter}, {Schnurr}, \& {Zinnecker}}]{Sana2014ApJS..215...15S}
{Sana}, H., {Le Bouquin}, J.~B., {Lacour}, S., {et~al.} 2014, \apjs, 215, 15

\bibitem[{{Sanhueza} {et~al.}(2017){Sanhueza}, {Jackson}, {Zhang},
  {Guzm{\'a}n}, {Lu}, {Stephens}, {Wang}, \&
  {Tatematsu}}]{Sanhueza_2017ApJ...841...97S}
{Sanhueza}, P., {Jackson}, J.~M., {Zhang}, Q., {et~al.} 2017, \apj, 841, 97

\bibitem[{{Schaefer} {et~al.}(2016{\natexlab{a}}){Schaefer}, {Hummel}, {Gies},
  {Zavala}, {Monnier}, {Walter}, {Turner}, {Baron}, {ten Brummelaar}, {Che},
  {Farrington}, {Kraus}, {Sturmann}, \&
  {Sturmann}}]{Schaefer2016AJ....152..213S}
{Schaefer}, G.~H., {Hummel}, C.~A., {Gies}, D.~R., {et~al.} 2016{\natexlab{a}},
  \aj, 152, 213

\bibitem[{{Schaefer} {et~al.}(2016{\natexlab{b}}){Schaefer}, {Hummel}, {Gies},
  {Zavala}, {Monnier}, {Walter}, {Turner}, {Baron}, {ten Brummelaar}, {Che},
  {Farrington}, {Kraus}, {Sturmann}, \& {Sturmann}}]{Schaefer2016}
{Schaefer}, G.~H., {Hummel}, C.~A., {Gies}, D.~R., {et~al.} 2016{\natexlab{b}},
  \aj, 152, 213

\bibitem[{{Shenar} {et~al.}(2015){Shenar}, {Oskinova}, {Hamann}, {Corcoran},
  {Moffat}, {Pablo}, {Richardson}, {Waldron}, {Huenemoerder}, {Ma{\'\i}z
  Apell{\'a}niz}, {Nichols}, {Todt}, {Naz{\'e}}, {Hoffman}, {Pollock}, \&
  {Negueruela}}]{shenar2015}
{Shenar}, T., {Oskinova}, L., {Hamann}, W.~R., {et~al.} 2015, \apj, 809, 135

\bibitem[{{Shenar} {et~al.}(2022){Shenar}, {Sana}, {Mahy}, {El-Badry},
  {Marchant}, {Langer}, {Hawcroft}, {Fabry}, {Sen}, {Almeida}, {Abdul-Masih},
  {Bodensteiner}, {Crowther}, {Gieles}, {Gromadzki}, {H{\'e}nault-Brunet},
  {Herrero}, {de Koter}, {Iwanek}, {Koz{\l}owski}, {Lennon}, {Ma{\'\i}z
  Apell{\'a}niz}, {Mr{\'o}z}, {Moffat}, {Picco}, {Pietrukowicz}, {Poleski},
  {Rybicki}, {Schneider}, {Skowron}, {Skowron}, {Soszy{\'n}ski},
  {Szyma{\'n}ski}, {Toonen}, {Udalski}, {Ulaczyk}, {Vink}, \&
  {Wrona}}]{Shenar2022NatAs...6.1085S}
{Shenar}, T., {Sana}, H., {Mahy}, L., {et~al.} 2022, Nature Astronomy, 6, 1085

\bibitem[{{Sota} {et~al.}(2014){Sota}, {Ma{\'\i}z Apell{\'a}niz}, {Morrell},
  {Barb{\'a}}, {Walborn}, {Gamen}, {Arias}, \&
  {Alfaro}}]{Sota2014ApJS..211...10S}
{Sota}, A., {Ma{\'\i}z Apell{\'a}niz}, J., {Morrell}, N.~I., {et~al.} 2014,
  \apjs, 211, 10

\bibitem[{Taniguchi {et~al.}(2015)Taniguchi, Kajisawa, Kobayashi, Shioya,
  Nagao, Capak, Aussel, Ichikawa, Murayama, Scoville, Ilbert, Salvato, Sanders,
  Mobasher, Miyazaki, Komiyama, Le~Fèvre, Tasca, Lilly, Carollo, Renzini,
  Rich, Schinnerer, Kaifu, Karoji, Arimoto, Okamura, Ohta, Shimasaku, \&
  Hayashino}]{Taniguchi2015}
Taniguchi, Y., Kajisawa, M., Kobayashi, M. A.~R., {et~al.} 2015, Publications
  of the Astronomical Society of Japan, 67, 104

\bibitem[{Thompson \& Morrison(2013)}]{Thompson_2013}
Thompson, G.~B. \& Morrison, N.~D. 2013, The Astronomical Journal, 145, 95

\bibitem[{{Tokovinin} {et~al.}(2013){Tokovinin}, {Fischer}, {Bonati},
  {Giguere}, {Moore}, {Schwab}, {Spronck}, \& {Szymkowiak}}]{Tokovinin2013}
{Tokovinin}, A., {Fischer}, D.~A., {Bonati}, M., {et~al.} 2013, \pasp, 125,
  1336

\bibitem[{{van Hamme}(1993)}]{vanHamme_1993AJ....106.2096V}
{van Hamme}, W. 1993, \aj, 106, 2096

\bibitem[{{van Leeuwen}(2007)}]{vanLeeuwen2007A&A...474..653V}
{van Leeuwen}, F. 2007, \aap, 474, 653

\bibitem[{{von Zeipel}(1924)}]{vonZeipel1924MNRAS..84..665V}
{von Zeipel}, H. 1924, \mnras, 84, 665

\bibitem[{{Wilson} {et~al.}(2010){Wilson}, {Van Hamme}, \&
  {Terrell}}]{Wilson2010ApJ...723.1469W}
{Wilson}, R.~E., {Van Hamme}, W., \& {Terrell}, D. 2010, \apj, 723, 1469

\bibitem[{{Wongwathanarat} {et~al.}(2013){Wongwathanarat}, {Janka}, \&
  {M{\"u}ller}}]{Wongwathanarat2013A&A...552A.126W}
{Wongwathanarat}, A., {Janka}, H.~T., \& {M{\"u}ller}, E. 2013, \aap, 552, A126

\end{thebibliography}

\begin{appendix}

\section{Calibrator of $\zeta$~Ori: HIP\,26108}\label{Calibrator_zeta}

The most challenging object for the calibration was $\zeta$~Ori
because the diameter of its calibrator HIP\,26108 
in the pipeline database (\texttt{GRAVI\_FAINT\_CALIBRATORS.fits})
was clearly wrong, 1.77718\,mas, and the squared visibility
was greater than one in several wavelength intervals. 
Therefore, we determined the diameter of the calibrator
based on the absolute flux from the photometric catalogues
in the \href{http://vizier.u-strasbg.fr/vizier/sed/doc/}{VizieR tool}
\citep{2014ASPC..485..219A}.
We used the standard Johnson photometric system
\citep{Ducati_2002yCat.2237....0D}
and measurements from
Hipparcos \citep{Anderson_2012AstL...38..331A},
Gaia DR3 \citep{Gaia_2020yCat.1350....0G},
2MASS \citep{Cutri_2003yCat.2246....0C},
WISE \citep{Cutri_2012yCat.2311....0C},
MSX \citep{Egan_2003yCat.5114....0E},
SkyMapper \citep{Keller2007PASA...24....1K},
Subaru/Suprime \citep{Taniguchi2015},
Akari \citep{Ishihara_2010A&A...514A...1I}, and
IRAS \citep{Neugebauer_1984ApJ...278L...1N}.
The data covered the spectral range from $0.42$ to $23.88\,\mu{\rm m}$.
In Table~\ref{SED_data}, the SED of $\varepsilon$~Ori is summarised.

\begin{figure}[h!]
\centering
\includegraphics[width=9cm]{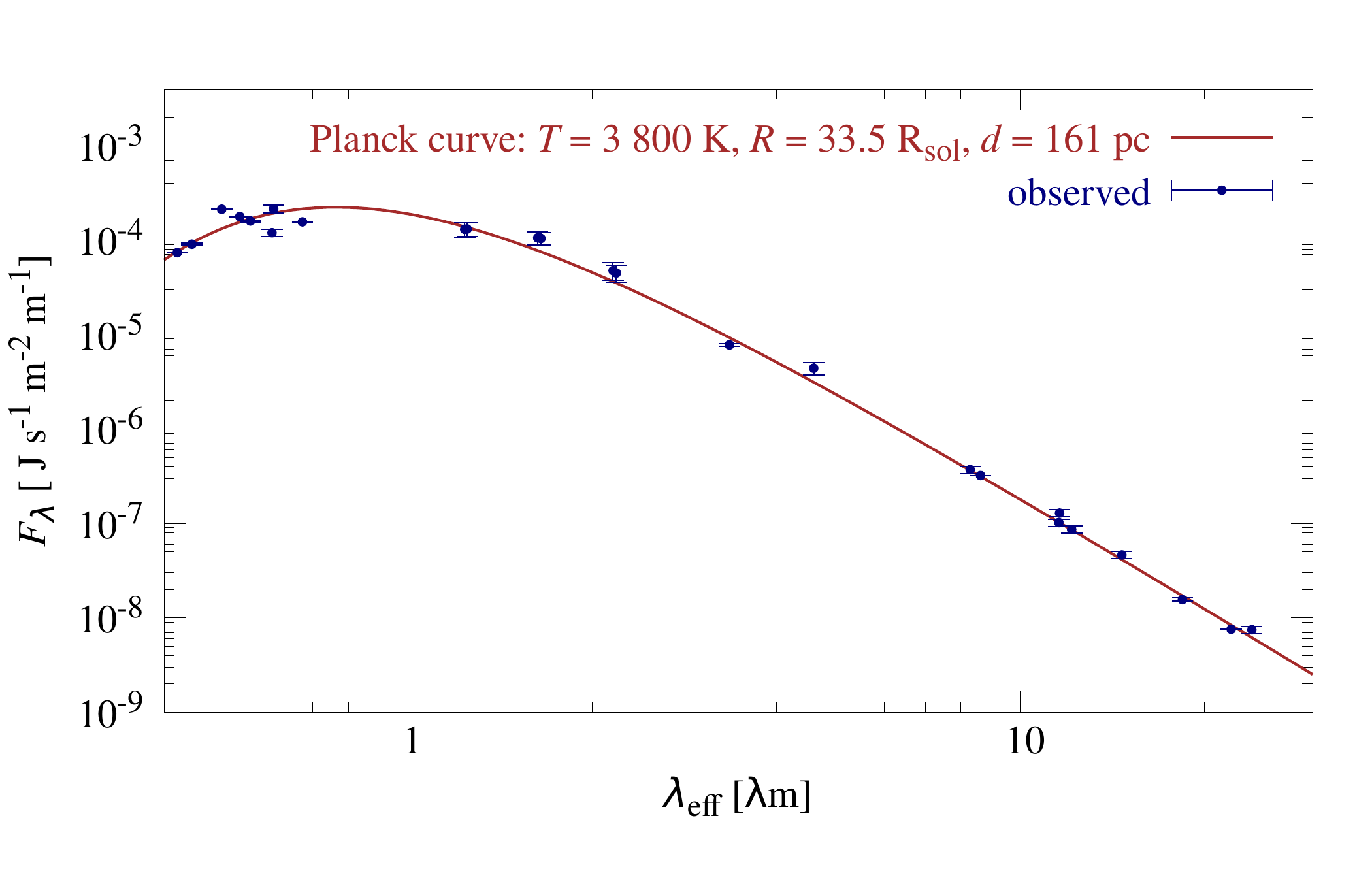}
\caption{
SED in terms of the monochromatic flux vs wavelength for the calibration star HIP 26108, which was used to determine the correct calibrator's radius $R = (33.52 \pm 0.53)$\,\Rnom\ and recalibrate the visibilities of $\zeta$~Ori.
}
\label{SED_calibrator} 
\end{figure}

For SED modelling, we used the reddening value 
$E_{B-V} = 0.014$\,mag,
i.e. extinction
$A_V = 0.0434$\,mag
\citep{Green_2019ApJ...887...93G}%
\footnote{\url{http://argonaut.skymaps.info/}},
and the parallax of
$(6.2 \pm 0.1)$\,mas
from Gaia, corresponding to
$d = (161.1 \pm 2.5)$\,pc.
Based on the spectral type K4III, 
we used the temperature $T=3\,800$\,K and $\log\,g=2.0$.
The fit with the lowest $\chi^2$ value 
(see Fig.~\ref{SED_calibrator}) led to the physical radius 
$R = (33.52 \pm 0.53)$\,\Rnom.
We thus recalibrated $\zeta$~Ori
using the calibrator's angular diameter,
$\theta = 2R/d = (1.935\pm 0.002)\,{\rm mas}$.

\begin{table}[h!]
\caption{
Absolute monochromatic fluxes for $\varepsilon$~Ori.
}
\label{SED_data}
\centering
\small
\renewcommand{\arraystretch}{1.0}
\begin{tabular*}{\hsize}{S[table-format=4.1]ccc}
\noalign{\smallskip}\hline\hline\noalign{\smallskip}
$\lambda$ \ [{\AA}] & {$F_\lambda$ [J s$^{-1}$\,m$^{-2}$\,m$^{-1}$]} & {$\sigma$ [J s$^{-1}$\,m$^{-2}$\,m$^{-1}$]}   & Filter       \\
\noalign{\smallskip}\hline\noalign{\smallskip}
3531   & \scnum{3.66609e-02}  & \scnum{3.22576e-03}  &  Johnson:U \\
3670   & \scnum{3.45552e-02}  & \scnum{7.86573e-04}  &  Johnson:U, Hvar \\
4360   & \scnum{1.92305e-02}  & \scnum{5.92874e-04}  & Johnson:B, Hvar \\
4442   & \scnum{1.85143e-02}  & \scnum{1.13899e-04}  &  Johnson:B \\
5450   & \scnum{9.07753e-03}  & \scnum{1.49252e-04}  &  Johnson:V, Hvar \\
5537   & \scnum{8.64778e-03}  & \scnum{1.36701e-04}  &  Johnson:V \\
6730   & \scnum{4.68045e-03}  & \scnum{4.11829e-04}  &  Gaia:G    \\
6938   & \scnum{4.03631e-03}  & \scnum{3.55151e-04}  &  Johnson:R \\
8780   & \scnum{1.64095e-03}  & \scnum{1.44385e-04}  &  Johnson:I \\
12390  & \scnum{4.32121e-04}  & \scnum{9.97331e-05}  &  2MASS:J   \\
12500  & \scnum{4.32235e-04}  & \scnum{9.95143e-05}  &  Johnson:J \\
16300  & \scnum{1.32848e-04}  & \scnum{1.91180e-05}  &  Johnson:H \\
16495  & \scnum{1.30820e-04}  & \scnum{1.86919e-05}  &  2MASS:H   \\
21638  & \scnum{5.47162e-05}  & \scnum{1.13176e-05}  &  2MASS:Ks  \\
21900  & \scnum{5.17698e-05}  & \scnum{1.06892e-05}  &  Johnson:K \\   
\noalign{\smallskip}\hline\noalign{\smallskip}
\end{tabular*}
\tablefoot{
$\lambda$~denotes the wavelength;
$F_\lambda$, the absolute monochromatic flux; and
$\sigma$, the uncertainty of the absolute flux.
Dereddening was applied to all measurements,
assuming $E(B-V) = 0.050\,{\rm mag}$.
}
\end{table}

\section{Radial velocities}\label{radial_velocities}

\begin{figure}[h!]
\centering
\begin{minipage}{0.49\textwidth}
    \centering
    \includegraphics[width=\linewidth]{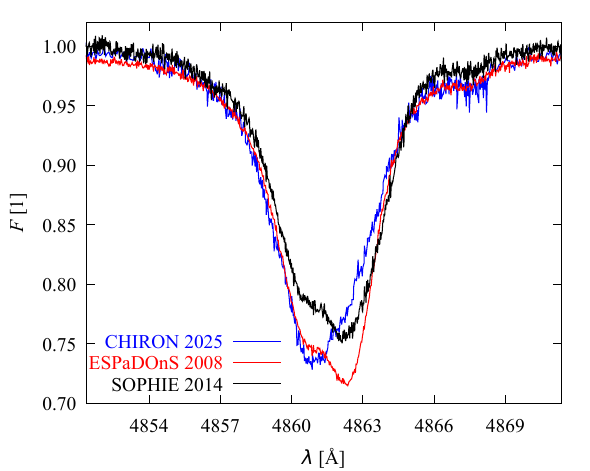}
    \caption{
         H$_\beta$ line exhibiting variable asymmetry due to partial emission filling in one of the wings.
    }
    \label{beta}
\end{minipage}
\hfill
\begin{minipage}{0.49\textwidth}
    \centering
    \includegraphics[width=\linewidth]{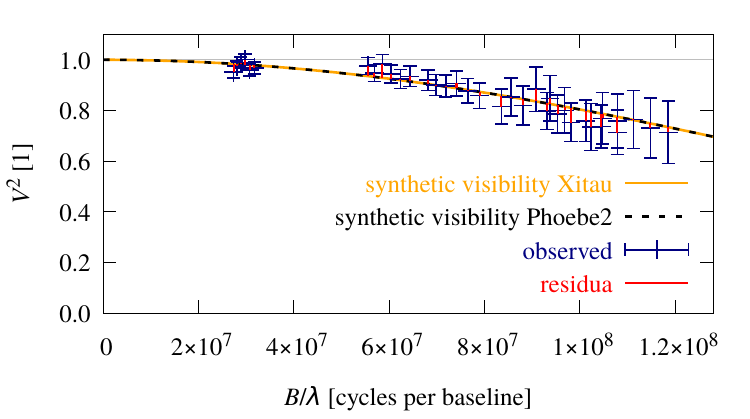}
    \caption{
Spherical non-rotating model of $\varepsilon$~Ori
based VLTI/PIONIER observations. The squared visibility vs the projected baseline $B/\lambda$
is plotted for two models,
Xitau \citep{Broz_2017ApJS..230...19B}
and Phoebe \citep{Broz_2025a}.
The resulting
$\chi^2_{\mathrm{VIS}} = 12.1$
is the same in both cases.
The number of degrees of freedom,
$\nu = N-M = 35$.
The resulting parameters are
$m = 23.52$\,\Mnom \ (free),
$d = 384$\,pc (fixed),
$T = 27\,000$\,K \citep[fixed according to][]{Puebla2016MNRAS.456.2907P},
$\log g = 3.0$ \citep[fixed according to][]{Puebla2016MNRAS.456.2907P},
$R = 25.40$\,\Rnom \ (derived),
$\theta = 0.615$\,mas (derived).
    }
    \label{eps_xitau_phoebe_VIS_12-1633_paper}
\end{minipage}
\end{figure}

\citet{Thompson_2013} measured RVs of the \ion{He}{i}~6678\,\AA\ 
lines on 132 electronic spectra from the Ritter Observatory
and found variations ranging from about $-10$ to $+20$\,\ks.
In an effort to understand the nature of possible RV changes, we measured RVs of 
various spectral lines available in the high-resolution spectra at our disposal
(CFHT, SOPHIE, and CTIO). The RVs were measured in {\tt reSPEFO}, comparing 
direct and flipped line profiles on the computer screen. The zero point of the
velocity scale was checked via measurements of a selection of telluric lines and 
also interstellar lines. The results of RV measurements are summarised 
in Table~\ref{rvind}. It is seen that the individual RVs from spectra taken
during a particular observing night agree very well, but there are undoubtedly real
RV changes from one observing season to another. We note, however, that the
RVs of the \ion{He}{ii} lines of higher ionisation show little evidence of
variability. This seems to contradict the presence of orbital
motion in a putative binary system. The observed changes can be caused by some
velocity fields in the stellar atmosphere and/or envelope.
As Fig.~\ref{beta} illustrates, H$_\beta$ line shows variable asymmetry due to partial filling of one or
another wing by the emission, which also seems to affect, to some extent,
the \ion{He}{i}\,5876 and 6678\,\AA\ lines. A more systematic study of the nature
of these changes seems desirable. At present, we conclude that 
the single-star models are preferable.

\begin{table}
\caption{Individual radial velocities (in km\,s$^{-1}$) measured in CFHT, 
SOPHIE, and CTIO electronic spectra.}
\label{rvind}
\small
\centering
\begin{tabular}{lcccSS}
\hline\hline\noalign{\smallskip}
Line & {CFHT} & {SOPHIE} & {SOPHIE} & {CTIO} & {CTIO} \\
\noalign{\smallskip}\hline\noalign{\smallskip}
\ion{He}{i} 6678\,\AA   & \text{--}  & 26.8 & 25.7 & 9.9  & 10.5\\ 
\ion{He}{i} 5876\,\AA   & 21.6 & 19.7 & 19.6 &  5.4 &  5.4\\
\ion{He}{i} 5015\,\AA   & 22.6 & 29.6 & 30.1 & 17.7 & 17.5\\
\ion{He}{i} 4922\,\AA   & 22.2 & 27.1 & 27.2 & 15.8 & 15.5\\
\ion{He}{i} 4713\,\AA   & 21.8 & 26.7 & 25.8 & 16.0 & 16.4\\
\ion{Si}{iii} 4574\,\AA & 16.6 & 25.3 & 23.8 & 13.9 & 13.4\\
\ion{Si}{iii} 4568\,\AA & 17.9 & 25.4 & 26.2 & 12.6 & 12.6\\
\ion{Si}{iii} 4552\,\AA & 20.5 & 28.4 & 27.1 & 12.4 & 13.8\\
\ion{He}{ii} 5411\,\AA  & 29.4 & 30.7 & 30.7 & 26.2 & 24.1\\
\ion{He}{ii} 4686\,\AA  & 42.3 & 44.7 & 45.4 & 35.2 & 37.6\\
H$_\beta$               & 55.1 & 50.1 & 50.5 & -22.4 & -24.3 \\
H$_\gamma$              & 34.5 & 35.8 & 33.6 & \text{--} & \text{--}\\
H$_\delta$              & 31.5 & 33.1 & 33.4 & \text{--} & \text{--}\\
H$_\varepsilon$         & 26.3 & 28.8 & 28.9 & \text{--} & \text{--}\\
H8                      & 25.9 & \text{--}   & \text{--}   & \text{--} & \text{--}\\
H9                      & 24.9 & \text{--}   & \text{--}   & \text{--} & \text{--}\\
\noalign{\smallskip}\hline\noalign{\smallskip}
\end{tabular}
\tablefoot{
Julian dates (UTC) of the observations were
2454754.1558,
2457015.4119,
2457015.4139,
2460637.7725,
and 2460637.7732.
}
\end{table}

\onecolumn
\section{Supplementary figures}\label{supplementary_figures}

\begin{figure*}[h!]
\centering
\includegraphics[width=0.49\textwidth]{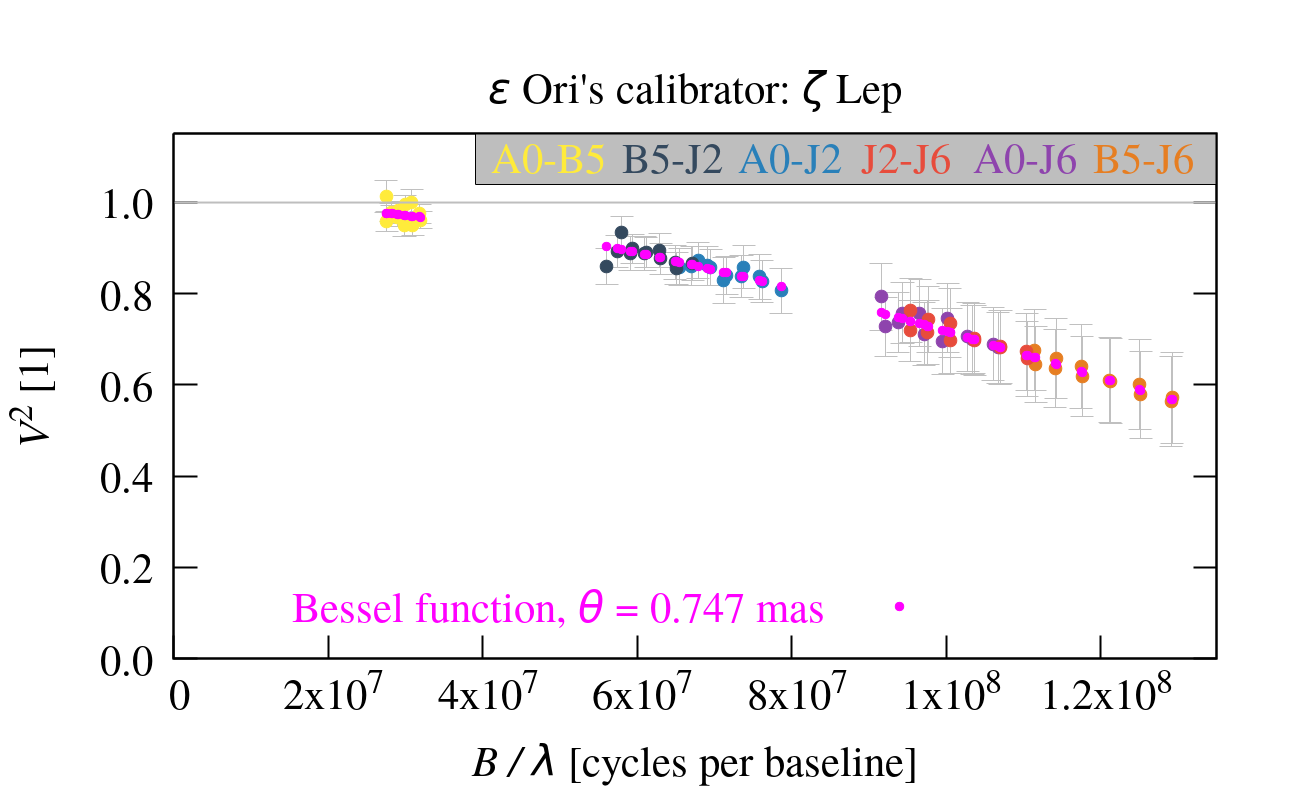}
\includegraphics[width=0.49\textwidth]{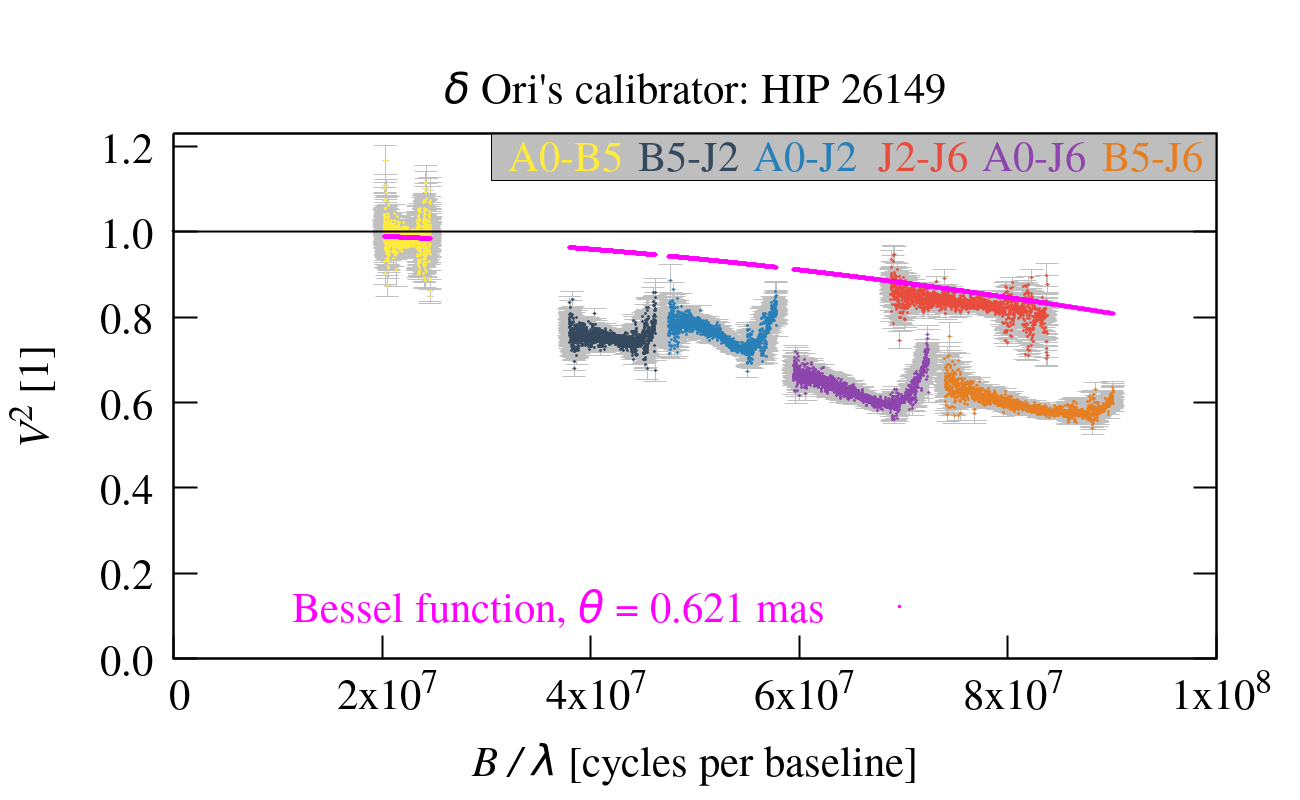}
\includegraphics[width=0.49\textwidth]{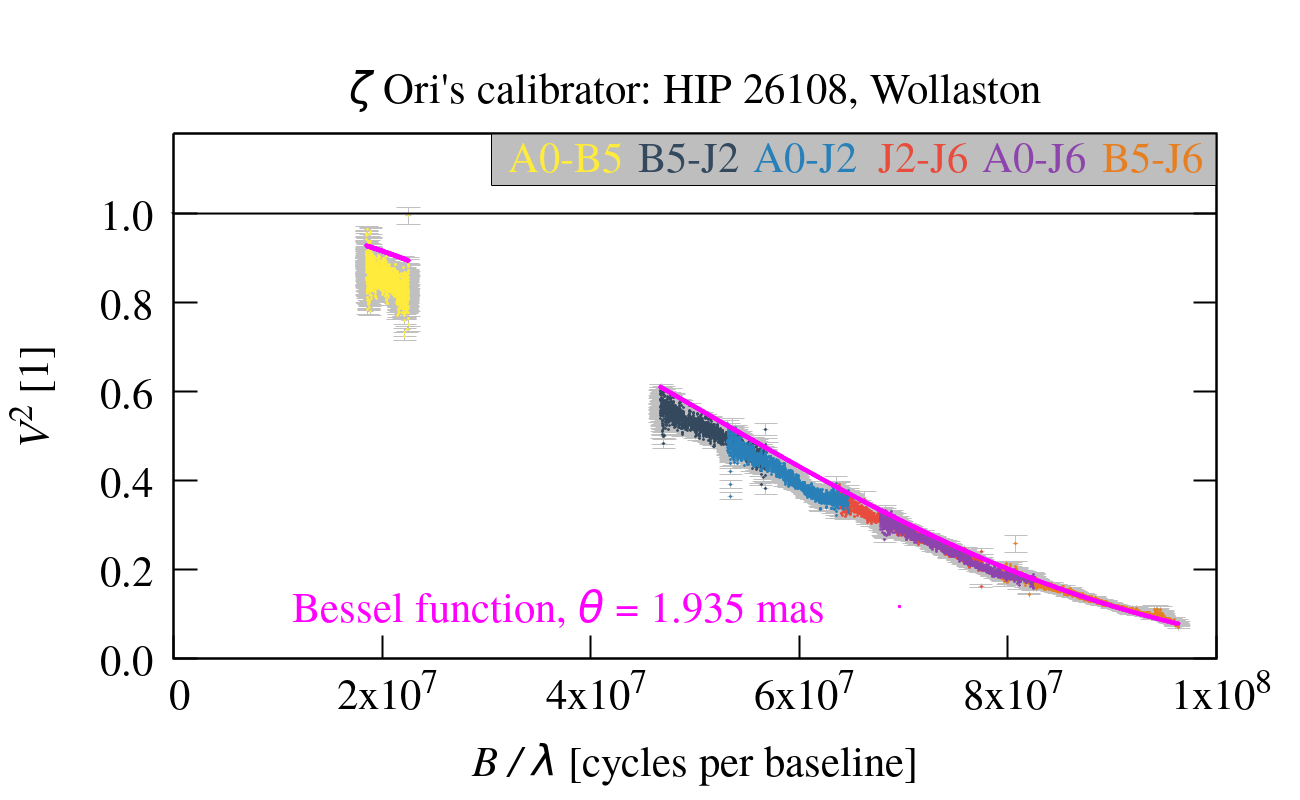}
\includegraphics[width=0.49\textwidth]{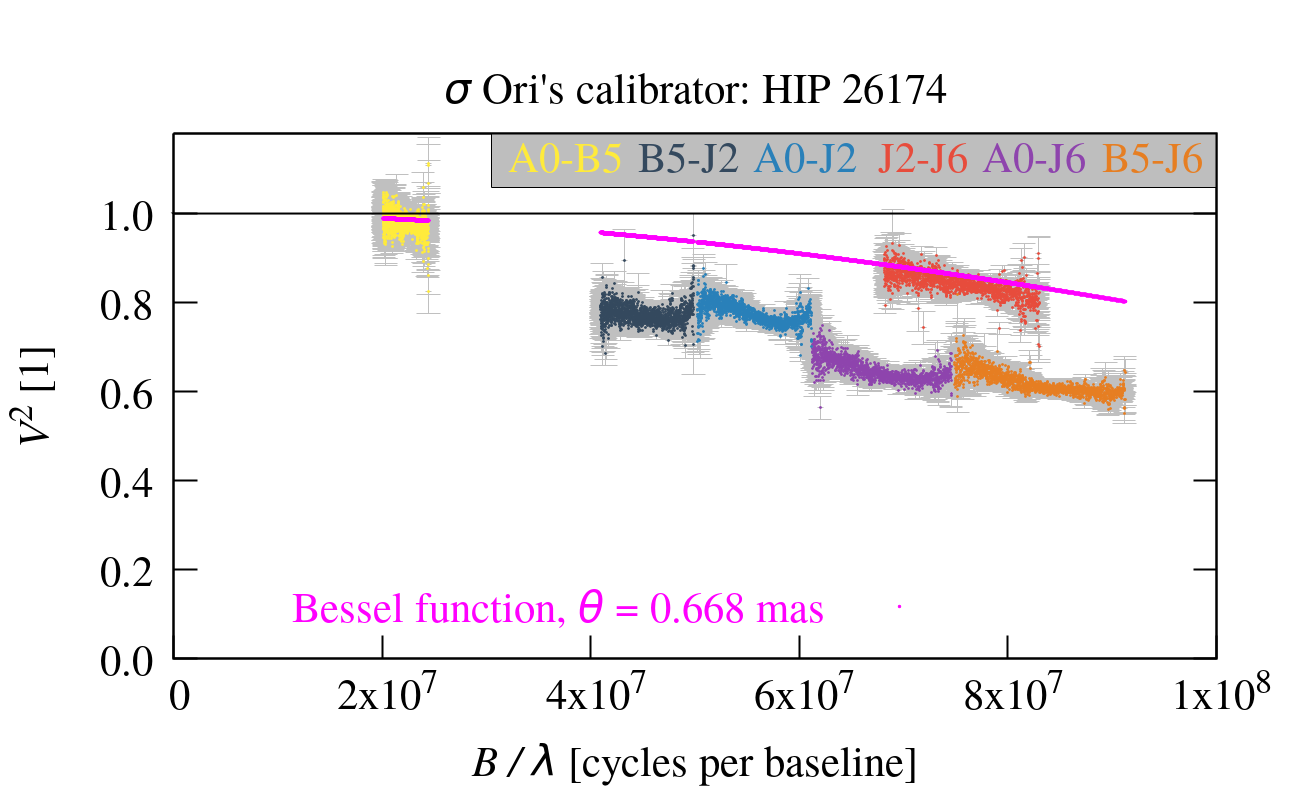}
\caption{
Examples of raw visibilities of the calibrators for
$\varepsilon$, $\delta$, $\zeta$, and $\sigma$~Ori
and the corresponding Bessel functions with the calibrator's diameters
plotted for reference (magenta).
Measurements are from nights: 
20 November 2023, 
22 November 2023,
8 January 2024,
and 23 November 2023, respectively.
Colours correspond to individual baselines.
}
\label{reduced_visibility_cal}
\end{figure*}

\begin{figure*}[h!]
\centering
\label{He}
\includegraphics[width=0.975\textwidth]{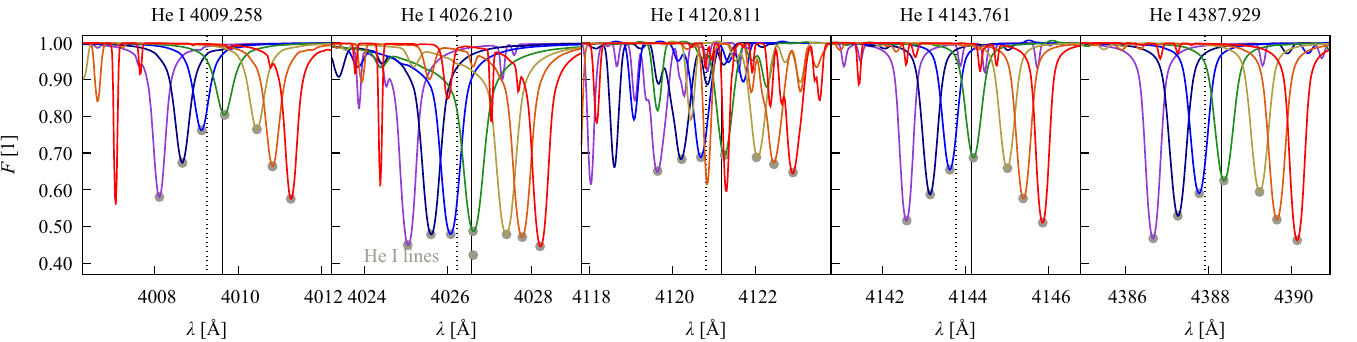}
\includegraphics[width=0.975\textwidth]{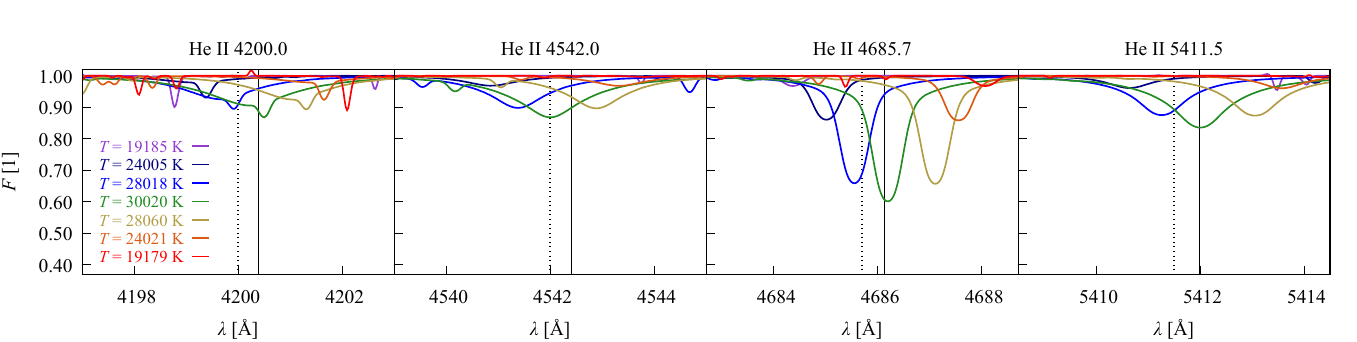}
\caption{
Synthetic \ion{He}{I} and \ion{He}{II} line profiles from our spectroscopic model,
corresponding to selected triangles with given temperatures and RVs.
They demonstrate where lines form in our model.
For a fast-rotating star,
the pole is substantially hotter than the equator.
The central line (${\rm RV} = 0$) corresponds to the pole,
other lines originate from the equator (${\rm RV}\ne 0$).
\ion{He}{I} and \ion{He}{II} lines form at different locations
on a fast or critically rotating star
due to strong temperature gradients.
\ion{He}{I} lines have the lowest intensity on the poles,
while \ion{He}{II} lines have the highest intensity at the poles.
\ion{He}{I} lines are less intense at the pole,
while \ion{He}{II} lines are less intense on the equator.
The grey points indicate the \ion{He}{I} lines for clarity.
}
\label{HeI_lines_plot}
\end{figure*}

\begin{figure}[h!]
\centering
\includegraphics[width=0.975\textwidth]{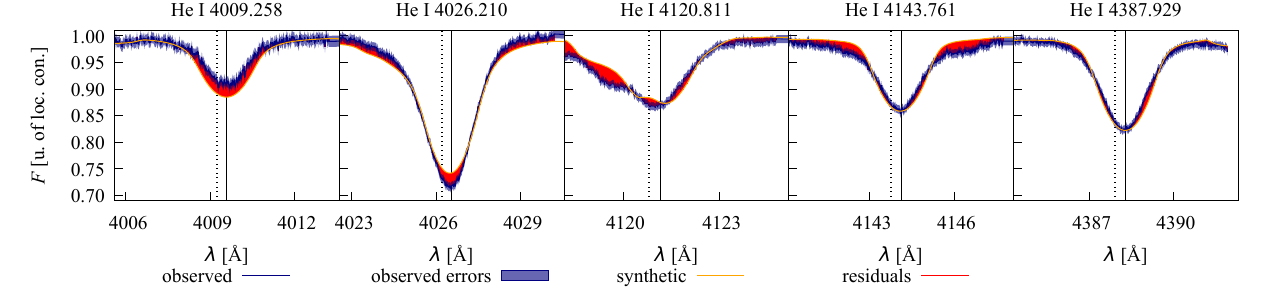}
\caption{
Same as Fig.~\ref{spe}, but for \ion{He}{I} lines.
The best-fit spectroscopic model still exhibits systematics
(in particular, 4009 and 4026\,\AA\ lines).
None of these are easy to correct, e.g. by changing $T$, $\log g$, or metallicity; see also \cite{Puebla2016}.}
\label{HeI_fit} 
\end{figure}

\begin{figure}[h!]
\centering
\begin{tabular}{@{}c@{}c@{}c@{}}
\includegraphics[width=0.33\textwidth]{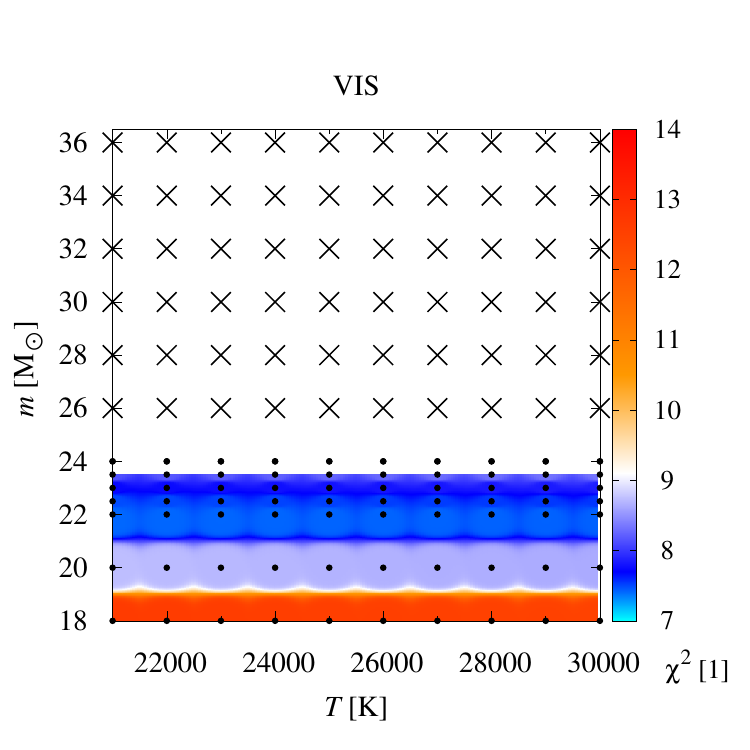} &
\includegraphics[width=0.33\textwidth]{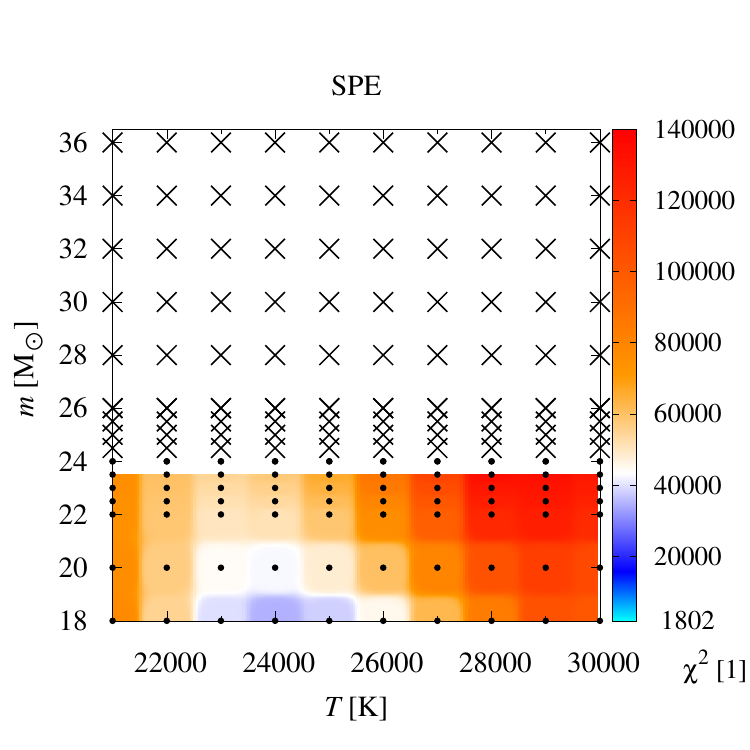} &
\includegraphics[width=0.33\textwidth]{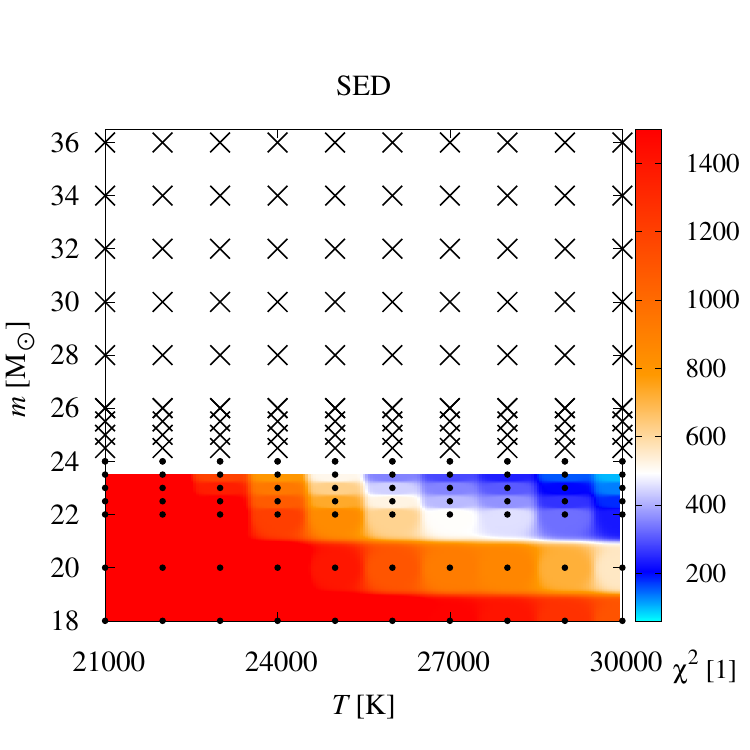} 
\\
\end{tabular}
\caption{Same as the second row of Fig.~\ref{maps}, but for the distance of 350\,pc, computed on a somewhat coarser grid. Blank regions mark combinations where $R > R_\mathrm{crit}$; crosses indicate grid points exceeding this limit. PHOEBE2 model also fails very near the critical boundary, which we explored with finer grids.
}
\label{maps350} 
\end{figure}

\begin{figure}[h!]
\centering
\begin{tabular}{@{}c@{}c@{}c@{}}
\includegraphics[width=0.33\textwidth]{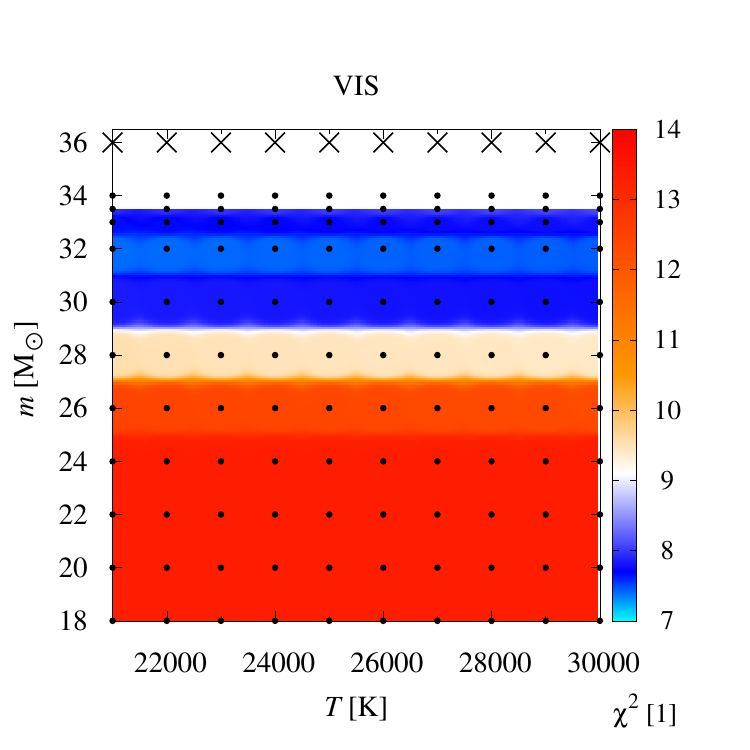} &
\includegraphics[width=0.33\textwidth]{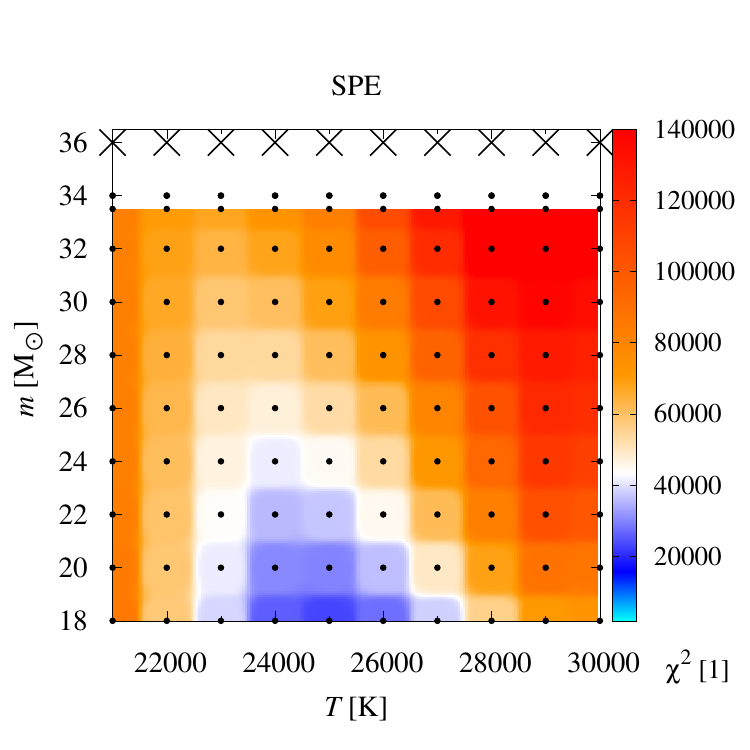} &
\includegraphics[width=0.33\textwidth]{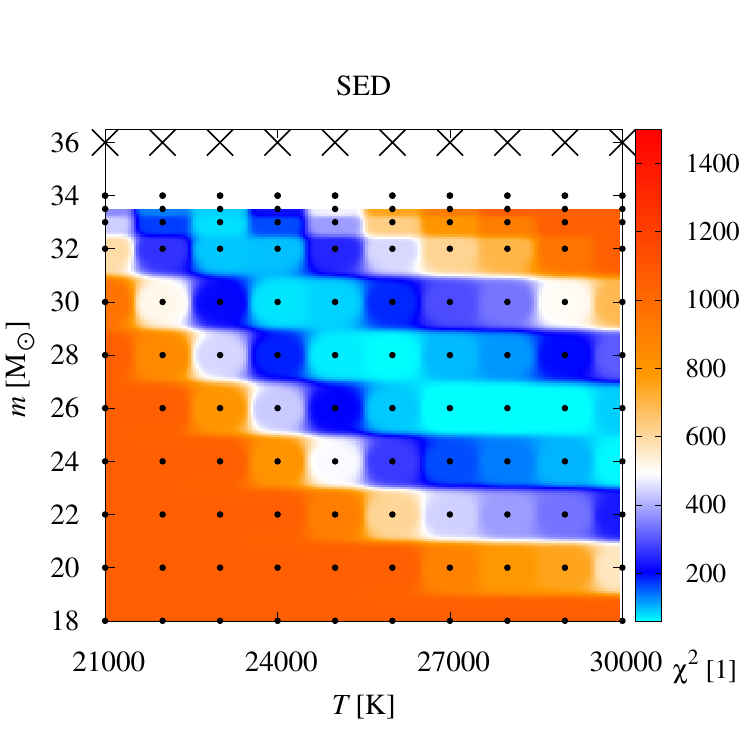} 
\\
\end{tabular}
\caption{As Fig.~\ref{maps350} but for the distance of 420\,pc.
}
\label{maps420} 
\end{figure}

\begin{figure}[h!]
\centering
\includegraphics[width=0.49\textwidth]{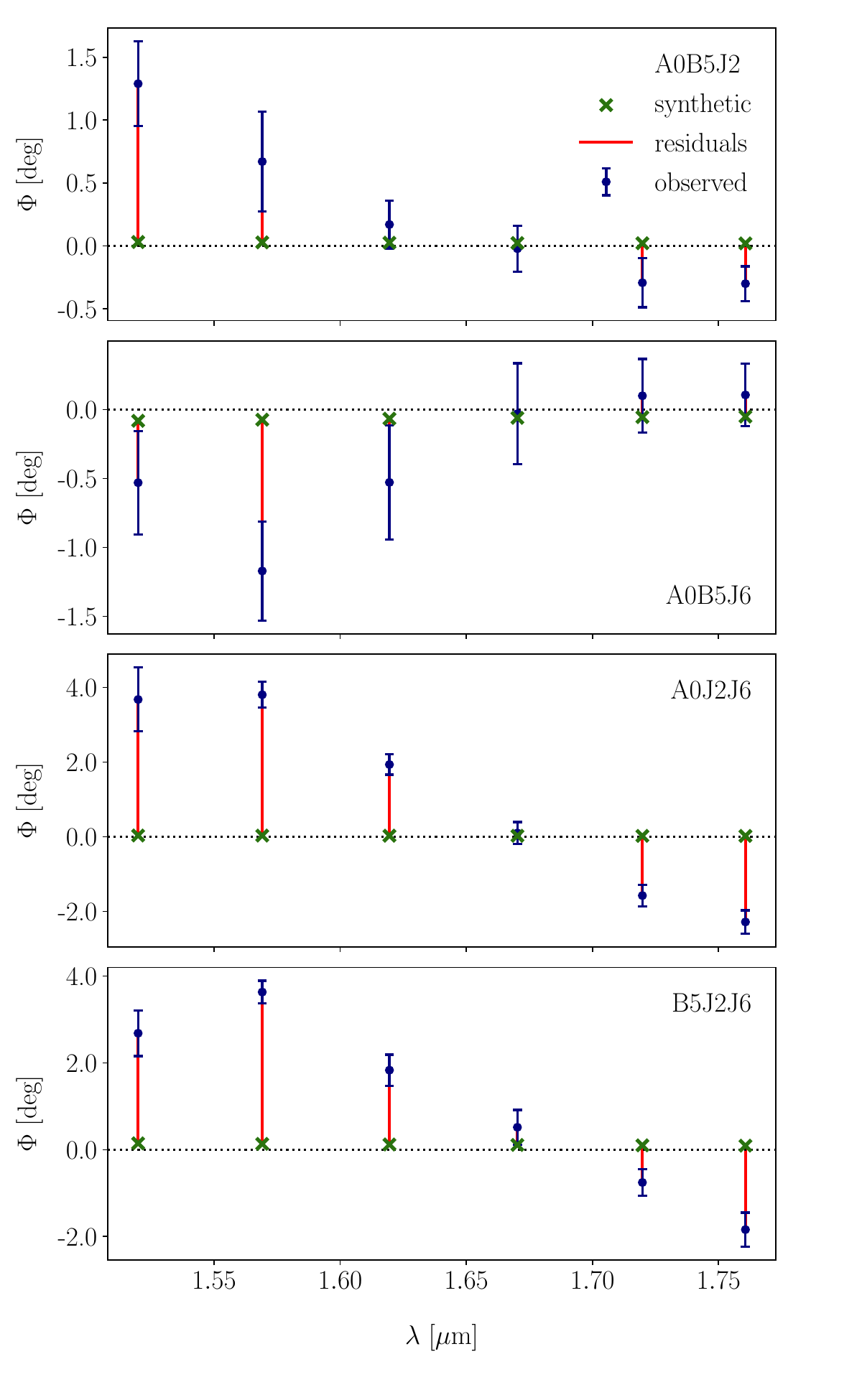}
\includegraphics[width=0.49\textwidth]{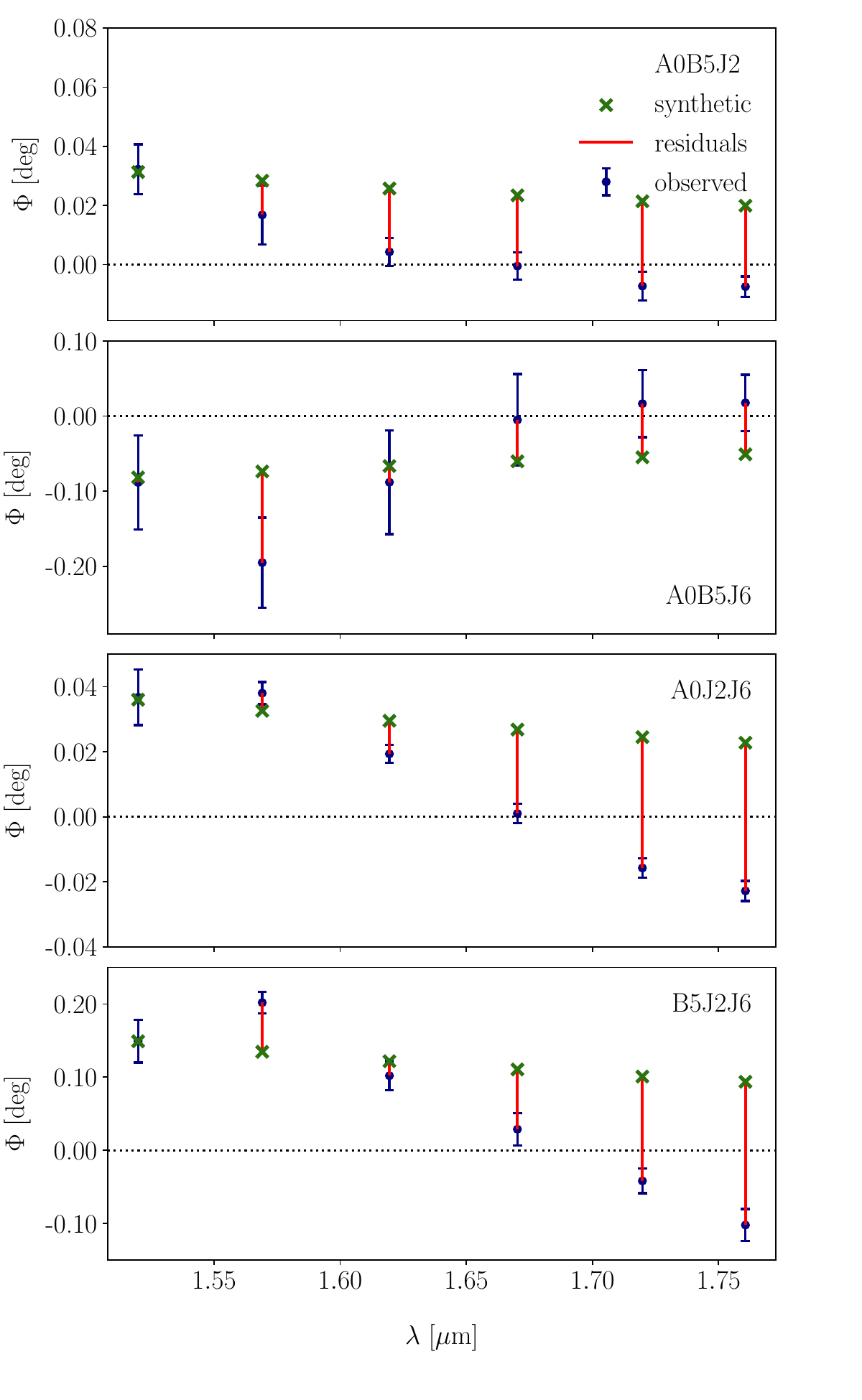}
\caption{
Model of $\varepsilon$~Ori showing a forward computation of the closure phases 
with fixed parameters from the visibility model (Fig.~\ref{eps_best_fit_free_vsini}).
The observed closure phases reach up to nearly 4\,deg,
while the synthetic ones are at most 0.15\,deg.
If we recalibrate the observed closure phases (left column)
by multiplying them by scaling factors 1/40, 1/6, 1/100, and 1/18, respectively for each of the triple baselines (right column),
we see similar trends in the observed and synthetic closure phases,
${\rm arg}\,T_3$ vs $\vec{B}/\lambda$.
We used this asymmetry of the flux to set 
the orientation of the star's equator, as seen by the observer, $\Omega$
(favouring 300\,deg over 120\,deg).
}
\label{CLO} 
\end{figure}

\end{appendix}
\end{document}